\documentclass[12pt]{article}
\usepackage{amsmath}
\usepackage{mathrsfs}
\usepackage{enumerate}
\begin{document}
\def\be{\begin{equation}}
\def\bea{\begin{eqnarray}}
\def\ee{\end{equation}}
\def\eea{\end{eqnarray}}
\def\d{\partial}
\def\eps{\varepsilon}
\def\la{\lambda}
\def\b{\bigskip}
\def\nn{\nonumber \\}
\def\p{\partial}
\def\t{\tilde}
\def\h{{1\over 2}}
\def\be{\begin{equation}}
\def\bea{\begin{eqnarray}}
\def\ee{\end{equation}}
\def\eea{\end{eqnarray}}
\def\b{\bigskip}
\def\u{\uparrow}
\def\AA{\mathscr{A}}
\def\DD{\mathscr{D}}
\def\FF{\mathscr{F}}
\def\LL{\mathscr{L}}
\def\CC{\mathscr{C}}
\def\MM{\mathscr{M}}
\newcommand{\comment}[2]{#2}

\setcounter{tocdepth}{2}

\makeatletter
\def\blfootnote{\xdef\@thefnmark{}\@footnotetext}  
\makeatother

\begin{center}
{\LARGE Maxwell's Equations in the Myers--Perry Geometry}
\\
\vspace{18mm}
{\bf   Oleg Lunin}
\vspace{14mm}

Department of Physics,\\ University at Albany (SUNY),\\ Albany, NY 12222, USA\\ 

\vskip 10 mm

\blfootnote{email: olunin@albany.edu}

\end{center}

\begin{abstract}

We demonstrate separability of the Maxwell's equations in the Myers--Perry--(A)dS geometry and derive explicit solutions for various polarizations. Application of our construction to the four--dimensional Kerr black hole leads to a new ansatz for the Maxwell field which has significant advantages over the previously known parameterization.

\b

\end{abstract}

\newpage

\addtocontents{toc}{\vskip-5pt}
\addtocontents{toc}{\protect\enlargethispage{\baselineskip}}
{
\small
\tableofcontents
\addtocontents{toc}{\vskip-5pt}
}

%

\newpage

\section{Introduction and summary}

Black holes are important laboratories for studying quantum gravity. The usefulness of these objects comes from combination of the rich phenomena associated with them and a relative simplicity of the underlying geometry. For example, dynamics of various fields in the background of the Schwarzschild black hole is fully understood, and yet it leads to the Hawking radiation \cite{Hawking} and to the black hole information paradox, one of the major challenges facing theoretical physics. Scattering from the higher--dimensional counterparts of the Schwarzschild black hole and from D--branes has also been studied in detail \cite{PreAdSCFT}, and important lessons extracted from such investigations inspired the formulation of the AdS/CFT correspondence \cite{AdSCFT}. Static black holes built from the D--branes have played the crucial role in the microscopic explanation of the Bekenstein entropy \cite{StrVafa}, a major step towards resolving the information paradox.

The rotating black holes are also relatively simple, but study of their dynamical properties is not as straightforward as in the static case. While the Schwarzschild geometry and its higher--dimensional generalizations have sufficient numbers of isometries to guarantee the full separation of variables in equations for all dynamical fields, this not the case even for the four--dimensional Kerr metric \cite{Kerr}. The problem can be seen already at the level of probe particles: the $U(1)\times U(1)$ isometry of the Kerr solution leads to two conserved quantities (energy and angular momentum of the probe), which are not sufficient to fully characterize motion in a four--dimensional space. However, particles in the Kerr geometry posses a third conserved quantity which cannot be attributed to a Noether charge associated with isometries \cite{Carter1}. All such Noether charges come from Killing vectors of the background geometry, while Carter's integral of motion is associated with an irreducible rank-two Killing 
tensor\footnote{The Kerr metric also admits several reducible Killing tensors constructed as products of Killing vectors.} \cite{Carter1,Carter2,Penrose}. This tensor also ensures full separation of variables in the Klein--Gordon equation. 

Since separation of variables in the Schwarzschild geometry follows from isometries, the decomposition into spherical harmonics 
persists for fields of arbitrary spin. In contrast, for the Kerr geometry, separation of variables for various fields has been worked out 
only on a case--by--case basis. For spinors, such separation of the Dirac equation follows from the existence of an anti--symmetric Killing--Yano tensor \cite{Yano,GRyano}. For the electromagnetic field and for gravitational waves, the separation was demonstrated in the classic work by Teukolsky \cite{Teuk}, which did not rely on the Killing(--Yano) tensors encoding the hidden symmetries of the Kerr geometry. The main goal of this article is the construction of separable electromagnetic waves in the higher--dimensional generalizations of the Kerr geometry, and as a byproduct of our general approach, we will clarify the relation between the 
Teukolsky's ansatz and the Killing(--Yano) tensors of the Kerr metric. We will also propose a modified ansatz for the gauge field, which leads to full separation of Maxwell's equations in four dimensions, even beyond the single polarization discussed in \cite{Teuk}.

The main motivation for studying black holes in higher dimensions comes from string theory, which has been very successful in counting microscopic states \cite{StrVafa,BMPV}, computing scattering amplitudes \cite{PreAdSCFT}, and getting insights into various quantum properties of black holes \cite{5dBH,fuzz} in $D>4$. For the Schwarzschild--Tangherlini geometry \cite{Tang}, separation of the 
Klein--Gordon and Maxwell equations is rather straightforward\footnote{We briefly discuss Maxwell's equations for such space in the Appendix \ref{SecAppSchw}.}, but the study of gravitational waves in such backgrounds is an active area of research \cite{5DGravWave}.   In contrast, the vast majority of efforts in studying rotating geometries has been dedicated to scalar and spinor fields \cite{PreKub,Kub1,Kub2,Kub3} with a few notable exceptions \cite{MaxwHighD}. Unfortunately, in this case, the full description of the 
Maxwell's equations, let alone gravitational waves, has been missing, and the goal of the present article is to close this gap in the literature on electromagnetic waves. A better understanding of the black hole excitations will give new handles on probing these fascinating objects. 

The approach pursued in this article is based on utilizing hidden symmetries of rotating black holes in higher dimensions. Such symmetries encoded in the Killing--(Yano) tensors have been explored in the past, both in higher--dimensional general relativity \cite{PreKub,Kub1,Kub2,Kub3} and in string theory \cite{CveticLarsen,KYTbh,StrYano,ChLKill}. This paper will connect the properties of such tensors, in particular, their eigenvectors, to separation of the Maxwell's equations in an arbitrary number of dimensions. While most of this article focuses on electromagnetic waves in the background of the Myers--Perry black hole \cite{MyersPerry}, the extensions of our results to the GLPP geometry \cite{GLPP}, which generalizes MP solution to non--zero value of the cosmological constant, is straightforward, and such extensions are presented in the appropriate sections. The scalar and spinor excitations of the GLPP black holes have been subjects of intensive studies \cite{PreKub,Kub1,Kub2,Kub3}\footnote{See also \cite{GLPPflwUp} for a general discussion of the GLPP black holes and their properties.}, and the results are summarized in a very nice recent review \cite{KubRev}. Under very mild assumptions, the GLPP solution with added NUT charge is the most general geometry admitting separation of variables in the wave and Dirac equations \cite{UniqueKY}, and the same uniqueness property is expected to hold for the Maxwell's equations discussed in this article. 

\bigskip

This paper has the following organization. In section \ref{SecKerr} we review Teukolsky's classic construction of electromagnetic field in the background of the Kerr black hole and use these results as an inspiration for formulating a new ansatz suitable for generalizations to higher dimensions. Teukolsky's  equations \cite{Teuk} are the necessary conditions for a specific ansatz to solve Maxwell's equations, and they appear to describe two polarizations. However, it turns out that one of such polarizations is completely determined in terms of the other (see \cite{ChandraBook} for a detailed discussion of this point), so only one configuration is independent. The second polarization of the electromagnetic field is governed by a scalar function that satisfies a non--separable partial differential equation. While Teukolsky's equations are extremely useful for getting insights into the dynamics of electromagnetic fields, it might be desirable to describe both polarizations of photons by separable equations, and we accomplish this goal in section \ref{SecKerr} by proposing a new ansatz for the gauge potential in the Kerr geometry. 

The main results of section \ref{SecKerr} are the reformulation (\ref{KerrMaxwFrames}) of the Teukolsky's construction and derivation of the new solution for the electromagnetic field (\ref{KerrNewAnstz}), (\ref{4dHighSpinGuess})--(\ref{4dSmryAnstz}) inspired by it. Specifically, we use the form (\ref{KerrMaxwFrames}) to motivate a new ansatz (\ref{KerrNewAnstz}) for the vector potential and derive the {\it most general} solutions of the Maxwell's equations consistent with such ansatz. All configurations naturally split into two classes, which we label as ``electric'' and ``magnetic'' polarizations, even though generically all fields are excited in both cases. Such separation into two distinct classes persists in all dimensions. The resulting configurations of the electromagnetic fields, which cover {\it all photon polarizations}, are summarized in section \ref{SectSubKerrWave}, where we also discuss a close similarity between separable solutions of the Maxwell's and Klein--Gordon equations. These observations might help in applying the constructions presented in this article to gravitational waves. We conclude the discussion of the four--dimensional geometries by generalizing all results to the Kerr--(A)dS black holes in subsection 
\ref{SecKerr4DAdS}.

\bigskip

Sections \ref{SecMyersPerry}-\ref{SecWaveMP} of this article are dedicated to extending the success in separating the Maxwell's equations to higher dimensions. We begin with reviewing rotating black holes and their symmetries in section \ref{SecMyersPerry}. In particular, we stress the importance of the Killing(--Yano) tensors and their eigenvectors, which play crucial role in the subsequent constructions. Since both the Myers--Perry \cite{MyersPerry} and the GLPP \cite{GLPP} black holes have different structures in even and odd dimensions, the four--dimensional pattern discussed in section \ref{SecKerr} is applicable only to half of the cases. To have a sample for the other half, we dedicate section \ref{SecKerr5D} to the detailed discussion of the Maxwell's equations in five dimensional geometries, and we explicitly construct {\it all three polarizations of a photon}. Apart from serving as a starting point for describing electromagnetic waves in odd dimensions, the five dimensional black hole is important for the role it has played in addressing the black hole information paradox \cite{StrVafa,BMPV,5dBH,fuzz}, so the results of section \ref{SecKerr5D} might be very useful in their own right. 

The final section of this paper uses insights from four and five dimensions to propose a separable ansatz for electromagnetic field in the Myers--Perry geometry and to derive the resulting equations 
(\ref{EvenDimGenSpin})--(\ref{EvenDimGenSpinLmb}), (\ref{GenOddElectr})--(\ref{OddDimGenSpinLmb}). As demonstrated in subsection \ref{SecSubCmprSchw}, these systems describe $(D-2)$ independent polarizations of a photon in $D$ dimensions, so any electromagnetic wave can be approximated by a linear combination of separable solutions constructed in this paper. The minor modifications in the gauge field caused by the cosmological constant in GLPP geometry are discussed in section \ref{SecKerrAdSMxwMP}. As in four and five dimensions, there is a remarkable similarity between equations governing dynamics of photons and scalar particles, so perhaps the structures discussed in this article can accommodate gravitational waves as well, as it happened in the case of the Teukolsky's system.

\section{All excitations of the Kerr black hole}
\renewcommand{\theequation}{2.\arabic{equation}}
\setcounter{equation}{0}
\label{SecKerr} 

Before analyzing rotating black holes in higher dimensions, we will review the well-known facts about solutions of the Maxwell's equations in the background of the Kerr black hole \cite{Kerr} and reformulate them in a form suitable for making generalizations. The main insight into electromagnetic waves in the Kerr geometry was gained in 1972, when Teukolsky demonstrated separability of Maxwell's equations \cite{Teuk}, and the full expressions for the gauge fields were obtained soon after this remarkable discovery \cite{ChandPaper}. While reviewing the Teukolsky's ansatz in section \ref{SecKerrReview}, we will see that some details of this construction crucially rely on the number of dimensions, so in section \ref{SecKerrNewAns} we will rewrite the Maxwell fields in terms of new variables, which allow extension to the general Myers--Perry black holes. As an added bonus, the new ansatz leads to full separation of the Maxwell's equations, while in the past this was demonstrated only for one polarization. Moreover, the new variables give simpler expressions for the gauge fields by removing various constraints, which were implicit in Teukolsky's construction. 

We begin with reviewing Teukolsky's classic construction in section \ref{SecKerrReview} and observing an interesting separation pattern for the gauge potentials (\ref{KerrMaxwFrames}), which has not been discussed in the literature. Using this feature as an inspiration, in section \ref{SecKerrNewAns} we formulate a new ansatz 
(\ref{KerrNewAnstz}) for the gauge field and find the most general configurations fitting such separation of variables. In section \ref{SecKerrCompare} the answers are compared with known results, in particular, we find that the electromagnetic fields and scalar excitations can be described in a unified fashion by a system 
(\ref{4dHighSpinGuess})--(\ref{4DfuncD}). This suggests that gravitational fields might follow the same pattern, although a detailed discussion of gravitational waves is beyond the scope of this article. Finally, section \ref{SecKerr4DAdS} extends the results to the Kerr-(A)dS black hole.

\subsection{Review of the known results}
\label{SecKerrReview}

In this subsection we review the Teukolsky's construction following the original articles \cite{Teuk} and a nice pedagogical exposition presented in \cite{ChandraBook}. Study of electromagnetic and gravitational 
waves in four dimensions has been mostly carried out using the Newman--Penrose (NP) formalism \cite{NPform}, in particular, applying this framework to electromagnetic fields, Teukolsky discovered separation of variables in a large class of backgrounds \cite{Teuk}. Thus we begin with recalling the description of electromagnetic field in the NP formalism, then we review Teukolsky's construction for the Kerr geometry, rewrite it in a more transparent form, and use the result as an inspiration for a better ansatz introduced in subsection \ref{SecKerrNewAns}. The technical details are presented in the Appendix  \ref{SecAppTeuk}. 

\bigskip

In this section we will mostly focus on electromagnetic fields in the background of the Kerr black hole, \cite{Kerr}
\bea\label{Kerr1}
ds^2=\frac{1}{\Sigma}\Big\{-\Delta [dt-as_\theta^2 d\phi]^2+s_\theta^2[(r^2+a^2)d\phi-a dt]^2\Big\}+
\Sigma\left[\frac{dr^2}{\Delta}+d\theta^2\right]\,,
\eea
and the extension to AdS-Kerr space will be briefly discussed in subsection \ref{SecKerr4DAdS}. Functions $\Delta$ and $\Sigma$ are defined by 
\bea
\Delta=r^2+a^2-2Mr,\quad \Sigma=r^2+(a c_\theta)^2. 
\eea
While separation of variables in the Klein--Gordon equation in the geometry (\ref{Kerr1}) is rather straightforward, the very meaning of separation for the fields with higher spin requires a clarification: which degrees of freedom should separate? For Maxwell's equations in the geometry (\ref{Kerr1}) the answer to this question was discovered by Teukolsky \cite{Teuk}, who showed that the relevant variables are the components of the field strength used in the Newman--Penrose formalism \cite{NPform}. 

In this formalism one begins with defining a vierbein of null vectors $({\bf l},{\bf n},{\bf m},\bar{\bf m})$ \cite{NPform}. All products of these four vectors vanish, with two exceptions:
\bea
l_\mu n^\mu=-1,\quad m_\mu{\bar m}^\mu=1.
\eea
The metric is written as
\bea\label{gMuNuKerr}
g_{\mu\nu}=-l_\mu n_\nu-l_\mu n_\nu+m_\mu{\bar m}_\nu+m_\nu{\bar m}_\mu,
\eea
and all dynamical fields are expanded in the $({\bf l},{\bf n},{\bf m},\bar{\bf m})$ basis. For example, the six components of the electromagnetic field tensor are encoded in three complex scalars $(\phi_0,\phi_1,\phi_2)$ as 
\bea\label{StrenFramesNP}
F_{\mu\nu}=2\left[\phi_1(n_{[\mu}l_{\nu]}+m_{[\mu}{\bar m}_{\nu]})+\phi_2l_{[\mu}m_{\nu]}+
\phi_0 {\bar m}_{[\mu}n_{\nu]}\right]+cc.
\eea
To clarify the physical meaning of the functions  $(\phi_0,\phi_1,\phi_2)$, we look at the combinations of $F$ and its dual:
\bea
F+i\star F&=&4\left[{\phi}_2l_{[\mu}{\bar m}_{\nu]}+
\phi_0 {\bar m}_{[\mu}n_{\nu]}\right]+
4{\phi}_1\left[n_{[\mu}l_{\nu]}-m_{[\mu}{\bar m}_{\nu]}\right],\nonumber\\
F-i\star F&=&4\left[{\bar\phi}_2l_{[\mu}{m}_{\nu]}+
{\bar\phi}_0 {m}_{[\mu}n_{\nu]}\right]+
4{\bar\phi}_1\left[n_{[\mu}l_{\nu]}-m_{[\mu}{\bar m}_{\nu]}\right].\nonumber
\eea
Thus complex functions $(\phi_0,{\phi}_1,{\phi}_2)$ describe the imaginary--self--dual part of the field strength, while $({\bar\phi}_0,{\bar\phi}_1,{\bar\phi}_2)$ parameterize the anti--self--dual part. Note that this construction is specific to four dimensions, where forms $F$ and $\star F$ have the same rank.

As demonstrated by Teukolsky, variables $\phi_0$ and $\phi_2$ decouple in Maxwell's equations for any type D vacuum metric, moreover, the resulting scalar PDEs admit separation of variables for the Kerr geometry if one chooses the ``canonical tetrad'':
\bea\label{FramesKerr}
&&\hskip -0.5cm l^\mu\d_\mu=\frac{r^2+a^2}{\Delta}\d_t+\d_r+\frac{a}{\Delta}\d_\phi,\quad
n^\mu\d_\mu=\frac{r^2+a^2}{2\Sigma}\d_t-\frac{\Delta}{2\Sigma}\d_r+\frac{a}{2\Sigma}\d_\phi,\\
&&\hskip -0.5cm m^\mu\d_\mu=\frac{1}{\sqrt{2}\rho}\left[{ia s_\theta}\d_t+\d_\theta+\frac{i}{s_\theta}\d_\phi\right],
\quad \rho=r+ia c_\theta,\quad \Sigma=\rho{\bar\rho},\quad \Delta=r^2+a^2-2Mr.\nonumber
\eea 
Specifically, introducing new functions
\bea\label{PsiPMKerr}
\psi_+=\phi_0, \qquad \psi_-={{\bar\rho}^2}{\phi_2}\,,
\eea
writing them as\footnote{Throughout this article we choose the sign of $\omega$ using the conventions of \cite{ChandraBook}.}
\bea\label{TeukAnsatz}
\psi=e^{i\omega t+i m\phi}R(r)S(\theta),
\eea
and substituting the result into some of the Maxwell's equations, one finds a system of ODEs with a separation constant 
$\lambda$ \cite{Teuk}:
\bea\label{TeukEqn}
&&\hskip -0.8cm\frac{1}{\Delta^s}\frac{d}{dr}\left[\Delta^{s+1}\frac{dR}{dr}\right]+
\left[\frac{K(K-2i s r)+2isMK}{\Delta}-4is\omega r-\Lambda-(a\omega+m)^2+m^2\right]R=0,\nn
&&\hskip -0.8cm\frac{1}{s_\theta}\frac{d}{d\theta}\left[s_\theta\frac{dS}{d\theta}\right]+
\left[(a\omega c_\theta+s)^2-\frac{(m+s c_\theta)^2}{s_\theta^2}+s(1-s)+\Lambda\right]S=0.
\eea
Here parameter $s$ takes values $\pm 1$ for functions $\psi_\pm$ defined by (\ref{PsiPMKerr}). Function $K$ was defined in \cite{Teuk}:
\bea
K=(r^2+a^2)\omega-am.
\eea
Note that equations (\ref{TeukEqn}) also apply to massless scalar fields for $s=0$, as well as to spinors and gravitons for $s=\pm\frac{1}{2}$ and $s=\pm 2$. The discussion of spin-half and spin-two fields is beyond the scope of this article.

While equations (\ref{TeukEqn}) are remarkably simple, unfortunately they describe only one of the two possible polarizations. The remaining mode is governed by $\phi_1$, and equation for it appears to be non--separable \cite{Teuk}. Moreover, functions $(S_\pm,R_\pm)$ defined by (\ref{TeukAnsatz}) and (\ref{PsiPMKerr}) are not independent, but rather they are subject to complicated differential constraints (\ref{Starob}), and we refer to Appendix \ref{SecAppChandr} for the discussion of this issue. Thus it is desirable to rewrite the expressions for the Maxwell's fields in terms of unconstrained variables which cover all independent polarizations. We will present such reformulation in the next subsection, but to motivate that ansatz, we need the explicit expressions for the gauge potential in Teukolsky's variables. This result exists in the literature, and it is quoted in the Appendix \ref{SecAppTeuk}. Unfortunately, the final expression (\ref{ChandraPotential}) is not very illuminating, but we found the frame components of the gauge field to be rather simple. Specifically, multiplying  equations (\ref{ChandraPotential}) by vectors (\ref{FramesKerr}) and performing some algebraic manipulations, we found
\bea\label{KerrMaxwFrames1}
&&l^\mu A_\mu=\frac{2ia}{\Delta}P_+{\tilde f}_++2l^\mu\d_\mu H_+,\quad 
n^\mu A_\mu=
-\frac{ia}{\Sigma}P_-{\tilde f}_-+2n^\mu \d_\mu H_-\,,\nn
&&m^\mu A_\mu=
-\frac{\sqrt{2}}{\rho}{\tilde g}_+S_++2m^\mu \d_\mu H_+,\quad 
{\bar m}^\mu A_\mu=
-\frac{\sqrt{2}}{\bar\rho}{\tilde g}_-S_-+2{\bar m}^\mu \d_\mu H_-\,.
\eea
Here functions 
\bea
{\tilde g}_\pm=e^{i\omega t+im\phi}g_\pm (r)\quad \mbox{and}\quad
{\tilde f}_\pm=e^{i\omega t+im\phi}f_\pm (\theta)
\eea
are solutions of the first--order ordinary differential equations (\ref{ChndrEqnForDg})\footnote{We also used relations (\ref{DoLoFrames}) to eliminate operators $(\DD_0,\DD^\dagger_0,\LL_0,\LL^\dagger_0)$ from (\ref{ChndrEqnForDg}). Furthermore, we defined ${\tilde S}_\pm=e^{i\omega t+im\phi} S_\pm$, 
${\tilde P}_\pm=e^{i\omega t+im\phi} P_\pm$.},
\bea\label{ChndrEqnForDgA}
&&\sqrt{2}\rho m^\mu\d_\mu {\tilde f}_+=c_\theta {\tilde S}_+,\quad
\sqrt{2}{\bar\rho}{\bar m}^\mu\d_\mu {\tilde f}_-=c_\theta {\tilde S}_-,\nn
&&\Delta l^\mu \d_\mu {\tilde g}_+=r {\tilde P}_+,\qquad -2\Sigma n^\mu\d_\mu {\tilde g}_-=r {\tilde P}_-
\eea
and $H_\pm$ satisfy the second--order PDE (\ref{ChndrEqnForH}). Furthermore, 
\bea\label{DefPpm}
P_-=R_-,\quad P_+=\Delta R_+.
\eea
Substitution of (\ref{ChndrEqnForDgA}) into (\ref{KerrMaxwFrames1}) leads to our main result in this subsection:
\bea\label{KerrMaxwFrames}
l^\mu A_\mu&=&\frac{2ia}{r}l^\mu\d_\mu [e^{i\omega t+im\phi}g_+f_+]+2l^\mu\d_\mu H_+\nn
n^\mu A_\mu&=&\frac{2ia}{r}n^\mu\d_\mu [e^{i\omega t+im\phi}g_-f_-]+2n^\mu \d_\mu H_-\\
m^\mu A_\mu&=&
-\frac{2ia}{(iac_\theta)}m^\mu\d_\mu [e^{i\omega t+im\phi}f_+g_+]+2m^\mu \d_\mu H_+\nn
{\bar m}^\mu A_\mu&=&
-\frac{2ia}{(iac_\theta)}{\bar m}^\mu\d_\mu [e^{i\omega t+im\phi}f_-g_-]+2{\bar m}^\mu \d_\mu H_-\nonumber
\eea
Although this is just a reformulation of the known expressions (\ref{ChandraPotential}), a very suggestive form of (\ref{KerrMaxwFrames}) will serve as an inspiration for the new constructions developed in the rest of this article. 

\bigskip

We conclude this subsection with comments about various ingredients appearing in (\ref{KerrMaxwFrames}). 
To determine $f_+$ and $g_+$, one should begin with solving Teukolsky's equations (\ref{TeukEqn}) for 
$S_+$ and $R_+$ and substitute the results into (\ref{DefPpm}) and (\ref{ChndrEqnForDgA}). Although a similar procedure can be repeated for $f_-$ and $g_-$, generically it would lead to an inconsistent result since the constraints (\ref{Starob}) would be violated. Thus a better option is to solve the constraints (\ref{Starob}) instead, even though this breaks the symmetry between the modes with $s=\pm 1$. Thus to recover solution (\ref{KerrMaxwFrames}), one should implement the following sequence:
\bea\label{PathTeuk}
\begin{array}{ccccccc}
(\ref{TeukEqn})_+&\rightarrow&(S_+,R_+)&\rightarrow&(f_+,g_+)&\rightarrow&(l^\mu A_\mu,m^\mu A_\mu)\\
&&\downarrow&&&&\\
&&(S_-,R_-)&\rightarrow&(f_-,g_-)&\rightarrow&(n^\mu A_\mu,{\bar m}^\mu A_\mu)
\end{array}
\eea 
Unfortunately this construction recovers only one polarization. The second polarization should come 
from functions $(H_+,H_-)$, which satisfy a second--order PDE  (\ref{ChndrEqnForH})\footnote{We quote this equation only for completeness, see Appendix \ref{SecAppTeuk} for the detailed discussion of notation.}
\bea\label{HeqnMain}
\DD^\dagger_0\frac{\Delta \DD_0 H_+}{{\bar\rho}^2}+\LL_1\frac{\LL^\dagger_0 H_+}{{\bar\rho}^2}
-\DD_0\frac{\Delta \DD^\dagger_0 H_-}{{\bar\rho}^2}-\LL_1^\dagger\frac{\LL_0 H_-}{{\bar\rho}^2}
=0.
\eea
It is not clear how to construct the mode corresponding to $H_\pm$ polarization. Note that according to (\ref{KerrMaxwFrames}) solutions with $H_+=H_-$ correspond to a pure gauge, so to find a physical mode we can set $H_-=-H_+$ in (\ref{HeqnMain}). Solving the resulting PDE for one function is still a challenge, and we will avoid it by introducing a new ansatz for the gauge field in the next subsection. In the subsequent sections we will also extend this new construction to higher dimensions. 

\subsection{New ansatz for the gauge field}
\label{SecKerrNewAns}

In this subsection we will introduce a new ansatz for the gauge field that cures the problems encountered in the standard construction: the missing polarization corresponding to the fields $H_\pm$ and a rather convoluted path (\ref{PathTeuk}) for recovering the one known polarization. Specifically our ansatz will lead to separable equations for both polarizations of the electromagnetic wave, and in contrast to the modes described by Teukolsky's equations (\ref{TeukEqn}) with $s=\pm 1$, ours will be unconstrained.

The inspiration for our ansatz comes from the expression (\ref{KerrMaxwFrames}): we will require that each mode can be separated as 
\bea\label{KerrNewAnstz}
&&l^\mu A_\mu=G_+(r)l^\mu\d_\mu \Psi,\qquad
n^\mu A_\mu=G_-(r)n^\mu\d_\mu\Psi,\\
&&m^\mu A_\mu=
F_+(\theta)m^\mu\d_\mu \Psi,\quad 
{\bar m}^\mu A_\mu=
F_-(\theta){\bar m}^\mu\d_\mu\Psi,\quad \Psi=e^{i\omega t+im\phi}R(r)S(\theta).\nonumber
\eea
At first sight this ansatz appears to be more restrictive than  (\ref{KerrMaxwFrames}) since we used only one set of functions $(R,S)$ instead of two ($(f_+,g_+)$ and $(f_-,g_-)$). However, as we discussed in the last subsection, functions $(f_-,g_-)$ are uniquely determined in terms of $(f_+,g_+)$ via some non--local relations, so the ansatz (\ref{KerrNewAnstz}) seems to be the most natural way of avoiding complicated differential constraints. One can hope that introduction of undetermined functions $(F_\pm,G_\pm)$ makes the ansatz (\ref{KerrNewAnstz}) sufficiently general. Note that the differential operator $l^\mu\d_\mu$ does not depend on $\theta$ (see (\ref{FramesKerr})), so the requirement $\d_\theta G_+=0$ seems rather natural. Similarly, the $\theta$--dependence in $n^\mu\d_\mu$ cancels between the two sides of the second equation in (\ref{KerrNewAnstz}), suggesting the condition $\d_\theta G_-=0$. The requirements $\d_r F_\pm=0$ appear to be natural for the same reason. 

Before analyzing the general properties of the ansatz (\ref{KerrNewAnstz}), it might be instructive to look at solutions without $(t,\phi)$ dependence:
\bea\label{OmMZero1}
\omega=0,\quad m=0.
\eea
Although they don't describe physically interesting waves, such simple configurations lead to insights into the structure of functions $(F_\pm,G_\pm)$. The lessons from configurations (\ref{OmMZero1}) extracted in subsection \ref{SectSubKerrStat} will be used in subsections \ref{SecSubKerrElectr} and  \ref{SecSubKerrMagn} to find the most general solution consistent with the ansatz (\ref{KerrNewAnstz}).
We go through some details of derivation to stress the uniqueness of the resulting solution, and readers not interested in these arguments can go directly to section \ref{SecKerrCompare}, where the results are summarized by equations (\ref{4dHighSpinGuess}) and (\ref{4DfuncD}).

\subsubsection{Electro-- and magnetostatics}
\label{SectSubKerrStat}
In this subsection we focus on the special configurations (\ref{OmMZero1}),
\bea\label{OmMZero}
\omega=0,\quad m=0,
\eea
and demonstrate that the separable ansatz (\ref{KerrNewAnstz}) leads to only two types of solutions. As we will see, in the $a=0$ limit one of these branches describes an electrostatic configurations, while the other one corresponds to a magnetostatic case. To distinguish between the two types of solutions, we will call them ``electric'' and ``magnetic'' polarizations, even though in the presence of $(a,\omega,m)$ electric and magnetic fields are switched on in both cases. These two branches have different functions $(G_\pm,F_\pm)$, so their combinations {\it do not} fit into the ansatz (\ref{KerrNewAnstz}). However, the separable ``electric'' and ``magnetic'' configurations form a basis in the space of static electromagnetic fields. 
\bigskip

In the special case (\ref{OmMZero}), the two pairs $(A_r,A_\theta)$ and $(A_t,A_\phi)$ decouple in Maxwell's equations, so we analyze them one--by--one. 
The first pair enters only through $F_{r\theta}$, which satisfies two equations:
\bea\label{FrThEqnKerr}
\d_r\left[\frac{\Delta s_\theta}{\Sigma} F_{r\theta}\right]=0,\quad 
\d_\theta\left[\frac{\Delta s_\theta}{\Sigma} F_{r\theta}\right]=0.
\eea
We used the metric encoded by (\ref{gMuNuKerr}), (\ref{FramesKerr}), as well as the expression for its determinant
\bea
\sqrt{-g}=(r^2+a^2c_\theta^2)s_\theta.\nonumber
\eea
Equations (\ref{FrThEqnKerr}) have only one--dimensional space of solutions parameterized by an arbitrary constant $C$:
\bea\label{aa11}
F_{r\theta}=C\frac{\Sigma}{\Delta s_\theta}\,,
\eea
so $(A_r,A_\theta)$ come either from integrating this expression or from a pure gauge: 
\bea\label{KerrPureGauge}
l^\mu A_\mu=l^\mu\d_\mu \Psi,\quad
n^\mu A_\mu=n^\mu\d_\mu\Psi,\quad
m^\mu A_\mu=m^\mu\d_\mu \Psi,\quad 
{\bar m}^\mu A_\mu={\bar m}^\mu\d_\mu\Psi.
\eea
Solutions described by (\ref{aa11}) and (\ref{KerrPureGauge}) do not describe physical excitations with separation parameters, such as the order of a spherical harmonic, so in the special case (\ref{OmMZero}) one should set
\bea
A_r=A_\theta=0,\quad A=A_t dt+A_\phi d\phi.
\eea
Substitution of this expression for $A$ into the left--hand side of (\ref{KerrNewAnstz}) leads to two restrictions on four functions $(F_\pm,G_\pm)$:
\bea\label{FpmSchw4}
F_-=-F_+,\quad G_-=-G_+.
\eea
To find further constraints on functions $(F_\pm,G_\pm)$, we begin with looking at a special case of the Schwarzschild black hole. As we will see, the extension to the rotating case would be rather straightforward, although the intermediate formulas are complicated. Introducing a convenient notation for the components of the Maxwell's equations,
\bea\label{MaxwNotation}
{\MM}^\mu\equiv \frac{e^{-i\omega t-im\phi}}{\sqrt{-g}}\d_\nu\left[\sqrt{-g}F^{\mu\nu}\right],
\eea
and setting $a=0$, we find
\bea\label{MaxwKerrStatic}
\MM^t&=&-\frac{1}{r}\left\{G_+{\dot R}\frac{(s_\theta S')'}{s_\theta}+rS\frac{d^2}{dr^2}[(r-2M)G_+{\dot R}]\right\}\,,\nn
{\MM}^\phi&=&-\frac{i}{r^4 s_\theta}\left\{r^2S'F_+\frac{d}{dr}\left[\frac{r-2M}{r}{\dot R}\right]+
R\frac{d}{d\theta}\left[\frac{1}{s_\theta}\frac{d}{d\theta}(s_\theta F_+S')\right]\right\}\,.
\eea
Here and below prime denotes the derivative with respect to $\theta$, and dot denotes the derivative with respect to $r$. Assuming that $G_+\ne 0$\footnote{We will come back to the option $G_+= 0$ after equation (\ref{SpecialKerrElectr}).}, we conclude that equation $\MM^t=0$ reduces to two ODEs:
\bea
\MM^t=0:&&\frac{(s_\theta S')'}{s_\theta}=-\lambda_1 S,\quad 
r\frac{d^2}{dr^2}[(r-2M)G_+{\dot R}]=\la_1 G_+{\dot R}\,.
\eea
Dimensional analysis ensures that introduction of a rotation parameter $a$ does not modify the first equation\footnote{The only possible correction is an expansion in powers of $a/M$, but such terms become singular in the $M=0$ limit corresponding to the flat space.}, thus we must impose the relation
\bea\label{SeqnKerrOne}
(s_\theta S')'+\lambda_1 {s_\theta}S=0
\eea
even for the rotating geometry. Although the counterparts of (\ref{MaxwKerrStatic}) for the Kerr metric are rather complicated, repeated use of equation (\ref{SeqnKerrOne}) leads to drastic simplifications in  one combination\footnote{Since function $S$ satisfies only the second--order differential equation (\ref{SeqnKerrOne}), coefficients in front of $S$ and $S'$ must vanish independently. Also, $S'=0$ is not an interesting option, so the square bracket in (\ref{tempEqnJul15}) must vanish.}:
\bea\label{tempEqnJul15}
l_\mu\MM^\mu\Big|_{S=0}=\frac{2a S'}{\Sigma^2}\left[{iRc^2_\theta}\frac{d}{d\theta}\frac{F_+}{c_\theta}+
s_\theta {\dot R}(a c_\theta G_+-i rF_+)\right]=0.
\eea
For every value of the separation constant $\la_1$ in (\ref{SeqnKerrOne}), function $R$ should satisfy a second order differential equation, a radial counterpart of (\ref{SeqnKerrOne}), and this should be the 
{\it only} restriction on the radial profile. In particular, the coefficients in front of $R$ and ${\dot R}$ in (\ref{tempEqnJul15}) must vanish independently, then
\bea\label{FGplusElectr}
F_+=-iC ac_\theta,\quad G_+=C r 
\eea
with some constant $C$. Without loss of generality, we can set $C=1$. Note that a priori the over--constrained system of two equations coming from (\ref{tempEqnJul15}) is not guaranteed to have solutions for $F_+(\theta)$ and $G_+(r)$, so existence of the solution (\ref{FGplusElectr}) serves as a highly nontrivial consistency check for our ansatz (\ref{KerrNewAnstz}). Substituting (\ref{SeqnKerrOne}) and (\ref{FGplusElectr}) into the remaining Maxwell's equations, we find only one additional relation\footnote{The fact that $\theta$ does not appear in equation (\ref{Jul17Reqn}) is an additional nontrivial consistency check of the ansatz (\ref{KerrNewAnstz}).}:
\bea\label{Jul17Reqn}
\frac{d}{dr}[\Delta \dot R]-\la_1 R=0.
\eea
To summarize, we found that the static electromagnetic field (\ref{OmMZero}) in the Kerr geometry (\ref{Kerr1}) admits a separable solution:
\bea\label{SpecialKerrElectr}
&&l^\mu A^{(el)}_\mu=r {\hat l} \Psi,\quad
n^\mu A^{(el)}_\mu=-r{\hat n}\Psi,\quad
m^\mu A^{(el)}_\mu=
-i ac_\theta {\hat m} \Psi,\quad 
{\bar m}^\mu A^{(el)}_\mu=
i ac_\theta {\hat{\bar m}}\Psi,\nn
&&\Psi=R(r)S(\theta),\quad (s_\theta S')'+\lambda_1 {s_\theta}S=0,\quad 
\frac{d}{dr}[\Delta \dot R]-\la_1 R=0.
\eea
To simplify this and subsequent formulas, we introduced a convenient notation ${\hat v}$ for a differential operator corresponding to any vector $v^\mu$:
\bea\label{DefVhat}
v^\mu\quad\rightarrow\quad {\hat v}\equiv v^\mu\d_\mu\, .
\eea
The polarization (\ref{SpecialKerrElectr}) will be called ``electric'' just to distinguish it from the alternative option (\ref{SpecialKerrMagn}), which is discussed below. The bifurcation into two branches, the counterparts of  (\ref{SpecialKerrElectr}) and  (\ref{SpecialKerrMagn}), will persist in all dimensions, and the names ``electric'' and ``magnetic'' are given just keep track of these polarizations. While (\ref{SpecialKerrElectr}) describes a pure electric field for $a=0$, and (\ref{SpecialKerrMagn}) gives a pure magnetic field in the same limit (see equations (\ref{ElMagStaticKerr})), for generic values of $(a,\omega,m)$ both polarizations have nontrivial 
${\bf E}$ and ${\bf B}$, so the words ``electric'' and ``magnetic'' should be viewed only as labels. 

Recall that to derive (\ref{SpecialKerrElectr}), we assumed that in the non--rotating limit, $a=0$, this solution gives a nontrivial $G_+$. This assumption is justified by the final answer, but we also notice that in the non--rotating limit solution (\ref{SpecialKerrElectr}) gives $F_+=0$, so the second component of (\ref{MaxwKerrStatic}) vanishes trivially. It is natural to expect existence of a second polarization with nontrivial $F_+$, and we will discuss it next.
\bigskip

If $G_+=0$ in the non--rotating limit, then equation $\MM^t=0$ in (\ref{MaxwKerrStatic}) is trivially satisfied, but the ansatz (\ref{KerrNewAnstz}), (\ref{FpmSchw4}) implies that $F_+$ cannot vanish, so functions $(R,S)$ are constrained by the second equation in (\ref{MaxwKerrStatic}).
For configurations with $F_+\ne 0$, the Maxwell's equation $\MM^\phi=0$ reduces to a system of two ODEs:
\bea\label{ThirdOrderKerr}
\MM^\phi=0:\quad
\frac{d}{d\theta}\left[\frac{1}{s_\theta}\frac{d}{d\theta}(s_\theta F_+S')\right]=\la_2F_+ S',\quad r^2\frac{d}{dr}\left[\frac{r-2M}{r}{\dot R}\right]=-\la_2 R.
\eea
Once again, the scaling argument implies that the differential equation for $S(\theta)$ remains unchanged, even for non--zero $a$. Unfortunately, in the present case, $S'$ satisfies a third--order differential equation,
which can be used to eliminate only $S'''$ from $l_\mu\MM^\mu$, while the remaining entries $(S,S',S'')$ are not expected to be independent. Thus a simple argument that led to (\ref{SeqnKerrOne}) does not apply. However, after elimination of $S'''$, we have two equations
\bea
\MM^t=0,\qquad \MM^\phi=0\nonumber
\eea
for three variables $(S,S',S'')$, and eliminating $S''$, we end up with one equation for two {\it independent} functions $(S,S')$. Long but straightforward manipulations lead to the conclusion that the coefficient in front of $S$ cannot vanish unless
\bea
F_+=\frac{C}{c_\theta}
\eea
for some constant $C$. Angular equation from (\ref{ThirdOrderKerr}) with this value of $F_+$ has three 
linearly--independent solutions: two of them satisfy a second order equation
\bea\label{SecondOrderKerr}
\frac{c_\theta^2}{s_\theta}\frac{d}{d\theta}\left[\frac{s_\theta}{c_\theta^2}S'\right]+\la_2 S=0,
\eea
and the third one is constant $S$. The latter case leads to ODEs without a separation constant, similar to the one encountered in (\ref{FrThEqnKerr}), so the angular equation in (\ref{ThirdOrderKerr}) can be replaced by (\ref{SecondOrderKerr}).

Equation (\ref{SecondOrderKerr}) is analogous to (\ref{SeqnKerrOne}), and repeating the logic applied to the latter equation, we arrive at a counterpart of (\ref{tempEqnJul15}):
\bea
l_\mu\MM^\mu\Big|_{S=0}=-\frac{2r S'{\dot R}}{\Sigma^2}(iCa+r G_+)\tan_\theta=0.
\eea
This determines
\bea
G_+=-\frac{iCa}{r},
\eea
and the remaining Maxwell's equations reduce to one relation
\bea\label{Jul17ReqnMgn}
r^2\frac{d}{dr}\left[\frac{\Delta}{r^2}{\dot R}\right]-\la_2 R=0.
\eea
Collecting all relevant formulas, we arrive at a ``magnetic'' counterpart of (\ref{SpecialKerrElectr}):
\bea\label{SpecialKerrMagn}
&&\hskip -0.8cm l^\mu A^{(mgn)}_\mu=\frac{ia}{r} {\hat l} \Psi,\quad
n^\mu A^{(mgn)}_\mu=-\frac{ia}{r}{\hat n}\Psi,\quad
m^\mu A^{(mgn)}_\mu=-\frac{1}{c_\theta}{\hat m} \Psi,\quad 
{\bar m}^\mu A^{(mgn)}_\mu=
\frac{1}{c_\theta} {\hat{\bar m}}\Psi,\nn
&&\hskip -0.8cm \Psi=R(r)S(\theta),\quad \frac{c_\theta^2}{s_\theta}\frac{d}{d\theta}\left[\frac{s_\theta}{c_\theta^2}S'\right]+\la_2 S=0,\quad 
r^2\frac{d}{dr}\left[\frac{\Delta}{r^2}{\dot R}\right]-\la_2 R=0.
\eea
As before, the fact that all Maxwell's equations have consistently reduced to a system of two ordinary differential equations is a highly nontrivial feature of our separable ansatz (\ref{KerrNewAnstz}).

\bigskip

To summarize, we have shown that application of the ansatz (\ref{KerrNewAnstz}) to static configurations (\ref{OmMZero}) is consistent only in three instances ((\ref{KerrPureGauge}), (\ref{SpecialKerrElectr}), (\ref{SpecialKerrMagn})), and the first case corresponds to a pure gauge. The interesting solutions, (\ref{SpecialKerrElectr}) and (\ref{SpecialKerrMagn}), describe two independent polarizations of the static field, 
and in the limit of Schwarzschild geometry ($a=0$), they give rise to electrostatic  (\ref{SpecialKerrElectr}) and magnetostatic (\ref{SpecialKerrMagn}) configurations:
\bea\label{ElMagStaticKerr}
a=0:&&A^{(el)}=\frac{\Delta}{r}\d_r(RS)dt,\quad (s_\theta S')'+\lambda_1 {s_\theta}S=0,\quad 
\frac{d}{dr}[\Delta \dot R]-\la_1 R=0.\\
&&A^{(mgn)}=-\frac{is_\theta}{c_\theta}\d_\theta(RS)d\phi,\quad 
\frac{c_\theta^2}{s_\theta}\frac{d}{d\theta}\left[\frac{s_\theta}{c_\theta^2}S'\right]+\la_2 S=0,\quad 
r^2\frac{d}{dr}\left[\frac{\Delta}{r^2}{\dot R}\right]-\la_2 R=0.\nonumber
\eea
Since such separation into two branches persists even for time--dependent fields, we will refer to the counterparts of (\ref{SpecialKerrElectr}) and  (\ref{SpecialKerrMagn}) as electric and magnetic polarizations, even though generically both electric and magnetic fields are excited.  
The two polarizations will be discussed separately in the next two subsections.

\subsubsection{Electric polarization}
\label{SecSubKerrElectr}

In this subsection we will extend the solution (\ref{SpecialKerrElectr}) to arbitrary values of $(m,\omega)$, relaxing the constraint (\ref{OmMZero}). Already in the special case (\ref{OmMZero}), the separable electromagnetic field (\ref{KerrNewAnstz}) splits into two distinct polarizations (\ref{SpecialKerrElectr}) and (\ref{SpecialKerrMagn}), so this feature must persist also for generic values of $(\omega,m)$. It is convenient to analyze the two branches one--by--one. Here we focus on extending the electric solution (\ref{SpecialKerrElectr}), and its magnetic counterpart will be analyzed in section \ref{SecSubKerrMagn}. 

\bigskip

The detailed derivation of the equations for the electric polarization is rather technical, and it is presented in the Appendix \ref{SecAppKerrElectr}. Here we just summarize the main logical steps to emphasize the uniqueness of the final result (\ref{KerrElectrFinal}). 
\begin{enumerate}
\label{Steps4D}
\item{To make the expressions more transparent, we first set $M=0$. Then the metric (\ref{gMuNuKerr}), (\ref{FramesKerr}) describes flat space in unusual coordinates, but remarkably, even in this case the separable ansatz (\ref{KerrNewAnstz}) leads to the {\it unique} set of equations for the electric polarization. A failure to add mass to this result would have indicated inconsistency of the proposal (\ref{KerrNewAnstz}), but fortunately the extension to black hole geometry is rather straightforward, and it is unique (see item 6).}
\item{To distinguish between electric and magnetic branches in the geometry with $M=0$, we observe that the electric solution (\ref{SpecialKerrElectr}) has $F_\pm=0$ if $a=0$. In the Appendix \ref{SecAppKerrElectr} we argue that the same property must hold even once $\omega$ and $m$ are switched on, and this leads to a simpler ansatz for the gauge field in the $M=a=0$ geometry: 
\bea\label{RestrctAnstzKerrElectr}
l^\mu A_\mu=G_+(r){\hat l} \Psi,\quad
n^\mu A_\mu=G_-(r){\hat n}\Psi,\quad
m^\mu A_\mu={\bar m}^\mu A_\mu=0.
\eea
Recall that $\Psi=e^{i\omega t+im\phi}R(r)S(\theta)$.}
\item{Substitution of the ansatz (\ref{RestrctAnstzKerrElectr}) into the Maxwell's equations in the $M=a=0$ geometry leads to an overly--constrained system of differential equations for functions $(G_\pm,R,S)$. In particular, there is only one function of angular variable, $S(\theta)$, and consistency of Maxwell's equations immediately leads to relation (\ref{KerrA0ElectSeqn}). Further analysis leads to the unique expressions for $G_\pm$ and the unique equation for $R(r)$. In the Appendix \ref{SecAppKerrElectr} we also extended this result to the Schwarzschild black hole, and the result is given by (\ref{KerrElectrSchw}).}
\item{To turn on parameter $a$, while still keeping $M=0$, we observe that, in the absence of mass, the metric (\ref{gMuNuKerr}), (\ref{FramesKerr}) has a $Z_2$ symmetry interchanging radial and angular coordinates 
($r\leftrightarrow  iac_\theta$). Requirement for the separable solution (\ref{KerrNewAnstz}) to transform covariantly under this symmetry determines $F_\pm$ and $G_\pm$ for arbitrary values of $a$. }
\item{Once functions $(F_\pm,G_\pm)$ are determined, substitution of the ansatz (\ref{KerrNewAnstz}) into the Maxwell's equations gives an over--constrained system for $(R(r),S(\theta))$. Although a priori existence of non--trivial solutions is not guaranteed, we found that all Maxwell's equations follow from a system of two ODEs with one separation constant. This is a highly nontrivial consistency check for the ansatz (\ref{KerrNewAnstz}).
}
\item{Once the unique system of equations for $M=0$ case is derived, the mass is added by a simple modification of the radial equation. We expect that such modification is also unique.}
\end{enumerate}

The detailed implementation of these six steps is presented in the Appendix \ref{SecAppKerrElectr}, here we just quote the result. To write it in a compact form and to compare with higher dimensions in subsequent sections, it is convenient to introduce a more symmetric notation for the standard frames (\ref{FramesKerr}):
\bea\label{KerrNewFrames}
l^\mu_+=l^\mu,\quad l^\mu_-=-\frac{2\Sigma}{\Delta} n^\mu,\quad m_+^\mu=\sqrt{2}\rho m^\mu,\quad 
m_-^\mu=\sqrt{2}{\bar\rho}{\bar m}^\mu\,.
\eea
The great advantage of these objects is the full separation of variables: $\theta$ does not appear in $l_\pm$, and $r$ does not appear in $m_\pm$:
\bea\label{KerrBasisImprvd}
&&l_\pm^\mu\d_\mu=\d_r\pm\left[\frac{r^2+a^2}{\Delta}\d_t+\frac{a}{\Delta}\d_\phi\right],\quad
m_\pm^\mu\d_\mu=\d_\theta\pm\left[{ia s_\theta}\d_t+\frac{i}{s_\theta}\d_\phi\right]\,.
\eea 
The small price to pay for this convenience is a more complicated form of the metric (\ref{gMuNuKerr}),
\bea\label{gMuNuKerrRscld}
g^{\mu\nu}=\frac{1}{2\Sigma}\left[\Delta l_+^\mu l_-^\nu+\Delta l_-^\mu l_+^\nu+
m_+^\mu{m}_-^\nu+m_-^\mu{m}_+^\nu\right],
\eea
but the advantages are more significant, especially for extending the construction (\ref{KerrNewAnstz}) to 
higher dimensions. In the frames (\ref{KerrNewFrames}), the main result of the Appendix \ref{SecAppKerrElectr} becomes
\bea\label{KerrElectrFinal}
&&l_\pm^\mu A^{(el)}_\mu=\pm\frac{r}{1\pm i\mu r}{\hat l}_\pm \Psi,\quad
m_\pm^\mu A^{(el)}_\mu=\mp\frac{i a c_\theta}{1\pm \mu ac_\theta}{\hat m}_\pm \Psi,\quad 
\Psi=e^{i\omega t+im\phi}R(r)S(\theta),
\nn
&&\frac{E_\theta}{s_\theta}\frac{d}{d\theta}\left[\frac{s_\theta}{E_\theta} S'\right]
+\left\{-\frac{2\Lambda}{E_\theta}+
(a\omega c_\theta)^2-\frac{m^2}{s_\theta^2}-C\right\}S=0,
\\
&&{E_r}\frac{d}{dr}\left[\frac{\Delta}{E_r} {\dot R}\right]
+\left\{\frac{2\Lambda}{E_r}+
(\omega r)^2+\frac{(am)^2}{\Delta}+\frac{2Mr\omega^2\Delta_0}{\Delta}
+\frac{4Mar\omega m}{\Delta}+C\right\}R=0.\nonumber
\eea
Here we defined convenient functions
\bea
E_r=1+(\mu r)^2,\quad
E_\theta=1-(\mu a c_\theta)^2,\quad \Delta=r^2+a^2-2M,\quad 
\Delta_0=r^2+a^2,
\eea
and introduced two separation parameters:
\bea
\Lambda=a\mu[m+a\omega-\frac{\omega}{a\mu^2}],\quad C=-{\Lambda}+2 am\omega+(a\omega)^2.
\eea
Note that separation of variables is governed by one constant $\mu$, which appears in several places. The angular equation is more traditional in the Schwarzschild limit (\ref{KerrElectrSchw}), although even in that case the separation constant $\la_1=\frac{\omega}{\mu}$ enters the radial equation in a complicated fashion.

Interestingly, electrostatic configurations can be recovered in two distinct limits: as $\omega$ goes to zero, one can either keep $\mu$ fixed or scale it as $\mu=\la \omega$. The latter case gives
\bea
l_\pm^\mu A_\mu=\pm r{\hat l}_\pm \Psi,\quad
m_\pm^\mu A_\mu=\mp{i a c_\theta}{\hat m}_\pm \Psi,\quad C=-\Lambda=\frac{1}{\la}\,,
\eea 
and in particular, it reproduces the previously found solution (\ref{SpecialKerrElectr}) if $m=0$. Fixing $\mu$ instead, one finds
\bea
l_\pm^\mu A_\mu=\pm\frac{r}{1\pm i\mu r}{\hat l}_\pm \Psi,\quad
m_\pm^\mu A_\mu=\mp\frac{i a c_\theta}{1\pm \mu ac_\theta}{\hat m}_\pm \Psi,\quad
C=-\Lambda=a\mu m.
\eea
In the $m=0$ limit this solution loses a separation parameter in equations for $R$ and $S$, so it is not very interesting, even for $m\ne 0$.\footnote{Recall that we have already encountered a similar situation with configuration (\ref{aa11}).}

\bigskip

To summarize, in this subsection we outlined the derivation of the {\it unique} separable solution (\ref{KerrElectrFinal}) for the electric polarization. For fixed $m$ and $\omega$, the system (\ref{KerrElectrFinal})  represents an eigenvalue problem for $\mu$, but the full analysis of the resulting modes is beyond the scope of this article. In the next subsection we will discuss the magnetic polarization.

\subsubsection{Magnetic polarization}
\label{SecSubKerrMagn}

The derivation of the magnetic polarization closely follows the six steps outlined on page \pageref{Steps4D}, and the details are given in the Appendix \ref{SecAppKerrMagn}. The final result reads
\bea\label{KerrMagnFinal}
&&l_\pm^\mu A^{(mgn)}_\mu=\pm\frac{ia}{r \pm i\mu a}{\hat l}_\pm \Psi,\quad
m_\pm^\mu A^{(mgn)}_\mu=\mp\frac{1}{c_\theta\mp \mu}{\hat m}_\pm \Psi,\quad 
\Psi=e^{i\omega t+im\phi}R(r)S(\theta),\nn
&&\frac{M_\theta}{s_\theta}\frac{d}{d\theta}\left[\frac{s_\theta}{M_\theta}\d_\theta S\right]
+\left\{-\frac{m^2}{s_\theta^2}-\frac{2\Lambda}{M_\theta}+(a\omega c_\theta)^2-C\right\}S=0,\\
&&{M_r}\frac{d}{dr}\left[\frac{\Delta}{M_r} R'\right]+\left\{
-\frac{2\Lambda a^2}{M_r}+\frac{(a m)^2}{\Delta}+(r\omega)^2+\frac{2Mr\omega^2\Delta_0}{\Delta}
+\frac{4Mar\omega m}{\Delta}+C\right\}R=0.\nonumber
\eea
As in the electric case, we defined convenient functions
\bea
M_r= r^2+(\mu a)^2,\quad M_\theta= c_\theta^2-\mu^2,\quad \Delta=r^2+a^2-2M,\quad 
\Delta_0=r^2+a^2,
\eea
and introduced two separation parameters:
\bea
\Lambda=\mu\left[a\omega+m-{a\omega\mu^2}\right],\qquad 
C=\frac{\Lambda}{\mu^2}+a\omega\left[a\omega+2m\right].
\eea
The solution (\ref{KerrMagnFinal}) simplifies in the Schwarzschild limit, and the result is given by  (\ref{KerrMagnSchw}). As in the electric case, the system (\ref{KerrMagnFinal})  should be viewed as an eigenvalue problem for $\mu$. 

\subsection{Summary and comparison to the known results}
\label{SecKerrCompare}

In this subsection we will compare the new solutions (\ref{KerrElectrFinal}) and (\ref{KerrMagnFinal}) with various existing results. We begin with discussing similarities between the new eigenvalue problems and the wave equation, then in subsection \ref{SecCmprTeuk} we will compare the new ansatz with Teukolsky's approach. 

\subsubsection{Electromagnetism and the wave equation}
\label{SectSubKerrWave}

Teukolsky's construction found striking similarities between equations for some components of electromagnetic field and the wave equation. Since these parallels also persisted for gravitons and neutrinos, it might be interesting to find similar patterns for the new ansatz (\ref{KerrNewAnstz}), even though a detailed discussion of spin--2 and spin--$\frac{1}{2}$ particles is beyond the scope of this paper. 

The wave equation in the Kerr geometry was studied by Carter \cite{Carter1}, and this work led to discovery of hidden symmetries parameterized by the Killing tensors. This symmetry structure will be discussed in detail in section \ref{SecSubKilling}, here we just observe that the wave equation in the metric (\ref{FramesKerr}) separates between $r$ and $\theta$ coordinate. Specifically, equation
\bea\label{4Dscalar}
\nabla_\mu\nabla^\mu \Psi=0,\quad \Psi=e^{i\omega t+im\phi}R(r)S(\theta)
\eea
reduces to a system of ODEs with one separation constant $\la$:
\bea\label{4dScalarEqn}
&&\frac{1}{s_\theta}\frac{d}{d\theta}\left[{s_\theta}S'\right]
+\left\{-\frac{m^2}{s_\theta^2}+(a\omega c_\theta)^2-\la\right\}S=0,\\
&&\frac{d}{dr}\left[{\Delta}{\dot R}\right]+\left\{
\frac{(a m)^2}{\Delta}+(r\omega)^2+\frac{2Mr\omega^2\Delta_0}{\Delta}
+\frac{4Mar\omega m}{\Delta}+\la\right\}R=0.\nonumber
\eea
Comparing this with (\ref{KerrElectrFinal}) and (\ref{KerrMagnFinal}), we conclude that all three systems can be written in a compact form
\bea\label{4dHighSpinGuess}
&&\frac{D_\theta}{s_\theta}\frac{d}{d\theta}\left[\frac{s_\theta}{D_\theta}\d_\theta S\right]
+\left\{-\frac{2\Lambda}{D_\theta}-(as_\theta)^2\left[\omega+\frac{m}{as_\theta^2}\right]^2+\Lambda\right\}S=0,\nn
&&{D_r}\frac{d}{dr}\left[\frac{\Delta}{D_r} R'\right]+\left\{
\frac{2\Lambda}{D_r}+\frac{(r^2+a^2)^2}{\Delta}\left[\omega+\frac{am}{r^2+a^2}\right]^2-
\Lambda\right\}R=0.
\eea
The difference between excitations appears only in the factors $(D_r,D_\theta)$ and in the expression for the parameter $\Lambda$:
\bea\label{4DfuncD}
\mbox{scalar}:&&D_r=1,\quad D_\theta=1,\quad \forall \Lambda;\nn
\mbox{electric}:&&D_r=1+(\mu r)^2,\
D_\theta=1-(\mu a c_\theta)^2,\ \Lambda=a\mu[m+a\omega-\frac{\omega}{a\mu^2}];\\
\mbox{magnetic}:&&D_r=1+\frac{r^2}{(\mu a)^2},\quad D_\theta=1-\frac{c_\theta^2}{\mu^2},\quad
\Lambda=-\frac{1}{\mu}\left[a\omega+m-{a\omega\mu^2}\right].\nonumber
\eea
The equations for the electric and magnetic excitations are interchanged under duality 
\bea\label{Duality4D}
\mu\rightarrow  -\frac{1}{a\mu}\,.
\eea
To complete the summary, we recall that the gauge fields are given by the first lines in (\ref{KerrElectrFinal}), (\ref{KerrMagnFinal}):
\bea\label{4dSmryAnstz}
&&\hskip -1.6cm l_\pm^\mu A^{(el)}_\mu=\pm\frac{r}{1\pm i\mu r}{\hat l}_\pm \Psi,\quad
m_\pm^\mu A^{(el)}_\mu=\mp\frac{i a c_\theta}{1\pm \mu ac_\theta}{\hat m}_\pm \Psi,\quad 
\Psi=e^{i\omega t+im\phi}R(r)S(\theta);\nn
&&\hskip -1.6cm l_\pm^\mu A^{(mgn)}_\mu=\pm\frac{ia}{r \pm i\mu a}{\hat l}_\pm \Psi,\quad
m_\pm^\mu A^{(mgn)}_\mu=\mp\frac{1}{c_\theta\mp \mu}{\hat m}_\pm \Psi,\quad 
\Psi=e^{i\omega t+im\phi}R(r)S(\theta),
\eea
but unlike equations (\ref{4DfuncD})--(\ref{4dHighSpinGuess}), these relations do not transform in a simple way under the duality (\ref{Duality4D}). 

It is natural to expect that the ``master equations'' (\ref{4dHighSpinGuess}) would hold even beyond scalar and vector fields, and that the spin would be encoded in functions $D_\theta$ and $D_r$. We leave exploration of this conjecture for future work. Regardless of the outcome of this investigation, the similarity between  (\ref{4dScalarEqn}) and (\ref{KerrElectrFinal}), (\ref{KerrMagnFinal}) is rather striking.

\subsubsection{Comparison to the Teukolsky's ansatz}
\label{SecCmprTeuk}

Let us now compare the new systems derived in section \ref{SecKerrNewAns} with the classic solutions by Teukolsky \cite{Teuk}. We will demonstrate that when the two ansatze overlap, they give identical results.

To derive separable expressions for the electromagnetic fields, Teukolsky worked in the first--order formalism and considered 
equations\footnote{Our discussion after expression (\ref{StrenFramesNP}) suggests that equations for 
$(\phi_0,\phi_2)$ follow naturally from combining (\ref{temp122}) into relations for $F\pm i\star F$.}
\bea\label{temp122}
dF=0,\quad d\star F=0.
\eea
Then separation of variables was imposed on particular components of the field strength in the basis  (\ref{FramesKerr}). To compare our new results with this classic discussion, we compute some components of $F_{\mu\nu}$ in the improved basis (\ref{KerrBasisImprvd}), where
\bea
l_+^\mu \d_\mu m^\nu_\pm=l_-^\mu \d_\mu m^\nu_\pm=0,\qquad
m_+^\mu \d_\mu l^\nu_\pm=m_-^\mu \d_\mu l^\nu_\pm=0\,.
\eea
Starting with expressions (\ref{KerrNewAnstz}), we find
\bea\label{KerrCompProj1}
l_+^\mu m_\pm^\nu F_{\mu\nu}&=&l_+^\mu\d_\mu(m_\pm^\nu A_\nu)-
m_\pm^\nu\d_\nu(l_+^\nu A_\nu)=(F_\pm-G_+){\hat l}_+{\hat m}_\pm\Psi,\nn
l_-^\mu m_\pm^\nu F_{\mu\nu}&=&(F_\pm-G_-){\hat l}_+{\hat m}_\pm\Psi.
\eea
The remaining two components of the field strength are rather complicated. To proceed, we evaluate various combinations of the prefactors for the electric and magnetic polarizations:
\bea\label{KerrCompProj2}
\mbox{electric}:&&F_\pm-G_+=\frac{-r\mp iac_\theta}{(1\pm \mu ac_\theta)(1+i\mu r)},\quad
F_\pm-G_-=\frac{r\mp iac_\theta}{(1\pm \mu ac_\theta)(1-i\mu r)},\nn
\mbox{magnetic}:&&F_\pm-G_+=\frac{r\pm iac_\theta}{(\mu\mp c_\theta)(r + i\mu a)},\quad
F_\pm-G_-=\frac{r\mp iac_\theta}{(\mu\mp c_\theta)(r - i\mu a)}.
\eea 
We conclude that the ansatz (\ref{KerrNewAnstz}) leads to multiplicative separation of the following 
expressions\footnote{Recall that according to (\ref{FramesKerr}), $\rho=r+iac_\theta$. Also note that 
real field configurations have $\Phi_3={\bar\Phi}_0$, $\Phi_4={\bar\Phi}_2$.}:
\bea
\Phi_0=\frac{1}{\rho}l_+^\mu m_+^\nu F_{\mu\nu},\quad 
\Phi_2=\frac{1}{\rho}l_-^\mu m_-^\nu F_{\mu\nu},\quad
\Phi_3=\frac{1}{\bar\rho}l_+^\mu m_-^\nu F_{\mu\nu},\quad 
\Phi_4=\frac{1}{\bar\rho}l_-^\mu m_+^\nu F_{\mu\nu}.
\eea
Teukolsky's classic solution imposed a separation of $\Phi_0$ and $\Phi_2$,
\bea
\Phi_0=R_+(r)S_+(\theta)e^{i\omega t+im\phi},\quad
\Phi_2=R_-(r)S_-(\theta)e^{i\omega t+im\phi},
\nonumber
\eea
and treated the resulting four functions $(S_+,S_-,R_+,R_-)$ as independent. As the result of this construction, no statements could be made about separation of 
\bea\label{tempJul18}
l_+^\mu l_-^\nu F_{\mu\nu},\qquad m_+^\mu m_-^\nu F_{\mu\nu}.
\eea
In contrast, our ansatz (\ref{KerrNewAnstz}) parameterizes the full configuration in terms of only one pair $(S,R)$, so even the non--separable components (\ref{tempJul18}) are written in terms of these functions, although the expressions are not very illuminating.  Equations (\ref{KerrCompProj1}) and (\ref{KerrCompProj2}) lead to explicit relation between our variables and the separable components used by Teukolsky:
\bea\label{MapToTeuk}
&&\Phi^{(el)}_0=-\frac{{\hat l}_+{\hat m}_+\Psi}{(1+\mu ac_\theta)(1 + i\mu r)},\quad 
\Phi^{(el)}_2=\frac{{\hat l_-}{\hat m_-}\Psi}{(1-\mu ac_\theta)(1 - i\mu r)};\\ 
&&\Phi^{(mgn)}_0=\frac{{\hat l}_+{\hat m}_+\Psi}{(\mu- c_\theta)(r + i\mu a)},\quad 
\Phi^{(mgn)}_2=\frac{{\hat l_-}{\hat m_-}\Psi}{(\mu+c_\theta)(r-i\mu a)}. \nonumber
\eea
It is clear that separable $\Psi$ produces separable $(\Phi_0,\Phi_2)$. Note, however, that the map (\ref{MapToTeuk}) becomes useful only for configurations for which the ansatze (\ref{KerrNewAnstz}) and (\ref{KerrMaxwFrames}) overlap. 

\subsection{Extension to the Kerr-(A)dS geometry}
\label{SecKerr4DAdS}

The results obtained in this section can be easily extended to four--dimensional rotating black hole in the presence of the cosmological constant. Away from the sources, such geometry solves Einstein's equations 
\bea
R_{\mu\nu}=3L g_{\mu\nu},
\eea
where $L$ is related to the cosmological constant\footnote{We reserve symbols $\la$ and $\Lambda$ for the eigenvalues associated with Maxwell's equations.}. The resulting Kerr--(A)dS metric is \cite{Carter2}\footnote{To compare with higher dimensions in subsequent sections, we use notation of \cite{GLPP}.}
\bea\label{KerrAdS4d}
ds^2&=&{\tilde g}_{tt}dt^2+
\frac{r^2+a^2}{1+L a^2}s_\theta^2[d{\tilde\phi}-La dt]^2+\frac{2Mr}{r^2+a^2 c_\theta^2}
\left[dt-\frac{as_\theta^2 d{\tilde \phi}}{1+La^2}\right]^2\nn
&&+(r^2+a^2c_\theta^2)\left[\frac{dr^2}{\Delta-Lr^2(r^2+a^2)}+\frac{d\theta^2}{1+La^2 c_\theta^2}\right]\,,\\
{\tilde g}_{tt}&=&-\frac{(1+La^2 c_\theta^2)(1-Lr^2)}{1+La^2},\quad \Delta=r^2-2Mr+a^2\,.\nonumber
\eea
Regularity of this metric near $\theta=0$ implies that coordinate ${\tilde \phi}$ has a standard periodicity
($0\le {\tilde\phi}<2\pi$), but to simplify some formulas below and especially to compare with higher dimensions in section \ref{SecKerrAdSMxwMP}, it is convenient to rescale the angular coordinate:
\bea\label{KerAdS4PhiPer}
\phi=\sqrt{1+La^2}{\tilde\phi},\quad 0\le \phi<2\pi \sqrt{1+La^2}\,.
\eea
To apply the ansatze (\ref{KerrElectrFinal}) and (\ref{KerrMagnFinal}) to electromagnetic waves in the geometry (\ref{KerrAdS4d}), we need the counterparts of the special vielbeins (\ref{KerrBasisImprvd}). The general method for constructing such objects will be discussed in detail in section \ref{SecAdSFramesMP}, here we just quote the result:
\bea
l_\pm^\mu\d_\mu=Q_r\d_r\pm\frac{1}{Q_r}\left[\frac{r^2+a^2}{\Delta}\d_t+\frac{a}{\Delta}\d_\phi\right],\quad
m_\pm^\mu\d_\mu=Q_\theta\d_\theta\pm\frac{1}{Q_\theta}\left[{ia s_\theta}\d_t+\frac{i}{s_\theta}\d_\phi\right].\nonumber
\eea 
Factors $(Q_r,Q_\theta)$ are defined by
\bea\label{Qfact4d}
Q_r=\sqrt{1-Lr^2\frac{r^2+a^2}{\Delta}},\quad Q_\theta=\sqrt{1+L(a c_\theta)^2}\,,
\eea
and the metric is still given by equation (\ref{gMuNuKerrRscld}).

Imposing the ansatze (\ref{4dSmryAnstz}) for the electric and magnetic polarizations, as well as separation (\ref{4Dscalar}) for the scalar,
and substituting the results into Maxwell and wave equations, we arrive at a counterpart of the ``master equations'' (\ref{4dHighSpinGuess}):
\bea\label{4dKerrAdSFull}
&&\frac{D_\theta}{s_\theta}\frac{d}{d\theta}\left[\frac{Q_\theta^2 s_\theta}{D_\theta}\d_\theta S\right]
+\left\{-\frac{2\Lambda}{D_\theta}-\frac{(as_\theta)^2}{Q_\theta^2}\left[\omega+\frac{m}{as_\theta^2}\right]^2+\Lambda\right\}S=0,\\
&&{D_r}\frac{d}{dr}\left[\frac{Q_r^2\Delta}{D_r} R'\right]+\left\{
\frac{2\Lambda}{D_r}+\frac{(r^2+a^2)^2}{Q_r^2\Delta}\left[\omega+\frac{am}{r^2+a^2}\right]^2-
\Lambda\right\}R=0.\nonumber
\eea
Functions $(D_r,D_\theta)$ and parameter $\Lambda$ are still given by (\ref{4DfuncD}). Note that the separation ansatz
\bea
\Psi=e^{i\omega t+im\phi}R(r)S(\theta)
\eea
contains the angular coordinate $\phi$ with a non--standard periodicity (\ref{KerAdS4PhiPer}), so $m$ appearing in equation (\ref{4dKerrAdSFull}) is not an integer, but it still takes discrete values. 

\bigskip

\bigskip

This concludes our discussion of the four--dimensional black holes. To summarize, we have reviewed Teukolsky's classic construction and rewrote it in a very suggestive form (\ref{KerrMaxwFrames}). This formula was used as an inspiration for the new ansatz (\ref{KerrNewAnstz}), which, in contrast to the classic construction, covers both polarizations of photons. We have demonstrated that the ansatz (\ref{KerrNewAnstz}) leads to only two options for the gauge potential, (\ref{KerrElectrFinal}) and (\ref{KerrMagnFinal}), which we labeled as ``electric'' and ``magnetic'' polarizations. Furthermore, we rewrote equations governing these polarizations, as well as massless scalar, in the unified form (\ref{4dHighSpinGuess})--(\ref{4DfuncD}), and this suggests that similar relations may hold for particles with higher spins. Finally, in subsection \ref{SecKerr4DAdS} all these constructions were generalized to describe the Kerr--AdS metric. The rest of this article  is dedicated to extension of the results obtained in this section to rotating black holes in arbitrary dimensions.

\section{Myers--Perry black hole and its symmetries}
\label{SecMyersPerry}
\renewcommand{\theequation}{3.\arabic{equation}}
\setcounter{equation}{0}

To extend the results obtained in the last section to rotating black holes in higher dimensions, we have to identify the key ingredients of the ansatz (\ref{KerrNewAnstz}) and uncover similar structures for other systems. In four dimensions, the ansatz (\ref{KerrNewAnstz}) for the gauge field relied on existence of a very special vierbein $(l^\mu,n^\mu,m^\mu,{\bar m}^\mu)$, so to extend the success of the construction introduced last section, it is important to find the counterpart of the expressions (\ref{FramesKerr}) in arbitrary dimensions. While it is possible to just guess the appropriate vielbein, a more constructive approach is based on characterizing the frames (\ref{FramesKerr}) by their algebraic properties and finding the generalization by {\it solving} appropriate equations. Such approach to special vielbeins was developed in \cite{ChLKill} based on earlier work \cite{PreKub,Kub1,Kub2,Kub3}, and in this section 
we will review the appropriate construction. Specifically, we introduce the geometry of the Myers--Perry black hole \cite{MyersPerry}, discuss it symmetries encoded in Killing--Yano tensors, and demonstrate that the higher dimensional counterparts of the vierbein (\ref{FramesKerr}) are uniquely determined by solving equations for such tensors. The resulting vielbeins, first constructed in \cite{ChLKill}, will then be used in subsequent sections to separate Maxwell's equations in higher--dimensional black holes. In section \ref{SecMPwave} we will also review separation of variables in the wave equation, which will be used later in the paper.  Finally, in section \ref{SecAdSFramesMP} all these constructions will be extended to the GLPP black holes \cite{GLPP} which generalize the Myers--Perry geometry to solutions of Einstein's equations with non--zero cosmological constant. 

\subsection{Killing--Yano tensors for the Myers--Perry black hole}
\label{SecSubKilling}

To extend the construction discussed in the last section to higher dimensions, we recall the higher--dimensional generalization of the Kerr geometry. The form of such Myers--Perry black hole \cite{MyersPerry}  differs between even and odd dimensions, so we begin with quoting the solution in even dimensions ($d=2n+2$) \cite{MyersPerry,Myers}:
\bea\label{MPeven}
ds^2&=&-dt^2+\frac{Mr}{FR}\Big(dt+\sum_{i=1}^n a_i\mu_i^2 d\phi_i\Big)^2+\frac{FR dr^2}{R-Mr}
+\sum_{i=1}^n(r^2+a_i^2)\Big(d\mu_i^2+\mu_i^2 d\phi_i^2\Big).\nn
&&+r^2d\alpha^2.
\eea
Here variables $(\mu_i,\alpha)$ are subject to a constraint
\bea
\alpha^2+\sum_{i=1}^n\mu_i^2=1,
\eea
and functions $F$, $R$ are defined by 
\bea\label{MPdefFR}
F=1-\sum_{k=1}^n\frac{a_k^2\mu_k^2}{r^2+a_k^2},\quad R=\prod_{k=1}^n (r^2+a_k^2).
\eea
To recover the standard Kerr geometry from the solution (\ref{MPeven}) one should set $n=1$ and make replacements
\bea
M\rightarrow 2M,\quad a_1\rightarrow -a.
\eea
Let us now discuss the symmetries of (\ref{MPeven}) following \cite{ChLKill}.

\bigskip

The metric (\ref{MPeven}) has an explicit $[U(1)]^{n+1}$ isometry which acts by constant shifts of $t$ and $\phi_i$:
\bea
x^\mu\rightarrow x^\mu+\eps V^\mu,\qquad V^\mu \d_\mu=B^t\d_t+\sum B^i\d_{\phi^i},\quad
B^\mu=\mbox{const},
\eea
and one can show that these symmetries exhaust all vectors satisfying the Killing equation
\bea\label{KVeqn}
\nabla_\mu V_\nu+\nabla_\nu V_\mu=0.
\eea
Although all geometric symmetries are encoded in Killing vectors satisfying (\ref{KVeqn})\footnote{One can also consider conformal Killing vectors, but they don't lead to new symmetries for the metric (\ref{MPeven}). However, the conformal Killing(--Yano) tensors play important role in the dynamics of particles and field in the backgrounds of charged black holes. We refer to \cite{ChLKill} for the detailed discussion.}, equations for particles and fields can have some ``hidden'' symmetries not covered by (\ref{KVeqn}). For example, it is such hidden symmetry that is responsible for separation  (\ref{TeukAnsatz})--(\ref{TeukEqn}) of the wave equation in the Kerr geometry into functions of $r$ and $\theta$. 

Study of hidden symmetries in general relativity was initiated by Carter \cite{Carter1,Carter2}, who demonstrated that separation (\ref{4dScalarEqn}) follows from existence of a symmetric Killing tensor of rank two, which satisfies a differential equation generalizing (\ref{KVeqn}):
\bea\label{KTeqn}
\nabla_{(\mu}K_{\nu\la)}=0.
\eea
While some such tensors can be constructed by combining two Killing vectors $(V,W)$,
\bea\label{KTreduce}
K_{\mu\nu}=V_\mu W_\nu+V_\nu W_\mu\,,
\eea
the Kerr geometry also admits an irreducible object, which cannot be written as (\ref{KTreduce}). 
To present the explicit expression for the Carter's tensor, we introduce convenient frames:
\bea\label{Kerr4DKillFrame}
ds^2&=&-e_t^2+e_r^2+e_\theta^2+e_\phi^2,\nonumber\\
e_{t}&=&\frac{\sqrt{\Delta}}{\rho}(dt-as_\theta^2 d\phi),\quad 
e_{\phi}=\frac{s_\theta}{\rho}\left[(r^2+a^2)d\phi-adt\right],\quad
e_{r}=\frac{\rho}{\sqrt{\Delta}}dr,\quad e_{\bf\theta}=\rho d\theta,\nonumber\\
\Delta&=&r^2+a^2-2mr,\quad \rho^2=r^2+a^2c_\theta^2,\quad
c_\theta=\cos\theta,\quad s_\theta=\sin\theta,
\eea
in which the Killing tensor becomes diagonal:
\bea\label{Kerr4DKT}
K&=&r^2\left[e_\phi^2+e_\theta^2\right]+(ac_\theta)^2\left[e_t^2-e_r^2\right]\,.
\eea
While a Killing tensor has a freedom of shifting by ``trivial'' terms (\ref{KTreduce}), the Kerr black hole also admits a more robust object, which is uniquely defined. A Killing--Yano tensor (KYT) is an anti--symmetric generalization of the Killing vector (\ref{KVeqn}), with defining relation
\bea
\nabla_\mu Y_{\nu_1\dots \nu_p}+\nabla_{\nu_1} Y_{\mu\nu_2\dots \nu_p}=0.
\eea
The Kerr black hole admits the unique rank-two KYT:
\bea\label{Kerr4DKYT}
Y=r e_\theta\wedge e_\phi+(ac_\theta) e_r\wedge e_t\,,
\eea
and once again the frames (\ref{Kerr4DKillFrame}) are very special: they are the closest analogs of eigenvectors that one can define for an antisymmetric tensor. 

The eigensystem of $K$ and $Y$ played an important role in constructing the higher--dimensional generalizations of the Killing(--Yano) tensor \cite{ChLKill}\footnote{An alternative approach, applicable only to neutral black holes, was introduced earlier in \cite{PreKub,Kub1,Kub2,Kub3}. This work is summarized in a very nice recent review \cite{KubRev}.}, and it will be crucial for extending the construction of section \ref{SecKerr} to higher dimensions. The ansatz (\ref{KerrNewAnstz}) relied on the particular frames (\ref{FramesKerr}), and now it is clear what made them special: $(l^\mu,n^\mu)$ are the ``light--cone versions''of the eigenvectors $(e_{t},e_r,e_\theta,e_\phi)$:
\bea
&&\hskip -0.8cm l^\mu\d_\mu=\frac{r^2+a^2}{\Delta}\d_t+\d_r+\frac{a}{\Delta}\d_\phi,\quad
n^\mu\d_\mu=\frac{r^2+a^2}{2\Sigma}\d_t-\frac{\Delta}{2\Sigma}\d_r+\frac{a}{2\Sigma}\d_\phi,\\
&&\hskip -0.8cm m^\mu\d_\mu=\frac{1}{\sqrt{2}\rho}\left[{ia s_\theta}\d_t+\d_\theta+\frac{i}{s_\theta}\d_\phi\right],
\quad \rho=r+ia c_\theta,\quad \Sigma=\rho{\bar\rho},\quad \Delta=r^2+a^2-2Mr.\nonumber
\eea 
Thus to extend the ansatz  (\ref{KerrNewAnstz}) to the Myers--Perry black hole, we should first find the eigenvectors of the Killing--Yano tensors in higher dimensions.
Fortunately this problem was solved in \cite{ChLKill}, and the answer reads\footnote{For compactness we write only the frames with upper indices $e_A^\mu$, which will be used in the subsequent sections. Explicit expressions for $e^A_\mu$ can be found in \cite{ChLKill}. To simplify expressions encountered in this article we made a replacement $x_i\rightarrow -x_i^2$ in comparison with \cite{ChLKill}.}
\bea\label{AllFramesMP}
e_t&=&-\sqrt{\frac{R^2}{FR(R-Mr)}}\left[\d_t-
\sum_k\frac{a_k}{r^2+a_k^2}\d_{\phi_k}\right],\quad 
e_r=\sqrt{\frac{R-Mr}{FR}}\d_r,\nn
e_i&=&-\sqrt{\frac{H_i}{d_i(r^2+x^2_i)}}\left[\d_t-\sum_k\frac{a_k}{a_k^2-x_i^2}\d_{\phi_k}
\right],\quad e_{x_i}=\sqrt{\frac{H_i}{d_i(r^2+x_i^2)}}\d_{x_i}\,.
\eea
Here we defined convenient expressions
\bea\label{MiscElliptic}
d_i=\prod_{k\ne i}(x_k^2-x_i^2),\ H_i=\prod_k(a_k^2-x_i^2),\
G_i= \prod_k(a^2_i-x_k^2),\
 c_i^2=\prod_{k\ne i}(a_i^2-a_k^2).
\eea
In terms of the new coordinates $(r,x_i)$, functions $F$ and $FR$ entering (\ref{AllFramesMP}) become
\bea\label{MisclEllipticEvev}
R=\prod_{k} (r^2+a_k^2),\quad
FR=\prod_k(r^2+x_k^2).
\eea
For completeness we also write the relation between the elliptic coordinates $\{x_k\}$ and the original variables 
$\{\mu_k\}$ 
\bea\label{EplsdEven}
(a_i\mu_i)^2=\frac{1}{c_i^2}\prod_{k=1}^{n} (a_i^2-x^2_k),\quad
0<x_1<a_1<\dots<x_n<a_n\,.
\eea
Note that, apart from the common overall factors, the components $(e_t^\mu,e_r^\mu)$ of the frames depend only on $r$, while the components $(e_i^\mu,e_{x_i}^\mu)$ depend only on $x_i$. As we will see, this crucial fact is responsible for separation of variables in Klein--Gordon and Maxwell equations. 
\bigskip

In terms of the frames (\ref{AllFramesMP}) the metric and the Killing tensor become
\bea\label{SymmKT}
&&ds^2=-(e^t)^2+(e^r)^2+\sum_k [(e^{x_k})^2+(e^k)^2],\nn
&&K_{MN}dx^Mdx^N=\Lambda_r[-(e^t)^2+(e^r)^2]+\sum_k \Lambda_k [(e^{x_k})^2+(e^k)^2],
\eea
where $\Lambda_r(r)$ and $\Lambda_k(x_k)$ are symmetric polynomials. The Killing--Yano tensors are summarized by a very nice formula\footnote{For the Myers--Parry black hole this compact result was first derived in \cite{Kub1,Kub2,Kub3} without relying on frames (\ref{AllFramesMP}), and in \cite{ChLKill} it was extended to charged geometries.}
\bea\label{Kub2}
Y^{2(n-k)}=\star\left[\wedge h^k\right].
\eea
Here $h$ has a very simple expression in terms of frames (\ref{AllFramesMP}) \cite{ChLKill}:
\bea\label{HInFrame}
h=re^r\wedge e^t+\sum_i x_i e^{x_i}\wedge e^i.
\eea
We refer to \cite{ChLKill} for further discussion of the Killing--Yano tensors and a special role played by their eigenvectors. It is important to stress that uniqueness of the KYT (\ref{Kub2})--(\ref{HInFrame}) also guarantees the uniqueness of the special frames (\ref{AllFramesMP}). 

\bigskip

We conclude this subsection by a brief discussion of the Myers--Perry black hole in odd dimensions. Instead of starting with (\ref{MPeven}) one should begin with
\bea\label{MPodd}
\hskip -0.3cm
ds^2=-dt^2+\frac{Mr^2}{FR}\Big(dt+\sum_{i=1}^n a_i\mu_i^2 d\phi_i\Big)^2+\frac{FR dr^2}{R-Mr^2}
+\sum_{i=1}^n(r^2+a_i^2)\Big(d\mu_i^2+\mu_i^2 d\phi_i^2\Big).
\eea
In this case the special frames are given by \cite{ChLKill}
\bea\label{AllFramesMPOdd}
e_t&=&-\sqrt{\frac{R^2}{FR(R-Mr^2)}}\left[ \d_t
-\sum_k\frac{a_k}{r^2+a_k^2}\d_{\phi_k}\right],\quad e_r=\sqrt{\frac{R-Mr^2}{FR}}\d_r,\nn
e_i&=&-\sqrt{\frac{H_i}{x^2_id_i(r^2+x^2_i)}}\left[\d_t-\sum_k\frac{a_k}{a_k^2-x_i^2}\d_{\phi_k}
\right],\quad e_{x_i}=\sqrt{\frac{H_i}{x_i^2 d_i(r^2+x_i^2)}}\d_{x_i},\nn
e_\psi&=&-\frac{\prod a_i}{r\prod x_k}\left[\d_t-\sum_k\frac{1}{a_k}\d_{\phi_k}
\right].
\eea
The relation (\ref{EplsdEven}) between Myers--Perry and ellipsoidal coordinates, as well as expression (\ref{MisclEllipticEvev}) for function $FR$ are modified\footnote{In contrast to the even-dimensional case, where $\mu_i$ were not constrained, now there is a relation $\sum \mu_i^2=1$, and, as a consequence, there only $n-1$ coordinates $x_i$.}:
\bea\label{EplsdOdd}
\mu_i^2=\frac{1}{c_i^2}\prod_{k=1}^{n-1} (a_i^2-x^2_k),\quad
R=\prod_{k}^n (r^2+a_k^2), \quad FR=r^2\prod_k(r^2+x^2_k).
\eea
The remaining relations (\ref{MiscElliptic}) still hold. As in the even--dimensional case, we emphasize 
a very special form of the relative coefficients in frames $e_a$: they depend only on $r$ in 
$e_t$, only on $x_i$ in $e_i$, and they are constant in $e_\psi$. The Killing and Killing--Yano tensors still have the form (\ref{SymmKT}), (\ref{Kub2})--(\ref{HInFrame}), although the metric acquires an extra term $(e_\psi)^2$, and we refer to \cite{ChLKill} for the detailed discussion. 

In the remaining part of this paper the special frames (\ref{AllFramesMP}) and (\ref{AllFramesMPOdd}) will be used to solve Maxwell's equations in the background of the Myers--Perry black hole, but before starting this discussion it is useful to review the separation of variables in the wave equation to stress some peculiarities associated with higher dimensions. 

\subsection{Separation of the wave equation}
\label{SecMPwave}

In this subsection we will analyze the wave equation in the Myers--Perry geometry, and the difference in the structure of frames (\ref{AllFramesMP}) and (\ref{AllFramesMPOdd}) suggests to separate the discussion of even and odd dimensions. We begin with the even--dimensional case.

The goal of this subsection is to study the wave equation:
\bea\label{WaveEqn}
\frac{1}{\sqrt{-g}}\d_\mu\left[\sqrt{-g}g^{\mu\nu}\d_\nu \Psi\right]=0.
\eea
While the expression for the matrix $g^{\mu\nu}$ is trivially encoded in the frames (\ref{AllFramesMP}), the evaluation of the determinant requires some algebra, and the result is
\bea
\sqrt{-g}=\frac{FR}{[\prod a_i]}\sqrt{\prod \frac{d_i}{c_i^2}}\,.
\eea
A full separation of variables in (\ref{WaveEqn}) is guaranteed by the existence of the family of the Killing tensors (\ref{SymmKT}), and we refer to \cite{ChLKill} for the detailed discussion of this approach based on symmetries\footnote{See also earlier mathematical work \cite{Moon,KalMil1} for the general discussion of the relationship between Killing tensors and separation of variables in the wave and Klein--Gordon equations.}. To compare to electromagnetic field in subsequent sections, we need the explicit form of ordinary differential equations for various pieces of $\Psi$, and although such expressions can be extracted from the conserved quantities associated with Killing tensors, it is easier to construct the equations directly from (\ref{WaveEqn}). 

We begin with rewriting the frames (\ref{AllFramesMP}) as
\bea\label{RscldFrmsMPeven}
e_t=-\frac{{\tilde e}_t}{\sqrt{FR}},\quad 
e_r=\frac{{\tilde e}_r}{\sqrt{FR}},\quad
e_i=-\frac{{\tilde e}_i}{\sqrt{d_i(r^2+x^2_i)}},\quad e_{x_i}=\frac{{\tilde e}_{x_i}}{\sqrt{d_i(r^2+x_i^2)}}
\,.
\eea
The coefficients in $({\tilde e}_t,{\tilde e}_r)$ depend only on $r$, while coefficients in 
$({\tilde e}_i,{\tilde e}_{x_i})$ depend only on $x_i$. The inverse metric becomes
\bea\label{gUpEven}
g^{\mu\nu}\d_\mu\d_\nu&=&\frac{1}{FR}[-({\tilde e}_t)^2+({\tilde e}_r)^2]+
\sum \frac{1}{{d_i(r^2+x^2_i)}}[({\tilde e}_i)^2+({\tilde e}_{x_i})^2]\\
&\equiv&\frac{1}{FR}{\tilde g}_r^{\mu\nu}\d_\mu\d_\nu+
\sum \frac{1}{{d_i(r^2+x^2_i)}}{\tilde g}_i^{\mu\nu}\d_\mu\d_\nu\,.\nonumber
\eea
Upon multiplication of this expression by $\sqrt{-g}$, the factor in front of 
${\tilde g}_r\equiv [-({\tilde e}_t)^2+({\tilde e}_r)^2]$ becomes $r$--independent, and the factor in front of 
${\tilde g}_i\equiv [({\tilde e}_i)^2+({\tilde e}_{x_i})^2]$ looses the $x_i$--dependence. This is one of the key properties leading to separation of the wave equation (\ref{WaveEqn}), which can be written as
\bea\label{Jul19temp}
\sqrt{\prod {d_i}}\,\d_\mu[{\tilde g}_r^{\mu\nu}\d_\nu\Psi]+
\sum \frac{FR \sqrt{\prod {d_k}}}{d_i(r^2+x^2_i)}\d_\mu[{\tilde g}_i^{\mu\nu}\d_\nu\Psi]=0.
\eea
Note that $\sqrt{\prod {d_i}}$ is a polynomial of degree $n-1$ in all $(x_k)^2$, and 
\bea
\frac{FR \sqrt{\prod {d_k}}}{d_i(r^2+x^2_i)}\nonumber
\eea
is a polynomial of degree $n-1$ in $r^2$ and in all $(x_k)^2$ with the exception of $k=i$. If we impose a separable ansatz,
\bea
\Psi=E\Phi(r)\left[\prod X_i(x_i)\right],\qquad E=e^{i\omega t+i\sum m_i\phi_i}\,,
\eea
then consistency of equation (\ref{Jul19temp}) implies that 
\bea\label{SeparWaveRad}
\d_\mu[{\tilde g}_r^{\mu\nu}\d_\nu (E\Phi)]=P_{n-1}[r^2]E\Phi,
\eea
where $P_{n-1}$ is an arbitrary polynomial of degree $n-1$. For $n=1$, $P_{n-1}$ reduces to a familiar 
{\it separation constant}, and this case was discussed in section \ref{SectSubKerrWave}: the parameter $\la$ appearing in (\ref{4dScalarEqn}) is a four--dimensional version of the polynomial $P_{n-1}$. 

Equation (\ref{4dScalarEqn}) also implies relations similar to (\ref{SeparWaveRad}) for functions $X_k$, and it constrains the coefficients of various polynomials. A detailed analysis presented in the 
Appendix \ref{SecAppMPWave} shows that $P_{n-1}[r^2]$ remains free, while all other polynomials are determined in terms of it:
\bea\label{SeparWaveX}
\d_\mu[{\tilde g}_k^{\mu\nu}\d_\nu (EX_k)]=-P_{n-1}[-x_k^2]EX_k\,.
\eea
For future reference we rewrite equations (\ref{SeparWaveRad}) and (\ref{SeparWaveX}) in a more explicit form:
\bea\label{SeparWaveEvenFull}
&&\frac{d}{dr}\left[(R-Mr)\frac{d\Phi}{dr}\right]+\frac{R^2}{R-Mr}
\left[\omega-
\sum_k\frac{a_k m_k}{r^2+a_k^2}\right]^2\Phi=P_{n-1}(r^2)\Phi,\nn
&&\frac{d}{dx_i}\left[H_i\frac{dX_i}{dx_i}\right]-
H_i\left[\omega-
\sum_k\frac{a_k m_k}{a_k^2-x_i^2}\right]^2X_i=-P_{n-1}(-x_i^2)X_i.
\eea
The set of equations (\ref{SeparWaveEvenFull}) should be viewed as an eigenvalue problem for the coefficients of the polynomial $P_{n-1}$. Separation of the Klein--Gordon equation is obtained as a straightforward extension of (\ref{SeparWaveEvenFull}), but we will not need these more cumbersome formulas.

\bigskip

We conclude this subsection by a brief discussion of the wave equation in odd dimensions. Using the frames (\ref{AllFramesMPOdd}), we find a counterpart of relation (\ref{gUpEven}):
\bea\label{gUpOdd}
g^{\mu\nu}\d_\mu\d_\nu=\frac{1}{FR}{\tilde g}_r^{\mu\nu}\d_\mu\d_\nu+
\sum \frac{1}{{d_i(r^2+x^2_i)}}{\tilde g}_i^{\mu\nu}\d_\mu\d_\nu+\frac{1}{r^2[\prod x_i^2]}
{\tilde g}_\psi^{\mu\nu}\d_\mu\d_\nu\,,
\eea
where ${\tilde g}_\psi^{\mu\nu}$ is a constant matrix with indices along $(t,\phi_i)$. Using the expression for the determinant of the metric,
\bea\label{OddDetMet}
\sqrt{-g}=\frac{FR}{r}\sqrt{\prod \frac{x_i^2 d_i}{c_i^2}},
\eea
and repeating the steps leading to (\ref{SeparWaveEvenFull}), we find
\bea\label{SeparWaveOddFullMain}
&&r\frac{d}{dr}\left[\frac{R-Mr^2}{r}\frac{d\Phi}{dr}\right]+\frac{R^2}{R-Mr^2}
\left[\omega-
\sum_k\frac{a_k m_k}{r^2+a_k^2}\right]^2\Phi=P_{n}[r^2]\Phi,\nn
&&{x_i}\frac{d}{dx_i}\left[\frac{H_i}{x_i}\frac{dX_i}{dx_i}\right]-
H_i\left[\omega-
\sum_k\frac{a_k m_k}{a_k^2-x_i^2}\right]^2X_i=-P_{n}[-x_i^2]X_i.
\eea
In contrast to the even--dimensional case, the polynomial $P_n$ has degree $n$, but it is subject to one constraint:
\bea\label{P0cnstrnt}
P_n[0]=\left[{\prod a_i}\right]^2\left[\omega-\sum_k\frac{m_k}{a_k}
\right]^2.
\eea
We refer to Appendix \ref{SecAppMPWave} for details. Equations (\ref{SeparWaveOddFullMain}) should be viewed as an eigenvalue problem for the coefficients of the polynomial $P_{n}$. 

\subsection{Extension to the Myers--Perry--(A)dS geometry}
\label{SecAdSFramesMP}

The results reviewed in this section can be easily extended to higher--dimensional solutions of Einstein's equations in the presence of the cosmological constant. Construction of such geometries was a result of a very impressive work \cite{GLPP}\footnote{The five--dimensional Kerr-AdS solution was found earlier in  \cite{KerrAdS5}.}, but once the final Gibbons--Lu--Page--Pope (GLPP) metrics are written, their form suggests that the separable frames can be obtained by a simple modification of
(\ref{AllFramesMP}) and (\ref{AllFramesMPOdd}). In this subsection we present the resulting frames (\ref{FramesMPAeven}), (\ref{AdSFramesMPOddAdS}) and derive the systems of ODEs (\ref{SeparWaveEvenAdS}),
(\ref{SeparWaveOddAdS}) governing the dynamics of separable solutions of the wave equation.

\bigskip

The GLPP solution describes rotating black holes in the presence of the cosmological constant, so away from the sources the metric solves the Einstein's equations 
\bea
R_{\mu\nu}=(D-1)L g_{\mu\nu},
\eea
where $L$ is related to the cosmological constant\footnote{We reserve symbols $\la$ and $\Lambda$ for the eigenvalues associated with Maxwell's equations.}. As in the Myers--Perry case, one should study the even and odd dimensional cases separately, and in $D=2n+2$ dimensions, the GLPP solution reads \cite{GLPP}\footnote{We made replacements $M\rightarrow M/2$ while quoting equations (3.1) and (3.5) of \cite{GLPP} to agree with the Myers--Perry notation.}
\bea\label{GLPPeven}
ds^2&=&-W(1-Lr^2)dt^2+\sum_{j=1}^{n}\frac{r^2+a_j^2}{1+La_j^2}\left[d{\tilde \phi}_j-\la a_j dt\right]^2+\frac{M}{U}\left[dt-\sum_{j=1}^{n}\frac{a_j \mu_j^2 d{\tilde \phi}_j}{1+La_j^2}\right]^2
\nn
&&+\sum_{j=1}^{n+1}\frac{r^2+a_j^2}{1+La_j^2}d\mu_j^2+\frac{L}{W(1-Lr^2)}
\left[\sum_{j=1}^{n+1}\frac{(r^2+a_j^2)\mu_j d\mu_j}{1+La_j^2}\right]^2+
\frac{Udr^2}{V-{M}}\,.
\eea
Here functions $(U,V,W)$ are defined by
\bea
U=r\left[\sum_{k=1}^{n+1}\frac{\mu_k^2}{r^2+a_k^2}\right]\prod_{j=1}^n (r^2+a_j^2),\quad
V=\frac{1-L r^2}{r}\prod_{j=1}^n(r^2+a_j^2),\quad
W=\sum_{k=1}^{n+1}\frac{\mu_k^2}{1+La_k^2}\,.\nonumber
\eea
Angular coordinates ${\tilde\phi}_i$ entering (\ref{GLPPeven}) have the standard periodicity, but to simplify expressions associated with electromagnetic field, it is convenient to define rescaled coordinates $\phi_i$:
\bea
\phi_i=\sqrt{1+La_i^2}{\tilde\phi}_i,\quad 0\le \phi_i<2\pi \sqrt{1+La_i^2}\,.
\eea
The easiest way to find the counterparts of the special frames (\ref{AllFramesMP}) for the geometry (\ref{GLPPeven}) is to separate the wave equation\footnote{One can also find the eigenvalues of the Killing--Yano tensors constructed in \cite{Kub1}.}. We refer to \cite{ChLKill} for the detailed discussion of this approach. The resulting special frames are related to (\ref{AllFramesMP}) by a very simple transformation
\bea\label{FramesMPAeven}
&&\hskip -1.5cm e_t=-\frac{1}{Q_r}\sqrt{\frac{R^2}{FR(R-Mr)}}\left[\d_t-
\sum_k\frac{a_k}{r^2+a_k^2}\d_{\phi_k}\right],\quad 
e_r=Q_r\sqrt{\frac{R-Mr}{FR}}\d_r,\nn
&&\hskip -1.5cm e_i=-\frac{1}{Q_i}\sqrt{\frac{H_i}{d_i(r^2+x^2_i)}}\left[\d_t-\sum_k\frac{a_k}{a_k^2-x_i^2}\d_{\phi_k}
\right],\quad e_{x_i}=Q_i\sqrt{\frac{H_i}{d_i(r^2+x_i^2)}}\d_{x_i}\,.
\eea
The dependence on the cosmological constant comes only through the ``dressing factors'' $(Q_r,Q_j)$, which are defined by
\bea\label{QfactDg4}
Q_r=\sqrt{1-Lr^2\frac{R}{R-Mr}},\quad Q_j=\sqrt{1+Lx_j^2}\,,
\eea
and the remaining notation used in (\ref{FramesMPAeven}) is described in section \ref{SecSubKilling}. Separation of the wave equation leads to a minor modification of (\ref{SeparWaveEvenFull}):
\bea\label{SeparWaveEvenAdS}
&&\frac{d}{dr}\left[(R-Mr)Q_r^2\frac{d\Phi}{dr}\right]+\frac{R^2}{Q_r^2(R-Mr)}
\left[\omega-
\sum_k\frac{a_k m_k}{r^2+a_k^2}\right]^2\Phi=P_{n-1}(r^2)\Phi\,,\nn
&&\frac{d}{dx_i}\left[H_iQ_i^2\frac{dX_i}{dx_i}\right]-
\frac{H_i}{Q_i^2}\left[\omega-
\sum_k\frac{a_k m_k}{a_k^2-x_i^2}\right]^2X_i=-P_{n-1}(-x_i^2)X_i\,.
\eea
As in the case of the frames (\ref{FramesMPAeven}), the cosmological constant enters the wave equation only through the ``dressing factors'' $(Q_r,Q_j)$.
\bigskip

In odd dimensions, the GLPP solution is still given by (\ref{GLPPeven}), but now the variables $\mu_i$ are constrained by the relation
\bea
\sum_{j=1}^n \mu_j^2=1,\nonumber
\eea
and the expression for the function $V$ is modified:
\bea
V=\frac{1-L r^2}{r}\prod_{j=1}^n(r^2+a_j^2).
\eea
Both features have been already encountered for the Myers--Perry black hole.

The special frames are obtained by a slight modification of (\ref{AllFramesMPOdd}), 
\bea\label{AdSFramesMPOddAdS}
e_t&=&-\frac{1}{Q_r}\sqrt{\frac{R^2}{FR(R-Mr^2)}}\left[ \d_t
-\sum_k\frac{a_k}{r^2+a_k^2}\d_{\phi_k}\right],\quad e_r=Q_r\sqrt{\frac{R-Mr^2}{FR}}\d_r,\nn
e_i&=&-\frac{1}{Q_i}\sqrt{\frac{H_i}{x^2_id_i(r^2+x^2_i)}}\left[\d_t-\sum_k\frac{a_k}{a_k^2-x_i^2}\d_{\phi_k}
\right],\quad e_{x_i}=Q_i\sqrt{\frac{H_i}{x_i^2 d_i(r^2+x_i^2)}}\d_{x_i},\nn
e_\psi&=&-\frac{\prod a_i}{r\prod x_k}\left[\d_t-\sum_k\frac{1}{a_k}\d_{\phi_k}
\right],
\eea
and the wave equation reduces to (\ref{SeparWaveOddFullMain}) with some additional factors of $(Q_r,Q_j)$ as in the even--dimensional case (\ref{SeparWaveEvenAdS}):
\bea\label{SeparWaveOddAdS}
&&r\frac{d}{dr}\left[\frac{Q_r^2\Delta}{r}\frac{d\Phi}{dr}\right]+\frac{R^2}{Q_r^2\Delta}
\left[\omega-
\sum_k\frac{a_k m_k}{r^2+a_k^2}\right]^2\Phi=P_{n}[r^2]\Phi,\quad \Delta=R-Mr^2,\nn
&&{x_i}\frac{d}{dx_i}\left[\frac{H_i Q_i^2}{x_i}\frac{dX_i}{dx_i}\right]-
\frac{H_i}{Q_i^2}\left[\omega-
\sum_k\frac{a_k m_k}{a_k^2-x_i^2}\right]^2X_i=-P_{n}[-x_i^2]X_i\,.
\eea
Although in this article we will mostly focus on the Myers--Perry metric with $L=0$, we will comment on Maxwell's equations in the GLPP geometry in section \ref{SecKerrAdSMxwMP}.

\bigskip

\bigskip

To summarize, in this section we have reviewed the structure of the Myers--Perry black hole and its symmetries. In particular, following \cite{ChLKill}, we have introduced the special frames (\ref{AllFramesMP}) and (\ref{AllFramesMPOdd}), which will play the central role in the rest of this article. We have also separated the wave equation in the background of the Myers--Perry black hole, and the final results (\ref{SeparWaveEvenFull}), (\ref{SeparWaveOddFullMain}) will serve as a guide for separating the 
Maxwell's equations. 

As we have seen, the symmetry structure of the Myers--Perry black hole differs between the even and odd dimensions, so these two cases should be discussed separately. The solutions for the even dimensions will be obtained by generalizing the construction presented in section \ref{SecKerrNewAns}, and in the next section we will discuss the five--dimensional black hole, which will serve as a similar starting point for odd dimensions. We will come back to the general Myers--Perry black hole in section \ref{SecWaveMP}.

\section{All excitation of the five--dimensional black hole}
\renewcommand{\theequation}{4.\arabic{equation}}
\setcounter{equation}{0}
\label{SecKerr5D}

In this section we will demonstrate separation of variables for Maxwell's equations in the background of a 
five--dimensional black hole and construct all three polarizations of the photons. Following the conventions of section \ref{SecKerrNewAns}, we separate polarizations into ``electric'' and ``magnetic', depending on their limit at $\omega=0$. As expected from the general properties of electromagnetic fields reviewed in Appendix \ref{SecAppSchw}, there is one electric polarization (which reduces to $A_t$ in the static limit), and two magnetic ones. 

\bigskip

The rotating five--dimensional black holes have been extensively used in the context of string theory \cite{BMPV,5dBH,CveticLarsen,KYTbh}, and in this special case the most common notation in the literature differs from the general parameterization (\ref{MPodd}). Specifically, instead of using two constrained variables $(\mu_1,\mu_2)$, one introduces a free angle $\theta$:
\bea
\mu_1=s_\theta,\quad \mu_2=c_\theta,
\eea
and uses special symbols for the coordinates $(\phi_1,\phi_2,a_1,a_2)$:
\bea
\phi_1=\phi,\quad \phi_2=\psi,\quad a_1=-a,\quad a_2=-b\,.
\eea
Note that in the five--dimensional case, there is only one $x$ coordinate in (\ref{AllFramesMPOdd}), which is related to $\theta$ in a simple way (see (\ref{EplsdOdd})):
\bea\label{X1asTheta5D}
x_1=\sqrt{(a c_\theta)^2+(b s_\theta)^2}\,.
\eea
To connect to the existing literature and to simplify the limits of vanishing $a$ and $b$,
we will use $\theta$ instead of $x_1$. Then the frames (\ref{AllFramesMPOdd}) become
\bea\label{AllFramesMPOdd5D}
e_t&=&\frac{R}{r\sqrt{\Sigma\Delta}}\left[ \d_t
-\sum_k\frac{a_k}{r^2+a_k^2}\d_{\phi_k}\right],\quad e_r=\sqrt{\frac{\Delta}{r^2\Sigma}}\d_r,
\quad e_{\theta}=\frac{1}{\sqrt{\Sigma}}\d_{\theta},\\
e_1&=&\frac{s_\theta c_\theta}{\Theta\sqrt{\Sigma}}\left[(a^2-b^2)\d_t+\frac{a}{s_\theta^2}\d_\phi
-\frac{b}{c^2_\theta}\d_\psi
\right],\quad 
e_\psi=\frac{1}{r\Theta}\left[ab\d_t+b\d_\phi+a\d_\psi
\right].\nonumber
\eea
To make these and subsequent formulas more compact, we  flipped signs of some frames and introduced notation inspired by the four--dimensional case (\ref{FramesKerr})
\bea\label{Delta5d}
\Delta=R-Mr^2,\quad \Sigma=r^2+(a c_\theta)^2+(b s_\theta)^2,\quad 
\Theta=\sqrt{(a c_\theta)^2+(b s_\theta)^2}\,.
\eea
Recall that in five dimensions the general definition (\ref{EplsdOdd}) gives
\bea\label{DeltaA5d}
R=(r^2+a^2)(r^2+b^2).
\eea
Mimicking the expression for the four--dimensional canonical vierbein (\ref{FramesKerr}), we combine the frames corresponding to $r$ and $\theta$ coordinates and define
\bea\label{LCframes5D}
l^\mu_\pm=\sqrt{\Sigma\Delta}\,(e^\mu_r\pm e^\mu_t),\quad 
m^\mu_\pm=\sqrt{\Sigma}\,(e^\mu_\theta\pm i e^\mu_1),\quad 
n^\mu=r\Theta\, e_\psi^\mu\,.
\eea
From now on we will work only with frames (\ref{LCframes5D}), so there should be no confusions between the frame and the space--time indices. In the rescaled frames (\ref{LCframes5D}), the inverse metric becomes
\bea\label{Metr5D}
g^{\mu\nu}\d^\mu\d^\nu=\frac{1}{\Sigma\Delta}l_+^\mu l_-^\nu\d_\mu\d_\nu+
\frac{1}{\Sigma}m_+^\mu m_-^\nu\d_\mu\d_\nu+\frac{1}{r\Theta}n^\mu n^\nu\d_\mu\d_\nu\,.
\eea

Note that components of $l_\pm^\mu$ depend only in $r$, $m_\pm^\mu$ are functions of $\theta$, and $n^\mu$ are constants. Thus, using (\ref{KerrNewAnstz}) as an inspiration,  it is very natural to propose the following ansatz for the gauge field:
\bea\label{5dNewAnstz}
l_\pm^\mu A_\mu=G_\pm (r)l_\pm^\mu\d_\mu \Psi,\qquad
m_\pm^\mu A_\mu=F_\pm(\theta)m_\pm^\mu\d_\mu\Psi,\quad n^\mu A_\mu=\la \Psi,
\eea
where $\Psi$ is a separated scalar function
\bea\label{Psi5D}
\Psi=e^{i\omega t+im\phi+in\psi}\Phi(r)S(\theta).
\eea
The rest of this section is dedicated to exploration of the ansatz (\ref{5dNewAnstz}). As in the four--dimensional case, we will demonstrate that Maxwell's equations uniquely determine the factors $(G_\pm (r),F_\pm(\theta))$ and lead to very simple equations for functions $(R,S)$. Readers not interested in justifications can go directly to the subsection \ref{SecSubSmry5D}, which summarizes our results. 

\subsection{Electro-- and magnetostatics}
\label{SectSub5dStat}

Following the logic of section \ref{SecKerrNewAns}, we begin with applying the ansatz (\ref{5dNewAnstz}) to the special configurations with
\bea\label{Special5D}
\omega=0,\quad m=n=0.
\eea
In this subsection we will derive the most general expression for $(F_\pm,G_\pm)$ and equations for 
$(\Phi,S)$, and demonstrate that separable configurations in the special case (\ref{Special5D}) must reduce to one of the two branches, (\ref{Special5DElectr}) or (\ref{Special5dMagn}). Then in the next subsection the restrictions  (\ref{Special5D}) will be relaxed following the logic outlined on page \pageref{Steps4D}, resulting in the final expressions (\ref{FullElectr5d}) and (\ref{FullMagn5d}) for the two branches.

\bigskip

In the special case (\ref{Special5D}), $A_r$ and $A_\theta$ decouple from the remaining components in Maxwell's equations, and using the determinant of the metric,
\bea
\sqrt{-g}=r s_\theta c_\theta\Sigma,
\eea
we find the unique expression for $F^{r\theta}$:
\bea
F^{r\theta}=\frac{\mbox{const}}{\sqrt{-g}}=\frac{\mbox{const}}{r s_\theta c_\theta\Sigma}\,.
\eea
Solutions of this type do not allow separation constants, so dropping a pure gauge, we can set 
$A_r=A_\theta=0$. This implies that in the special case (\ref{Special5D}), the ansatz (\ref{5dNewAnstz}) has
\bea
G_-=-G_+,\quad F_-=-F_+.
\eea
As in section \ref{SecKerrNewAns}, we define the components of Maxwell's equations by a counterpart of (\ref{MaxwNotation}):
\bea\label{MaxwNotation5d}
{\MM}^\mu\equiv \frac{e^{-i\omega t-im\phi-in\psi}}{\sqrt{-g}}\d_\nu\left[\sqrt{-g}F^{\mu\nu}\right].
\eea
Then looking at the special case (\ref{Special5D}) and setting $a=b=0$, while keeping the ratio $a/b$ fixed, we find
\bea\label{MaxwKerrStatic5D}
\MM^t&=&-\frac{1}{r^2}\left\{G_+{\dot \Phi}\frac{(s_{2\theta} S')'}{s_{2\theta}}+rS\frac{d}{dr}\left(\frac{1}{r}\frac{d}{dr}[(r^2-M)G_+{\dot \Phi}]\right)\right\}.
\eea
For configurations with $G_+\ne 0$, Maxwell's equation $\MM^t=0$ reduces to two ODEs\footnote{Solutions with $G_+=0$ will be discussed after equation (\ref{Kerr5DGpE0}).}:
\bea
\MM^t=0:\ \frac{(s_{2\theta} S')'}{s_{2\theta}}=-\lambda_1 S,\quad 
r\frac{d}{dr}\left(\frac{1}{r}\frac{d}{dr}[(r^2-M)G_+{\dot \Phi}]\right)=\la_1 G_+{\dot \Phi}.
\eea
Interestingly, the ratio $a/b$ does not enter these equations, as we will see, this is a peculiar feature of the electric polarization, which is not shared by its magnetic counterpart. As in section \ref{SectSubKerrStat}, dimensional analysis ensures that the angular equation is not modified in the presence of rotation parameters, this leads to applicability of equation 
\bea\label{SeqnElectr5D}
(s_{2\theta} S')'+\lambda_1 {s_{2\theta}}S=0
\eea
to all electric modes without dependence on cyclic coordinates (see (\ref{Special5D})). Turning on the rotations and using relation (\ref{SeqnElectr5D}) to eliminate $S''$ and $S'''$ from Maxwell's equations, we arrive at the five--dimensional counterpart of (\ref{tempEqnJul15}):
\bea\label{tempEqnJul15d5}
(l_+^\mu-l_-^\mu)\MM_\mu\Big|_{S=0}=\frac{2R S'}{r\Theta\Sigma^2}\left[{2i\Phi \Theta^3}\frac{d}{d\theta}\frac{F_+}{\Theta}+
(a^2-b^2)s_{2\theta} {\dot \Phi}(\Theta G_+-i rF_+)\right]=0.
\eea
As in four dimensions, by requiring function $\Phi$ to satisfy a second--order differential equation with separation constant $\la_1$, we conclude that coefficients in front of $\Phi$ and ${\dot\Phi}$ in (\ref{tempEqnJul15d5}) must vanish, leading to a counterpart of (\ref{FGplusElectr})\footnote{In the degenerate case $a=\pm b$ some freedom in $G_+$ still remains, but we will not discuss it here.}
\bea\label{FGplusElectr5d}
F_+=-i\Theta,\quad G_+=r.
\eea
Substitution of (\ref{SeqnElectr5D}) and (\ref{FGplusElectr5d}) into Maxwell's equations leads to the {\it unique} solution for the electric polarization of the special configurations (\ref{Special5D}):
\bea\label{Special5DElectr}
&&l_\pm^\mu A^{(el)}_\mu=\pm r {\hat l}_\pm \Psi,\quad
m_\pm^\mu A^{(el)}_\mu=
\mp i \Theta {\hat m}_\pm \Psi,\quad 
n^\mu A^{(el)}_\mu=0,\quad \Psi=\Phi(r)S(\theta),\nn
&&(s_{2\theta} S')'+\lambda_1 {s_{2\theta}}S=0,\quad 
\frac{1}{r}\frac{d}{dr}\left[\frac{\Delta}{r} \dot R\right]-\la_1 R=0.
\eea
While the electric polarization is very similar in even and odd dimensions, the structures of magnetic polarizations for these two cases are very different, and we will now discuss such modes for the five--dimensional black hole.

\bigskip

We recall that the solution (\ref{Special5DElectr}) has been rigorously derived from equation (\ref{SeqnElectr5D}), which was based on only one assumption: $G_+\ne 0$ in (\ref{MaxwKerrStatic5D}). Thus to describe the magnetic polarizations, we must require $G_+$ to vanish in the non--rotating limit:
\bea\label{Kerr5DGpE0}
a=b=0\quad\Rightarrow\quad G_+=0.
\eea
We also set $M=0$. Nontrivial solutions with vanishing $G_+$ must have non--zero $F_+$, and for such configurations one combination of Maxwell's equations is especially simple:
\bea\label{MagnStatic5dInterm}
\frac{bs_\theta^2}{a}\MM^\phi+c_\theta^2\MM^\psi=\frac{2ib\,\Phi}{r^4 s_{2\theta}}\frac{d}{d\theta}\left[\frac{s_{2\theta}F_+S'}{\Theta}\right]+\frac{\la S}{ar^3}\frac{d}{dr}[r{\dot\Phi}]+
\frac{\la\Theta^2\Phi}{ar^4s_{2\theta}}\frac{d}{d\theta}\left[\frac{s_{2\theta}}{\Theta^2}S'\right]+
\frac{4\la ab^2 S\Phi}{(r\Theta)^4}.\nn
\eea
Although the last expression contains $a$ and $b$, it is applicable only to the non--rotating limit with an arbitrary ratio $b/a$. The definition (\ref{5dNewAnstz}) of the constant $\la$ implies that $\la\sim a$ in the non--rotating limit, so all terms in the right hand side of (\ref{MagnStatic5dInterm}) approach finite values as $a$ goes to zero. Consistency of separation leads to equations
\bea\label{MagnStatic5dTwo}
&&\frac{2ib}{s_{2\theta}}\frac{d}{d\theta}\left[\frac{s_{2\theta}F_+S'}{\Theta}\right]+\frac{\la\la_2S}{a}+
\frac{\la\Theta^2}{as_{2\theta}}\frac{d}{d\theta}\left[\frac{s_{2\theta}}{\Theta^2}S'\right]+
\frac{4\la ab^2 S}{\Theta^4}=0,\nn
&&\frac{1}{r^3}\frac{d}{dr}[r{\dot\Phi}]=\la_2 \Phi.
\eea
The second relation is expected: since the limit $a=b=M=0$ removes all length scales from the metric, only the power law solution for $\Phi$ is possible.

Using the first equation in (\ref{MagnStatic5dTwo}) to eliminate $S''$ and $S'''$ from $\MM^\phi$, and 
recalling that coefficients in from of $S$ and $S'$ must vanish separately,
we find an over--constrained system of differential equations for $F_+$. Although the algebra is tedious, the result is very simple: up to an irrelevant multiplicative constant, 
\bea
F_+=\frac{iB}{\Theta},\qquad B\equiv \sqrt{a^2+b^2}.
\eea
Parameter $B$ is introduced just to keep $F_+$ finite in the non--rotating limit. Equation $\MM^\phi=0$ also determines $\la_2$ in terms of $\la$ and $B$, so equations (\ref{MagnStatic5dTwo}) become
\bea\label{A0eqns5D}
\frac{\Theta^2}{s_{2\theta}}\frac{d}{d\theta}\left[\frac{s_{2\theta}}{\Theta^2}S'\right]+
\left[{\tilde\la}^2+\frac{2\tilde\la a b}{\Theta^2}\right]S=0,\quad
\frac{1}{r^3}\frac{d}{dr}[r{\dot\Phi}]={\tilde\la}^2 \Phi,\quad {\tilde\la}=\frac{\la}{B}.
\eea
This completes our discussion of the $a=b=0$ case.

To turn on the rotation parameters, while still keeping (\ref{Special5D}) and $M=0$, we observe that on the dimensional grounds, equation for $S(\theta)$ can depend only on the ratio $a/b$. This implies that the angular equation in (\ref{A0eqns5D}) and $F_+$ are not modified, then by substituting $S''$ and $S'''$ into Maxwell's equations, we find an over--constrained system of ODEs for $G_+$ and $\Phi$. Similar systems were discussed in section \ref{SectSubKerrStat}, so here we skip the intermediate formulas and write the final 
result\footnote{To simplify notation, we replaced ${\tilde\la}$ by $\la$.}:
\bea\label{Special5dMagn}
&&\hskip -1.0cm l_\pm^\mu A^{(mgn)}_\mu=\pm\frac{B}{r} {\hat l} \Psi,\quad
m_\pm^\mu A^{(mgn)}_\mu=\pm\frac{iB}{\Theta}{\hat m}_\pm \Psi,\quad 
n^\mu A^{(mgn)}_\mu=B\la\Psi,\quad
\Psi=R(r)S(\theta),\nn
&&\hskip -1.0cm \frac{\Theta^2}{s_{2\theta}}\frac{d}{d\theta}\left[\frac{s_{2\theta}}{\Theta^2}S'\right]+
\left[{\la}^2+\frac{2\la a b}{\Theta^2}\right]S=0,\quad 
\frac{d}{dr}\left[\frac{\Delta}{r^3}{\dot \Phi}\right]-\left[\frac{\la^2}{r}-\frac{2\la a b }{r^2}\right]\Phi=0.
\eea
Scale $B$ is introduced to ensure a smooth limit as $a$ and $b$ go to zero, later we will remove this unnecessary parameter. 

\bigskip

We stress that our derivation ensures that in the special case (\ref{Special5D}), any nontrivial solution\footnote{We excluded the pure gauge: $G_\pm=F_\pm$, $\la=0$.} of the form (\ref{5dNewAnstz}) {\it must} reduce to either (\ref{Special5DElectr}) or (\ref{Special5dMagn}), and a direct check shows that both systems work even for the black hole geometry, i.e., for arbitrary values of $M$. 

Although we encountered only two systems, (\ref{Special5DElectr}) and (\ref{Special5dMagn}), they describe three different polarizations. To see this, we write more explicit expressions for the gauge field in Schwarzschild geometry and compare the results with the general analysis presented in the Appendix \ref{SecAppSchw}. Setting $a=b=0$ in (\ref{Special5DElectr}) and (\ref{Special5dMagn}), we find electrostatic, 
\bea\label{ElStat5d}
A^{(el)}&=&\frac{\Delta}{r^3}\d_r(\Phi S)dt,\quad 
\frac{(s_{2\theta} S')'}{s_{2\theta}}+\lambda_1 S=0,\quad 
\frac{1}{r}\frac{d}{dr}\left[\frac{\Delta}{r} \dot \Phi\right]-\la_1 \Phi=0,
\eea
and magnetostatic,
\bea\label{MagStat5d}
&&A^{(mgn)}=\frac{B\la}{\Theta^2}(bs_\theta^2 d\phi+a c_\theta^2 d\psi)\Phi S+
\frac{Bs_{2\theta}}{2\Theta^2}(b d\psi-ad\phi)\d_\theta(\Phi S),\\
&&\frac{\Theta^2}{s_{2\theta}}\frac{d}{d\theta}\left[\frac{s_{2\theta}}{\Theta^2}S'\right]+
\left[{\la}^2+\frac{2\la a b}{\Theta^2}\right]S=0,\quad 
\frac{d}{dr}\left[\frac{\Delta}{r^3}{\dot \Phi}\right]-\frac{\la^2}{r}\Phi=0.\nonumber
\eea
configurations. 

The solutions of the eigenvalue problem for the angular equation in (\ref{ElStat5d}) are parameterized by a positive integer $k$, and the profiles are given in terms of the hypergeometric function $F$:
\bea
\la_1=4k(k+1),\quad S=F\Big[-k,k+1;1;c_\theta^2\Big],\quad
\Phi\sim F\Big[-k,k+1;1;\frac{r^2}{M}\Big].
\eea 
For $M=0$ the regular radial function reduces to $\Phi=r^{2k}$. Every value of $k$ leads to the unique angular and radial profiles, so equation (\ref{ElStat5d}) describes one mode. 

In contrast, the magnetic modes (\ref{MagStat5d}) describe two different polarizations for every allowed radial profile. The angular equation has a $Z_2$ symmetry $(a,\la)\rightarrow (-a,-\la)$, which implies that 
all eigenvalues come in pairs $(\la,-\la)$. Of course, the corresponding profiles, $S_\la$ and $S_{-\la}$, are different, but they are related by a formal replacement $a\rightarrow -a$. Eigenvalues $(\la,-\la)$ have identical radial profiles, but the structures of the corresponding $A^{(mgn)}$ are very different, so the system (\ref{MagStat5d}) describes two distinct magnetic polarizations. The same is true even for the full solution  (\ref{Special5dMagn}), although the analysis is less transparent. 

\bigskip

To summarize, we have shown that in the special case (\ref{Special5D}), the ansatz (\ref{5dNewAnstz}) admits only two classes of solutions, (\ref{Special5DElectr}) and (\ref{Special5dMagn}). These systems describe three independent polarizations, as expected from the general analysis presented in Appendix \ref{SecAppSchw}. In the next subsection we will relax the (\ref{Special5D}) and present full solutions for three polarizations of electromagnetic field in the background of a five--dimensional Myers--Perry black hole. 

\subsection{General electromagnetic field in five dimensions}

The results of the last subsection can be extended to the general electromagnetic field satisfying separability condition  (\ref{5dNewAnstz}) by implementing the steps outlined on page \pageref{Steps4D}. In comparison to the four--dimensional case, the algebra is slightly more involved, and some details are presented in the 
Appendix \ref{SecApp5D}. Here we just stress the uniqueness of the resulting solution and present the final expressions.

The separable ansatz for electromagnetic fields (\ref{5dNewAnstz}) leads to two types of solutions, and following the established notation, we will call them ``electric'' and ``magnetic'' polarizations. The electric solution reads
\bea\label{FullElectr5d}
&&l_\pm^\mu A^{(el)}_\mu=\pm\frac{r}{1\pm i\mu r}{\hat l}_\pm\Psi ,\quad
m_\pm^\mu A^{(el)}_\mu=\mp\frac{i\Theta}{1\pm \mu\Theta}{\hat m}_\pm\Psi,\quad 
n^\mu A^{(el)}_\mu=0,\nn
&&\frac{E_\theta}{s_{2\theta}}\frac{d}{d\theta}\Big[\frac{s_{2\theta}}{E_\theta}S'\Big]+
\Big[\frac{2\Lambda}{E_\theta}+\omega^2\Theta^2-\frac{n^2}{c^2_\theta}-
\frac{m^2}{s_\theta^2}+C\Big]S=0,\\
&&\frac{E_r}{r}\frac{d}{dr}\Big[\frac{\Delta}{r E_r}{\dot\Phi}\Big]+
\Big[-\frac{2\Lambda}{E_r}+(\omega r)^2+\frac{m^2 d_a}{R_a}+\frac{n^2 d_b}{R_b}
+\frac{M R W^2}{\Delta}-C\Big]\Phi=0.\nonumber
\eea
Here we introduced convenient functions
\bea\label{Electr5dMfunc}
E_r=1+(\mu r)^2,\quad E_\theta=1-(\mu\Theta)^2,\quad W=\omega+\frac{am}{R_a}+\frac{bn}{R_b},
\quad R_c=r^2+c^2\,.
\eea
Recall that function $\Psi$ is given by (\ref{Psi5D}), while $\Delta$ and $R$ are defined by (\ref{Delta5d}) and (\ref{DeltaA5d}). We also introduced parameters $d_a$ and $d_b$,
\bea
d_a=-d_b=a^2-b^2,
\eea
which will simplify the comparison of (\ref{FullElectr5d}) with its higher--dimensional counterparts. 

Equations (\ref{FullElectr5d}) contain two constants $(\Lambda,C)$, which are completely determined by the separation parameter $\mu$. The limit of vanishing $\omega$ requires careful treatment, and to make this case more transparent, we present $(\Lambda,C)$ in terms of a new parameter $\la$, which remains finite in the limit:
\bea\label{Electr5dLmbd}
&&\Lambda=\frac{1}{\la^3}\left[{\la^2}-(a\omega)^2-a\omega m\right]
\left[{\la^2}-(b\omega)^2-b\omega n\right]-\frac{abmn\omega^2}{\la^3},\\
&&C=\frac{\omega^2}{\la^2}(ab\omega+bm+an)^2-\omega\left[\omega(a^2+b^2)+2am+2 bn\right]
,\quad 
\la=\frac{\omega}{\mu}.\nonumber
\eea
For $\omega=0$ we find
\bea
\mu=0,\quad \Lambda={\la},\quad C=0.
\eea
The angular equation in (\ref{FullElectr5d}) should be viewed as an eigenvalue problem for $\mu$, then   equation for $\Phi$ gives the appropriate radial profile. 

\bigskip

Detailed analysis of the magnetic polarization presented in the Appendix \ref{SecApp5DMagn} leads to the unique extension of the special solution (\ref{Special5dMagn}):
\bea\label{FullMagn5d}
&&l^\mu_\pm A^{(mgn)}_\mu=\pm \frac{1}{r\pm i \mu}{\hat l}_\pm \Psi,\quad
m^\mu_\pm A^{(mgn)}_\mu=\pm\frac{i}{\Theta\mp \mu}{\hat m}_\pm \Psi,\quad 
n^\mu A^{(mgn)}_\mu=\la \Psi,\nn
&&\frac{M_\theta}{s_{2\theta}}\frac{d}{d\theta}\left[\frac{s_{2\theta}}{M_\theta}S'\right]+
\left[\frac{2\Lambda}{M_\theta}+\omega^2\Theta^2-\frac{m^2}{s_\theta^2}-\frac{n^2}{c_\theta^2}+
C\right]S=0,\\
&&\frac{M_r}{r}\frac{d}{dr}\Big[\frac{\Delta}{r M_r}\dot\Phi\Big]+
\Big[-\frac{2\Lambda}{M_r}-C+\frac{m^2 d_a}{R_a}+\frac{n^2 d_b}{R_b}+
(\omega r)^2+\frac{MR W^2}{\Delta}\Big]\Phi=0.\nonumber
\eea
Here we used the expression (\ref{Psi5D}) for function $\Psi$ and defined
\bea\label{Magn5dMfunc}
M_\theta=\Theta^2-\mu^2,\quad M_r=-(r^2+\mu^2),\quad
W=\omega+\frac{am}{R_a}+\frac{bn}{R_b}\,.
\eea
The constants $(\Lambda,C)$ appearing in (\ref{FullMagn5d}) are expressed in terms of $(\mu,\la)$ as
\bea\label{Magn5dLmbd}
C&=&\la^2-2\omega(am+bn)-\omega^2(a^2+b^2),\nn
\Lambda&=&ab\la+\omega\mu^3-\mu[am+bn+(a^2+b^2)\omega],
\eea
and $\mu$ is given by
\bea\label{Magn5dLmbd1}
\mu=\frac{1}{\la}\left[ab\omega+an+bm\right].
\eea
As before, for fixed values of $(\omega,m,n)$, the angular equation in (\ref{FullMagn5d}) should be viewed as an eigenvalue problem for $\la$, then equation for $\Phi$ gives the corresponding radial profile. A straightforward modification of the arguments presented after equation (\ref{MagStat5d}) leads to the conclusion that solution (\ref{FullMagn5d}) describes two magnetic polarizations. 

\bigskip

Equations (\ref{FullElectr5d}) and (\ref{FullMagn5d}) constitute our main result for the five--dimensional black hole, and they describe all three polarizations of photons in the Myers--Perry geometry. There are striking similarities between differential equations appearing in (\ref{FullElectr5d}) and (\ref{FullMagn5d}), and in the next subsection we will demonstrate that the wave equation fits the same pattern. 

\subsection{Summary and comparison to the wave equation}
\label{SecSubSmry5D}

Let us now summarize the results of this section.
Differential equations appearing in (\ref{FullElectr5d}) and (\ref{FullMagn5d}) can be written in a uniform fashion:
\bea\label{EqnPtrn5d}
&&\frac{D_\theta}{s_{2\theta}}\frac{d}{d\theta}\left[\frac{s_{2\theta}}{D_\theta}S'\right]+
\left[\frac{2\Lambda}{D_\theta}+\omega^2\Theta^2-\frac{m^2}{s_\theta^2}-\frac{n^2}{c_\theta^2}+
C\right]S=0,\\
&&\frac{D_r}{r}\frac{d}{dr}\Big[\frac{\Delta}{r D_r}\dot\Phi\Big]+
\Big[-\frac{2\Lambda}{D_r}+(\omega r)^2+\frac{m^2 d_a}{R_a}+\frac{n^2 d_b}{R_b}-C+
\frac{MR W^2}{\Delta}\Big]\Phi=0.\nonumber
\eea
Here $d_a=-d_b=a^2-b^2$, and various functions are defined by
\bea\label{EqnPtrn5d1}
R_c=r^2+c^2,\quad R=R_a R_b,\quad \Delta=R-Mr^2,\quad W=\omega+\frac{am}{R_a}+\frac{bn}{R_b}.
\eea
Electric and magnetic polarizations differ by the explicit form of the functions $(D_r,D_\theta)$ given by (\ref{Electr5dMfunc}), (\ref{Magn5dMfunc}), and by the expressions for the constants $(\Lambda,C)$ in terms of the control parameter $\la$ (see (\ref{Electr5dLmbd}) and (\ref{Magn5dLmbd})). In this short subsection we will demonstrate that the wave equation fits the same pattern (\ref{EqnPtrn5d}). Our final result is summarized by the system (\ref{EqnPtrn5d}), (\ref{EqnPtrn5d1}), (\ref{5DfuncD})--(\ref{5dSmryAnstz2}).

\bigskip

Separation of the wave equation in the Myers--Perry geometry was discussed in section \ref{SecMPwave}, and for the odd--dimensional spacetime the result is given by (\ref{SeparWaveOddFullMain}). In five dimensions, there is only one angular coordinate $x_1$, and it is related to $\theta$ by equation (\ref{X1asTheta5D}). Substitution of $x_1$ in terms of $\theta$ into (\ref{SeparWaveOddFullMain}) leads to a system of ODEs governing the dynamics of a massless scalar:
\bea\label{SeparWave5d1}
&&r\frac{d}{dr}\left[\frac{\Delta}{r}\frac{d\Phi}{dr}\right]+\frac{R^2W^2}{\Delta}\Phi=P_1[r^2] \Phi,\\
&&{x_1}\frac{d}{dx_1}\left[\frac{H_1}{x_1}\frac{dX_1}{dx_1}\right]-
H_1\left[\omega+\frac{a m}{(a^2-b^2)s_\theta^2}+\frac{b n}{(b^2-a^2)c_\theta^2}
\right]^2X_1=-P_1[-x_1^2] X_1\,.\nonumber
\eea
The linear polynomial $P_1$ is subject to the constraint (\ref{P0cnstrnt}),
\bea
P_1[0]=\left[a b \right]^2\left[\omega+\frac{m}{a}+\frac{n}{b}
\right]^2\,,
\eea
so it is convenient to write it as $P_1[z]=\sigma z+P_1[0]$, where $\sigma$ is arbitrary parameter. 
Recalling the expression (\ref{X1asTheta5D}) for $x_1$ in terms of $\theta$, as well as definition of $H_1$,
\bea
H_1=(a^2-x_1^2)(b^2-x_1^2)=-(a^2-b^2)^2 s_\theta^2 c_\theta^2,\nonumber
\eea
equations (\ref{SeparWave5d1}) can be rewritten as
\bea\label{SeparWave5d2}
&&\frac{1}{r}\frac{d}{dr}\left[\frac{\Delta}{r}\frac{d\Phi}{dr}\right]+\frac{1}{r^2}\left\{\frac{R^2 W^2}{\Delta}
-P_1[0]\right\}\Phi=\sigma\Phi,\\
&&\frac{1}{s_{2\theta}}\frac{d}{d\theta}\left[s_{2\theta}\frac{dX_1}{d\theta}\right]+
\frac{1}{x_1^2}\left\{-s_\theta^2 c_\theta^2\left[(a^2-b^2)\omega+
\frac{a m}{s_\theta^2}-\frac{bn}{c_\theta^2}\right]^2-P_1[0]\right\}X_1=-\sigma X_1.\nonumber
\eea 
To compare this with the system (\ref{EqnPtrn5d}) describing electromagnetic field, we expand the brackets in the differential equations (\ref{SeparWave5d2}) and isolate all poles and residues:
\bea\label{SeparWave5d3}
&&\frac{1}{r}\frac{d}{dr}\left[\frac{\Delta}{r}\frac{d\Phi}{dr}\right]+\left\{
(\omega r)^2+\frac{m^2 d_a}{R_a}+\frac{n^2 d_b}{R_b}+\frac{MR W^2}{\Delta}-{\tilde C}
\right\}\Phi=\sigma\Phi,\nn
&&\frac{1}{s_{2\theta}}\frac{d}{d\theta}\left[s_{2\theta}\frac{dX_1}{d\theta}\right]+
\left\{\omega^2\Theta^2-
\frac{m^2}{s_\theta^2}-\frac{n^2}{c_\theta^2}+{\tilde C}\right\}X_1=-\sigma X_1.
\eea
Here ${\tilde C}$ is a constant defined by
\bea\label{Ctilde5d}
{\tilde C}=-\omega^2(a^2+b^2)-2\omega (am+bn).
\eea
We conclude that equations (\ref{SeparWave5d3}) fit the pattern (\ref{EqnPtrn5d}) with $D_r=D_\theta=1$ and an arbitrary  $\Lambda$. 

\bigskip

To summarize, the ODEs governing the separable solutions of the wave and the Maxwell's equations have the form (\ref{EqnPtrn5d}), and various polarizations are specified by functions $(D_r,D_\theta)$ and parameters $(C,\Lambda,\la)$\footnote{We simplified the expressions 
(\ref{FullElectr5d})--(\ref{Electr5dLmbd}) and  (\ref{FullMagn5d})--(\ref{Magn5dLmbd1}) for the electric and magnetic polarizations.}:
\bea\label{5DfuncD}
\mbox{scalar}:&&D_r=1,\quad D_\theta=1,\quad \forall \Lambda,\quad  \forall C;\nn
\mbox{electric}:&&D_r=1+(\mu r)^2,\
D_\theta=1-(\mu \Theta)^2,\quad
C=(\mu a b {\tilde\Omega})^2+{\tilde C},\nn
&&\Lambda={\omega}{\mu^3}
(\frac{1}{\mu^2}-\frac{am}{\omega}-a^2)(\frac{1}{\mu^2}-\frac{bn}{\omega}-b^2)-
\frac{\mu^3 abmn}{\omega};\\
\mbox{magnetic}:&&D_r=-1-\frac{r^2}{\mu^2},\quad D_\theta=-1+\frac{\Theta^2}{\mu^2},\quad
C=\frac{(a b {\tilde\Omega})^2}{\mu^2}+{\tilde C},\nn
&&\Lambda=\frac{\omega}{\mu^3}(\mu^2-\frac{am}{\omega}-a^2)(\mu^2-\frac{bn}{\omega}-b^2)-
\frac{abmn}{\mu^3\omega},\quad \la=\frac{a b\tilde\Omega}{\mu}\,.\nonumber
\eea
Here ${\tilde C}$ is given by (\ref{Ctilde5d}) and ${\tilde\Omega}$ is defined by
\bea
\tilde\Omega=\omega+\frac{m}{a}+\frac{n}{b}\,.
\eea
To complete this summary, we recall that the gauge fields are given by the first lines in equations (\ref{FullElectr5d}) and (\ref{FullMagn5d}):
\bea\label{5dSmryAnstz}
&&\hskip -1.6cm l_\pm^\mu A^{(el)}_\mu=\pm\frac{r}{1\pm i\mu r}{\hat l}_\pm\Psi ,\quad
m_\pm^\mu A^{(el)}_\mu=\mp\frac{i\Theta}{1\pm \mu\Theta}{\hat m}_\pm\Psi,\quad 
n^\mu A^{(el)}_\mu=0,\nn
&&\hskip -1.6cm l^\mu_\pm A^{(mgn)}_\mu=\pm \frac{1}{r\pm i \mu}{\hat l}_\pm \Psi,\quad
m^\mu_\pm A^{(mgn)}_\mu=\pm\frac{i}{\Theta\mp \mu}{\hat m}_\pm \Psi,\quad 
n^\mu A^{(mgn)}_\mu=\la \Psi,
\eea
and separable solutions have
\bea\label{5dSmryAnstz2}
\Psi=e^{i\omega t+im\phi+in\psi}\Phi(r)S(\theta).
\eea
As we already mentioned in the four--dimensional case, it would be interesting to see if the  ``master equations'' (\ref{SeparWave5d1}) would hold for the fields with spin higher than one. Note that extension of equations (\ref{SeparWave5d1}) to the Kerr-AdS case is rather straightforward: one has to add factors $Q_r$ and 
$Q_\theta$ as in equations (\ref{4dKerrAdSFull}). To avoid repetition, we postpone the discussion of the Kerr--AdS metrics until section \ref{SecKerrAdSMxwMP}, where the result will be written for all dimensions.

\bigskip

\bigskip

To summarize, in this section we {\it derived the most general separable solution} of Maxwell's equations  in the background of the five--dimensional black hole. The final result is given by the system (\ref{EqnPtrn5d}), (\ref{EqnPtrn5d1}), (\ref{5DfuncD})--(\ref{5dSmryAnstz2}). In the next section we will use the five--dimensional answers to {\it guess} the solution in all odd dimensions and check that the resulting ansatz indeed satisfies the Maxwell's equations. The solutions derived in section \ref{SecKerrNewAns} will be used as a similar starting point for the even--dimensional case.

\section{Electromagnetic waves in the Myers--Perry geometry}
\renewcommand{\theequation}{5.\arabic{equation}}
\setcounter{equation}{0}
\label{SecWaveMP}

After deriving the expressions for separable electromagnetic fields in four and five dimensions, here we will use the resulting expressions to {\it guess} the answer for higher dimensions and check it. Note that, unlike the results of sections \ref{SecKerr} and \ref{SecKerr5D}, solutions discussed here are not claimed to be unique, but we will see that they reproduce all $(D-2)$ independent polarizations in $D$ dimensions, at least in the static limit.

As we saw in section \ref{SecMyersPerry}, the structures associated with the Myers--Perry black hole in even and odd dimensions are rather different, so it is natural to discuss these two cases separately. We will use the electromagnetic waves found in section \ref{SecKerr} as a motivation for the ansatz in even dimensions, and the solutions found in section \ref{SecKerr5D} will serve as a guide for the odd--dimensional case. 

\subsection{Even dimensions}
\label{SecMPevenMxw}

We begin with recalling the ansatz (\ref{KerrNewAnstz}) used in four dimensions. In section
\ref{SecKerrNewAns} we imposed this form of the gauge field and derived the expressions for $(G_\pm,F_\pm)$ and differential equations for functions $R$ and $S$. To extend the ansatz (\ref{KerrNewAnstz}) to higher dimensions, we recall that each of the rescaled frames ${\tilde e}_A^\mu$ defined by (\ref{AllFramesMP}) and (\ref{RscldFrmsMPeven}) is a function of only one argument:
\bea
{\tilde e}_t&=&\sqrt{\frac{R^2}{R-Mr}}\left[\d_t-
\sum_k\frac{a_k}{r^2+a_k^2}\d_{\phi_k}\right],\quad 
{\tilde e}_r=\sqrt{{R-Mr}}\, \d_r,\nn
{\tilde e}_i&=&\sqrt{{H_i}}\left[\d_t-\sum_k\frac{a_k}{a_k^2-x_i^2}\d_{\phi_k}
\right],\quad {\tilde e}_{x_i}=\sqrt{{H_i}}\d_{x_i}\,.\nonumber
\eea 
To mimic the ansatz (\ref{KerrNewAnstz}), we define the ``light--cone'' combinations:
\bea\label{GenFramesEvenD}
&&l_\pm^\mu\d_\mu=\frac{R}{\sqrt{\Delta}}\left\{\frac{\Delta}{R}\d_r\pm \left[\d_t-
\sum_k\frac{a_k}{r^2+a_k^2}\d_{\phi_k}\right]\right\},\quad \Delta=R-Mr,\nn
&&\left[m_\pm^{(j)}\right]^\mu
\d_\mu=\sqrt{{H_j}}\left\{\d_{x_j}\pm i\left[\d_t-\sum_k\frac{a_k}{a_k^2-x_j^2}\d_{\phi_k}
\right]\right\}\,.
\eea
Then the natural generalization of the ansatz (\ref{KerrNewAnstz}) is
\bea\label{NewAnstzEven}
l_\pm^\mu A_\mu=G_\pm(r)l_\pm^\mu\d_\mu \Psi,\quad
\left[m_\pm^{(j)}\right]^\mu A_\mu=F^{(j)}_\pm(x_j)\left[m_\pm^{(j)}\right]^\mu
\d_\mu \Psi,
\eea
where $\Psi$ is taken to have the form
\bea\label{MasterPsiEven}
\Psi=e^{i\omega t+\sum im_k\phi_k}\Phi(r)\prod X_j(x_j).
\eea
Rather than undertaking a general study of the ansatz (\ref{NewAnstzEven}), we use the four--dimensional results to {\it guess} the form of $(G_\pm,F^{(j)}_\pm)$ and the differential equations for $\Phi$ and $X_j$. We then check that the resulting system solves Maxwell's equations in the metric (\ref{MPeven}) for all even dimensions. As in the four--dimensional case, the results split into electric and magnetic polarizations. 

\bigskip

The electric polarization in $D=2n+2$ dimensions is specified in terms of one separable scalar function 
$\Psi$ as\footnote{Note that we made a replacement $\mu\rightarrow\frac{1}{\mu}$ in comparison to the electric polarization in four dimensions defined in (\ref{4dSmryAnstz}). Such modified notation ensures that the electric and magnetic polarizations are described by the same differential equations.}
\bea\label{AnsGenEvenElectr}
l^\mu_\pm A^{(el)}_\mu=\pm\frac{\mu r}{\mu\mp i r}{\hat l}_\pm \Psi,\quad
[m_\pm^{(j)}]^\mu A^{(el)}_\mu=\pm\frac{i\mu x_j}{\mu\pm x_j}{\hat m}^{(j)}_\pm \Psi,
\eea
and Maxwell's equations reduce to a system of ODEs\footnote{
In practice, an alternative form (\ref{EvenDimGenSpin})--(\ref{EvenDimGenSpinLmb}) of the system (\ref{GenEvenElectr}) might be more useful, but expressions (\ref{GenEvenElectr}) stress the structure of poles in differential equations, and they arise naturally in the process of derivation.}
\bea\label{GenEvenElectr}
&&{E_j}\frac{d}{dx_j}\left[\frac{H_j}{E_j}X_j'\right]+
\left\{\frac{\Lambda}{E_j}-Q[-x_j^2]+P_{n-1}[-x_j^2]-\omega^2 (ix_j)^{2n}\right\}X_j=0,\\
&&{E_r}\frac{d}{dr}\left[\frac{\Delta}{E_r}{\dot\Phi}\right]-
\left\{\frac{\Lambda}{E_r}-Q[r^2]+{P}_{n-1}[r^2]-\omega^2 r^{2n}-\frac{MrRW^2}{\Delta}\right\}\Phi=0.\nonumber
\eea
Here we defined four functions
\bea\label{TempEjEven}
\begin{array}{l}
E_j=1-(\frac{x_j}{\mu})^2\\ E_r=1+(\frac{r}{\mu})^2
\end{array}\,,\qquad
Q[y]=\sum \frac{{\tilde c}_k (a_km_k)^2}{a_k^2+y},\quad
W=\omega-
\sum_k\frac{a_k m_k}{r^2+a_k^2},
\eea
and one separation constant $\Lambda$:
\bea\label{GenLambdaEvenElctr}
\Lambda=-\frac{2}{\mu}\left[\prod{\Lambda_i}\right]\left[-{\omega}+
\sum\frac{a_im_i}{\Lambda_i}\right],
\quad \Lambda_j\equiv a_j^2-{\mu^2}\,.
\eea
Parameters ${\tilde c}_k$ entering (\ref{TempEjEven}) are given by\footnote{These parameters are related to $c_k$ introduced in (\ref{MiscElliptic}) by ${\tilde c}_k=\pm c_k$, where sign depends on the number of dimensions.}
\bea\label{cKevenRev}
{\tilde c}_k=\prod_{m\ne k}(a_m^2-a_k^2).
\eea
Polynomial $P_{n-1}$ appearing in (\ref{GenEvenElectr}) has degree $(n-1)$ in its argument, and $n$ coefficients of this polynomial are subject to one linear constraint which will be discussed below. 
The separation constants are the free coefficients of the polynomial $P_{n-1}[y]$ and the parameter $\mu$, so as expected, there are $n$ free coefficients.  

\bigskip
The magnetic polarization is parameterized by a scalar function $\Psi$ 
\bea\label{AnsGenEvenMagn}
l^\mu_\pm A^{(mgn)}_\mu=\pm\frac{1}{r\pm i\mu}{\hat l}_\pm \Psi,\quad
[m_\pm^{(j)}]^\mu A^{(mgn)}_\mu=\mp\frac{i}{x_j\pm \mu}{\hat m}^{(j)}_\pm \Psi\,,
\eea
whose coordinate dependence is given by (\ref{MasterPsiEven}). The dynamics is still governed by the system (\ref{GenEvenElectr}) with various ingredients defined by (\ref{TempEjEven})--(\ref{cKevenRev}).
As in the electric case, the coefficients of $P_{n-1}$ are subject to one constraint, which will be discussed below. Verification of the solutions (\ref{AnsGenEvenElectr}) and (\ref{AnsGenEvenMagn}) with differential equation (\ref{GenEvenElectr}) is straightforward but tedious, and in the Appendix \ref{SecAppMPeven} we outline the procedure focusing on the special case 
$\omega=m_i=0$. 

Note that for generic values of $\mu$, the ansatze (\ref{AnsGenEvenElectr}) and (\ref{AnsGenEvenMagn}) are related by a gauge transformation and rescaling of $A_\mu$. To see this, we rewrite (\ref{AnsGenEvenElectr}) as
\bea\label{ElctPolarzReWrt}
l^\mu_\pm A^{(el)}_\mu=\left[i\mu\pm\frac{\mu^2}{r\pm i\mu}\right]{\hat l}_\pm \Psi,\quad
[m_\pm^{(j)}]^\mu A^{(el)}_\mu=\left[i\mu\mp\frac{i\mu^2}{x_j\pm \mu}\right]{\hat m}^{(j)}_\pm \Psi.
\eea
The constant terms in the square brackets correspond to a pure gauge, and the remaining fractions give the rescaled version of (\ref{AnsGenEvenMagn}). In spite of this equivalence, it is convenient to keep both (\ref{AnsGenEvenElectr}) and (\ref{AnsGenEvenMagn}) for making comparison with four dimensions and for taking the limits $\mu\rightarrow 0$ and $\mu\rightarrow \infty$. The first limit is simple in (\ref{AnsGenEvenMagn}), while the second limit is more natural in the ``electric gauge'' (\ref{AnsGenEvenElectr}).

\bigskip

We conclude the discussion of the even--dimensional case by comparing the system (\ref{GenEvenElectr}) with differential equations (\ref{SeparWaveEvenFull}) originating from the wave equation
\bea\label{WaveEqnEvenAnlz}
&&\frac{d}{dx_i}\left[H_i\frac{dX_i}{dx_i}\right]-
H_i\left[\omega-
\sum_k\frac{a_k m_k}{a_k^2-x_j^2}\right]^2X_j=-{P}_{n-1}[-x_j^2]X_j\,,\\
&&\frac{d}{dr}\left[\Delta \frac{d\Phi}{dr}\right]+\frac{R^2 W^2}{\Delta}
\Phi={P}_{n-1}[r^2]\Phi\,.\nonumber
\eea
Here ${P}_{n-1}$ is a polynomial with arbitrary coefficients. 
To compare (\ref{WaveEqnEvenAnlz}) with (\ref{GenEvenElectr}), we analyze the poles and residues of two expressions appearing in (\ref{WaveEqnEvenAnlz}):
\bea
&&-H_j\left[\omega-
\sum_k\frac{a_k n_k}{a_k^2-x_j^2}\right]^2=-\omega^2(ix_j)^{2n}-
\sum_k \frac{{\tilde c}_k(a_k m_k)^2}{a_k^2-x_j^2}+{\hat P}_{n-1}[-x_j^2]\,,\nn
&&\frac{R^2 W^2}{\Delta}=\frac{MrRW^2}{\Delta}+RW^2=\frac{MrRW^2}{\Delta}+(\omega r^n)^2+
\sum_k \frac{{\tilde c}_k(a_k m_k)^2}{a_k^2+r^2}+{\hat P}_{n-1}[r^2].\nonumber
\eea
These two expressions contain {\it the same} polynomial ${\hat P}_{n-1}$ of degree $(n-1)$.
Shifting the polynomial $P_{n-1}$ appearing in (\ref{GenEvenElectr}) by ${\hat P}_{n-1}$, we can rewrite the systems (\ref{GenEvenElectr}) and (\ref{WaveEqnEvenAnlz}) in the unified form: 
\bea\label{EvenDimGenSpin}
&&{D_j}\frac{d}{dx}\left[\frac{H_j}{D_j}X_j'\right]+\left\{\frac{2\Lambda}{D_j}-H_j W_j^2-
\Lambda +P_{n-2}[-x_j^2] D_j\right\}X_j=0,
\nn
&&{D_r}\frac{d}{dr}\left[\frac{\Delta}{D_r}{\dot\Phi}\right]-\left\{\frac{2\Lambda}{D_r}-
\frac{R^2 W_r^2}{\Delta}-
\Lambda +P_{n-2}[r^2]  D_r\right\}\Phi=0,\\
&&\Omega=\omega-\sum\frac{m_ia_i}{\Lambda_i},\quad 
W_j=\omega-\sum\frac{m_k a_k}{a_k^2-x_j^2},\quad W_r=\omega-\sum\frac{m_k a_k}{a_k^2+r^2}.
\nonumber
\eea
The difference between two polarizations of the electromagnetic field and the scalar equation appears only in the expressions for functions $(D_j,D_r)$ and for parameter $\Lambda$:
\bea\label{EvenDimGenSpinLmb}
\mbox{scalar}:&&D_r=D_j=1,\quad \forall \Lambda;\nn
\mbox{vector}:&&
\begin{array}{l}
D_j=1-\frac{x_j^2}{\mu^2}\\ D_r=1+\frac{r^2}{\mu^2}
\end{array}\,,\quad\quad
\Lambda=\frac{\Omega}{\mu}
\prod \Lambda_k,\quad \Lambda_i=(a_i^2-{\mu^2}).
\eea
The expressions for the gauge field are given by (\ref{AnsGenEvenElectr}) and (\ref{AnsGenEvenMagn}), and the separation of the ``master function'' $\Psi$ is given by (\ref{MasterPsiEven}). The vector version of equations (\ref{EvenDimGenSpin})--(\ref{EvenDimGenSpinLmb}) describes both for the electric and the magnetic polarizations.

Note that the constraints on polynomials $P_{n-1}$ mentioned earlier are already taken into account in (\ref{EvenDimGenSpin}): they imply that the last terms in equations for $X_j$ and $\Phi$ are proportional to $D_j$ and $D_r$. Thus the free parameters in (\ref{EvenDimGenSpin})--(\ref{EvenDimGenSpinLmb}) are 
$\mu$ and $(n-1)$ coefficients of $P_{n-2}$. As expected, this leads to $n$ arbitrary separation constants. 

\bigskip

To summarize, separation of variables for the electromagnetic field and for the massless scalar in even dimensions is described by (\ref{EvenDimGenSpin})--(\ref{EvenDimGenSpinLmb}) with free polynomial $P_{n-2}$ and with gauge potential given by (\ref{AnsGenEvenElectr}) and (\ref{AnsGenEvenMagn}).

\subsection{Odd dimensions}
\label{SecMPoddMxw}

The waves in odd dimensions are expected to follow the pattern discussed in section \ref{SecKerr5D}, and to extend this construction, we need to generalize the ansatz (\ref{5dNewAnstz}). We begin with extending (\ref{LCframes5D}) to higher dimensions by defining the counterparts of the special frames (\ref{GenFramesEvenD}) as linear combinations of (\ref{AllFramesMPOdd}):
\bea\label{GenFramesOddD}
&&l_\pm^\mu\d_\mu=\frac{R}{\sqrt{\Delta}}\left\{\frac{\Delta}{R}\d_r\pm \left[\d_t-
\sum_k\frac{a_k}{r^2+a_k^2}\d_{\phi_k}\right]\right\},\qquad \Delta=R-Mr^2,\nn
&&\left[m_\pm^{(j)}\right]^\mu
\d_\mu=\sqrt{{H_j}}\left\{\d_{x_j}\pm i\left[\d_t-\sum_k\frac{a_k}{a_k^2-x_j^2}\d_{\phi_k}
\right]\right\}\,,\\
&&n^\mu\d_\mu=\d_t-\sum_k\frac{1}{a_k}\d_{\phi_k}.\nonumber
\eea
In terms of these frames the inverse metric is 
\bea
g^{\mu\nu}\d_\mu\d_\nu=\frac{1}{FR}l^\mu_+l^\nu_-\d_\mu\d_\nu+
\left[\frac{\prod a_i}{r\prod x_k}\right]^2n^\mu n^\nu\d_\mu\d_\nu+
\sum_j \frac{\left[m_+^{(j)}\right]^\mu\left[m_-^{(j)}\right]^\mu\d_\mu\d_\nu}{x_j^2 d_i(r^2+x_j^2)}\,.
\eea
A natural generalization of the ansatz (\ref{5dNewAnstz}) is
\bea\label{NewAnstzOdd}
l_\pm^\mu A_\mu=G_\pm(r)l_\pm^\mu\d_\mu \Psi,\quad
\left[m_\pm^{(j)}\right]^\mu A_\mu=F^{(j)}_\pm(x_j)\left[m_\pm^{(j)}\right]^\mu
\d_\mu \Psi,\quad n^\mu A_\mu=\la\Psi
\eea
with separable function $\Psi$:
\bea
\Psi=e^{i\omega t+\sum im_k\phi_k}\Phi(r)\prod X_j(x_j).\nonumber
\eea
As in the even dimensional case, we will not look for the most general solution of the form (\ref{NewAnstzOdd}), but rather guess the expressions for $(G_\pm,F_\pm)$ using the 
five--dimensional case as a guide, and check all Maxwell's equations. Some details of such verification are presented in the Appendix \ref{SecAppMPodd}. 

\bigskip

The five--dimensional example suggests a separation into electric and magnetic modes with the gauge potential given by\footnote{As in (\ref{AnsGenEvenElectr}), we made a replacement $\mu\rightarrow\frac{1}{\mu}$ in comparison to the electric polarization  (\ref{5dSmryAnstz}) in five dimensions. This ensures that the electric and magnetic polarizations are equivalent for $\mu\ne 0,\infty$.}
\bea\label{GenOddElectr}
&&\hskip -1.5cm l^\mu_\pm A^{(el)}_\mu=\pm\frac{\mu r}{\mu \mp ir}{\hat l}_\pm \Psi,\quad
[m_\pm^{(j)}]^\mu A^{(el)}_\mu=\pm\frac{i\mu x_j}{\mu\pm x_j}{\hat m}^{(j)}_\pm \Psi\,\quad
n^\mu A^{(el)}_\mu=0;\\
&&\hskip -1.5cm l^\mu_\pm A^{(mgn)}_\mu=\pm\frac{1}{r\pm i\mu}{\hat l}_\pm \Psi,\quad
[m_\pm^{(j)}]^\mu A^{(mgn)}_\mu=\mp\frac{i}{x_j\pm \mu}{\hat m}^{(j)}_\pm \Psi\,\quad
n^\mu A^{(mgn)}_\mu=\la \Psi.\nonumber
\eea
Rewriting the electric polarization as in (\ref{ElctPolarzReWrt}), we conclude that for generic values of 
$\mu$, the two ansatze (\ref{GenOddElectr}) are equivalent up to a gauge transformation and a rescaling.  
The analysis of Maxwell's equations is very similar to the one discussed in the last subsection, so we present only the final result. The gauge fields (\ref{GenOddElectr}), as well as the massless scalar field, are described by the system of ODEs
\bea\label{OddMPmaster}
&&\frac{D_j}{x_j}\frac{d}{dx_j}\left[\frac{H_j}{x_j D_j}X_j'\right]+\left\{\frac{2\Lambda}{D_j}-
\frac{H_jW_j^2}{x_j^2}+\frac{\AA D_j}{x_j^2}{\tilde\Omega}^2+P_{n-2}[-x_j^2]D_j\right\}=0,\nn
&&\frac{D_r}{r}\frac{d}{dr}\left[\frac{\Delta}{r D_r}{\dot\Phi}\right]+\left\{\frac{2\Lambda}{D_r}+
\frac{R^2W_r^2}{r^2\Delta}-\frac{\AA D_r}{r^2}{\tilde\Omega}^2+P_{n-2}[r^2] D_r\right\}\Phi=0.
\eea
Here functions $(W_j,W_r)$ and constants $(\AA,\Omega,{\tilde\Omega})$ are given by
\bea\label{AAdef}
&&W_j=\omega-\sum_k\frac{m_k a_k}{a_k^2-x_j^2},\quad 
W_r=\omega-\sum_k\frac{m_k a_k}{a_k^2+r^2},\\
&&\AA=\left[\prod a_k\right]^2,\quad\Omega=\omega-\sum_k\frac{m_ka_k}{\Lambda_k},\quad 
{\tilde\Omega}=\omega-\sum_k \frac{m_k}{a_k}.\nonumber
\eea
As in even dimensions, the difference between scalar and vector excitations is encoded in functions $(D_j,D_r)$ and parameters $(\Lambda,\Lambda_i)$:
\bea\label{OddDimGenSpinLmb}
\mbox{scalar}:&&D_r=D_j=1,\quad \forall \Lambda;\\
\mbox{vector}:&&
\begin{array}{l}
D_j=1-\frac{x_j^2}{\mu^2}\\ D_r=1+\frac{r^2}{\mu^2}
\end{array}\,,\quad
\Lambda=\frac{\Omega}{\mu^3}
\prod \Lambda_k,\quad \Lambda_i=(a_i^2-{\mu^2}).\nonumber
\eea
Magnetic polarization (\ref{GenOddElectr}) also has a parameter $\la$, which is given by
\bea\label{GenOddLmbda}
\la=\frac{\tilde\Omega}{\mu}\,.
\eea
The last relation implies that $\la\Psi=-\frac{i}{\mu}n^\mu\d_\mu\Psi$, so the two branches described by (\ref{GenOddElectr}) are indeed related by a gauge transformation.

Note that, in spite of appearance, the curly brackets in  (\ref{OddMPmaster}) are regular at $x_j=0$ and $r=0$. For example, as $r$ approaches zero, we find
\bea
\frac{R^2W_r^2}{r^2\Delta}-\frac{\AA D_r}{r^2}{\tilde\Omega}^2\sim 
\frac{1}{r^2}\left[R\frac{R}{R-Mr^2}{\tilde\Omega}^2-\frac{\AA D_r}{r^2}{\tilde\Omega}^2+O(r^2)\right].
\nonumber
\eea
Recalling the expression (\ref{EplsdOdd}) for $R$, we conclude that the last line is indeed finite.

\bigskip

Although the ansatz (\ref{GenOddElectr}) depends on a continuous parameter $\mu$, it can describe at most $D-2$ independent polarizations of electromagnetic field in $D$ dimensions, and other values of $\mu$ must correspond to linear combinations of such building blocks. In section \ref{SecSubCmprSchw} we will demonstrate that all $D-2$ independent polarizations are indeed recovered, at least in the non--rotating limit. 

\subsection{Summary and extension to the Myers--Perry--(A)dS geometry}
\label{SecKerrAdSMxwMP}

To summarize, in the last two subsections we have constructed various configurations of the electromagnetic field specified by parameter $\mu$, and in the next subsection we will demonstrate that these ansatze reproduce all $(D-2)$ independent polarizations in an arbitrary number of dimensions. Here we summarize the results of the last two subsections and extend the construction to the Myers--Perry--(A)dS geometry.

\bigskip

The final answer for even dimensions is given by the ansatze (\ref{AnsGenEvenElectr}) and (\ref{AnsGenEvenMagn}), as well as the ``master equation'' (\ref{EvenDimGenSpin}) with ingredients defined by (\ref{EvenDimGenSpinLmb}). The extension to the Myers--Perry--(A)dS (GLPP) geometry discussed in section \ref{SecAdSFramesMP} is straightforward: one should start with ansatze (\ref{AnsGenEvenElectr}) and (\ref{AnsGenEvenMagn}) using the frames (\ref{FramesMPAeven}) and their ``light--cone'' combinations:
\bea
&&l_\pm^\mu\d_\mu=\frac{R}{\sqrt{\Delta}}\left\{\frac{Q_r\Delta}{R}\d_r\pm 
\frac{1}{Q_r}\left[\d_t-
\sum_k\frac{a_k}{r^2+a_k^2}\d_{\phi_k}\right]\right\},\quad \Delta=R-Mr,\nn
&&\left[m_\pm^{(j)}\right]^\mu
\d_\mu=\sqrt{{H_j}}\left\{Q_j\d_{x_j}\pm \frac{i}{Q_j}\left[\d_t-\sum_k\frac{a_k}{a_k^2-x_j^2}\d_{\phi_k}
\right]\right\}\,.
\eea
Then the Maxwell's equations and the wave equation reduce to a simple modification of the system (\ref{EvenDimGenSpin}):
\bea\label{EvenDimAdSKerr}
&&{D_j}\frac{d}{dx}\left[\frac{Q_j^2H_j}{D_j}X_j'\right]+\left\{\frac{2\Lambda}{D_j}-
\frac{H_j W_j^2}{Q_j^2}-
\Lambda +P_{n-2}[-x_j^2] D_j\right\}X_j=0,
\nn
&&{D_r}\frac{d}{dr}\left[\frac{Q_r^2\Delta}{D_r}{\dot\Phi}\right]-\left\{\frac{2\Lambda}{D_r}-
\frac{R^2 W_r^2}{Q_r^2\Delta}-
\Lambda +P_{n-2}[r^2]  D_r\right\}\Phi=0.
\eea
Various ingredients appearing in these equations are still given by (\ref{EvenDimGenSpinLmb}) and the last line of (\ref{EvenDimGenSpin}). These results can be verified applying the procedure used for the Myers--Perry black hole in section \ref{SecMPevenMxw}.  

\bigskip

The final answer for odd dimensions is given by the ansatze (\ref{GenOddElectr}) and  the ``master equation'' (\ref{OddMPmaster}) with ingredients defined by (\ref{AAdef}) and (\ref{OddDimGenSpinLmb}). The extension to the GLPP geometry is again straightforward: the frames used in  (\ref{GenOddElectr}) should be replaced by the linear combinations of (\ref{AdSFramesMPOddAdS}):
\bea
&&l_\pm^\mu\d_\mu=\frac{R}{\sqrt{\Delta}}\left\{\frac{Q_r\Delta}{R}\d_r\pm 
\frac{1}{Q_r}\left[\d_t-
\sum_k\frac{a_k}{r^2+a_k^2}\d_{\phi_k}\right]\right\},\quad \Delta=R-Mr^2,\nn
&&\left[m_\pm^{(j)}\right]^\mu
\d_\mu=\sqrt{{H_j}}\left\{Q_j\d_{x_j}\pm \frac{i}{Q_j}\left[\d_t-\sum_k\frac{a_k}{a_k^2-x_j^2}\d_{\phi_k}
\right]\right\}\,,\\
&&n^\mu \d_\mu=-\frac{\prod a_i}{r\prod x_k}\left[\d_t-\sum_k\frac{1}{a_k}\d_{\phi_k}
\right].\nonumber
\eea
The Maxwell's equations and the wave equation reduce to a modified version of the system (\ref{OddMPmaster}):
\bea
&&\frac{D_j}{x_j}\frac{d}{dx_j}\left[\frac{Q_j^2H_j}{x_j D_j}X_j'\right]+\left\{\frac{2\Lambda}{D_j}-
\frac{H_jW_j^2}{x_j^2Q_j^2}+\frac{\AA D_j}{x_j^2}{\tilde\Omega}^2+P_{n-2}[-x_j^2]D_j\right\}=0,\nn
&&\frac{D_r}{r}\frac{d}{dr}\left[\frac{Q_r^2\Delta}{r D_r}{\dot\Phi}\right]+\left\{\frac{2\Lambda}{D_r}+
\frac{R^2W_r^2}{r^2Q_r^2\Delta}-\frac{\AA D_r}{r^2}{\tilde\Omega}^2+P_{n-2}[r^2] D_r\right\}\Phi=0.
\eea
The definitions (\ref{AAdef}) and the identifications (\ref{OddDimGenSpinLmb}) still hold. 

\subsection{Reduction to the Schwarzschild--Tangherlini geometry} 
\label{SecSubCmprSchw}

In this subsection we consider the waves in the non--rotating black holes by taking the appropriate limits of various solutions derived earlier in this section, and compare the results with the general discussion presented in the Appendix \ref{SecAppSchw}. In particular, this will clarify the interpretation of the polarizations covered by the ansatze (\ref{AnsGenEvenElectr}), (\ref{AnsGenEvenMagn}), (\ref{GenOddElectr}). Since the discussion of the Appendix \ref{SecAppSchw} treats odd and even dimensions on the same footing, to establish the relation to this description, it is sufficient to look at one of the cases, and we will focus on even dimensions. 
\bigskip

The non--rotating limit of the frames (\ref{AllFramesMP}) requires some care. It is clear that the Schwarzschild--Tangherlini geometry is obtained by sending all rotation parameters to zero, and given the ranges (\ref{EplsdEven}), coordinates $x_i$ should be sent to zero as well.
Thus we will write
\bea\label{abyLmbd}
a_i=\la b_i,\quad x_i=\la y_i,
\eea 
and send $\la$ to zero while keeping $(b_i,y_i)$ fixed. This leads to an apparent problem in $m_\pm^{(j)}$ defined by (\ref{GenFramesEvenD}), and to cure it, one needs to recall the inverse metric in terms of the frames (\ref{GenFramesEvenD}):
\bea\label{ST533}
g^{\mu\nu}\d_\mu\d_\nu=\frac{1}{FR}l^\mu_+l^\nu_-\d_\mu\d_\nu+
\sum_j \frac{\left[m_+^{(j)}\right]^\mu\left[m_-^{(j)}\right]^\mu\d_\mu\d_\nu}{ d_i(r^2+x_j^2)}.
\eea
Then the relevant limits are
\bea\label{STframesLimit}
&&l_\pm^\mu\d_\mu\rightarrow \frac{R}{\sqrt{\Delta}}\left\{\frac{\Delta}{R}\d_r\pm 
\d_t\right\},\quad \Delta=R-Mr,\quad R=FR=r^{2n}\,,\\
&&\left[\tilde m_\pm^{(j)}\right]^\mu\d_\mu\equiv \frac{\left[m_\pm^{(j)}\right]^\mu}{\sqrt{d_i(r^2+x_j^2)}}\d_\mu\rightarrow
\frac{1}{r}\left[\sqrt{\frac{H_i}{\la d_i}}\right]_{\la=0}
\left\{\d_{y_j}\mp \sum_k\frac{ib_k}{b_k^2-y_j^2}\d_{\phi_k}\right\}\,.\nonumber
\eea
The square bracket in the last expression remains finite in the limit. 

\bigskip
\noindent
{\bf Electric polarization}

In the $\la=0$ limit, the ansatz (\ref{AnsGenEvenElectr}) for the electric polarization becomes\footnote{In contrast to the magnetic polarizations discussed below, here we do not send $\mu$ to zero in the $\la=0$ limit.}
\bea\label{tempAA7}
l^\mu_\pm A^{(el)}_\mu=\pm\frac{\mu r}{\mu\mp ir}{\hat l}_\pm \Psi,\quad
[{\tilde m}_\pm^{(j)}]^\mu A^{(el)}_\mu=0.
\eea
To compare this with the general electric solution (\ref{TangElectr}) in the Schwarzschild--Tangherlini geometry, we rewrite (\ref{TangElectr}) in a different gauge:
\bea\label{TangElectrTmp}
&&A^{(el)}_t=e^{i\omega t}\left[f(r)-i\omega{\dot g}(r)\right]Y,\quad 
A^{(el)}_r=\omega^2 e^{i\omega t} g(r) Y,\quad A^{(el)}_i=0,\\
&&\frac{1}{\sqrt{h}}\d_i[\sqrt{h}h^{ij}\d_j Y]=-\la_1 Y,\quad
{\dot g}=-\frac{r^2}{\la_1 H}\dot f,\quad
\frac{1}{r^d}\d_r[r^d \dot f]-
\frac{\la_1 f}{r^2 H}-\frac{\la_1\omega^2 g}{r^2 H}=0.\nonumber
\eea
Observing that in the $\la=0$ limit the angular ``electric'' factors in  
(\ref{EvenDimGenSpin})--(\ref{EvenDimGenSpinLmb}) reduce to $D_j=1$, we conclude that the angular equation (\ref{EvenDimGenSpin}) is the same as in the scalar case, in full agreement with equation (\ref{TangElectrTmp}) for $Y$. The radial equations also agree between (\ref{EvenDimGenSpin}) and (\ref{TangElectrTmp}), and for transparency we focus on $\omega=0$. In this case $\mu=\infty$ and the radial equation in (\ref{EvenDimGenSpin}) becomes
\bea
\frac{d}{dr}\left[\Delta {\dot\Phi}\right]-\left[\Lambda+P_{n-2}[r^2]\right]\Phi=0.
\eea  
Moreover, $\Lambda=0$, and in the absence of scales associated with $a_i$, the polynomial $P_{n-2}$ must have the form 
\bea\label{tempAA23}
P_{n-2}[r^2]=\alpha r^{2(n-2)}
\eea
with constant $\alpha$. Substitution of (\ref{TangElectrTmp}) into (\ref{AnsGenEvenElectr}) gives the relation between $f$ and $\Phi$:
\bea
l^\mu_\pm A^{(el)}_\mu=\pm\frac{e^{i\omega t}r^{2n}}{\sqrt{\Delta}}f Y=\pm r \sqrt{\Delta}\d_r \Psi\quad
\Rightarrow\quad f=\frac{\Delta}{r^{2n-1}}{\dot\Phi}.
\eea
Recalling that $H=1-\frac{M}{r^{d-1}}=\frac{\Delta}{r^{d}}$ and $d=2n$, we find two ingredients appearing in the radial equation in (\ref{TangElectrTmp}):
\bea
&&\hskip -1.5cm \frac{1}{r^d}\frac{d}{dr}[r^d \dot f]=\frac{1}{r^d}\frac{d}{dr}[r\frac{d}{dr}(\Delta\dot\Phi)-(d-1)\Delta\dot\Phi]=
\frac{1}{r^d}\left[r\frac{d}{dr}-(d-2)\right]P_{n-2}[r^2]\Phi\nn
&&\qquad =\frac{1}{r}\frac{d}{dr}\left[\frac{P_{n-2}[r^2]}{r^{d-2}}\Phi\right],\nn
&&\hskip -1.5cm \frac{f}{r^2 H}=\frac{r^{d-2}}{r^{2n-1}}{\dot\Phi}=\frac{\dot\Phi}{r}.\nonumber
\eea
Combination of the last two relations leads to the equation for $f$ expected from (\ref{TangElectrTmp}):
\bea
\frac{1}{r^d}\d_r[r^d \dot f]-
\frac{\alpha f}{r^2 H}=0.
\eea
The case of nonzero $\omega$ works in a similar way, although the algebra is more involved. Thus the electric polarization in the Schwarzschild--Tangherlini geometry is fully recovered form our ansatz (\ref{tempAA7}).

\bigskip
\noindent
{\bf Magnetic polarizations}

For the magnetic polarization, the $\la=0$ limit with (\ref{abyLmbd}) leads to nontrivial configurations only if $\mu$ goes to zero. Defining $\nu=(\mu/\la)$, we find
\bea\label{tempAA3}
l^\mu_\pm A^{(mgn)}_\mu=0,\quad
[{\tilde m}_\pm^{(j)}]^\mu A^{(mgn)}_\mu=\mp\frac{i}{y_j\pm {\nu}}{\hat {\tilde m}}^{(j)}_\pm \Psi.
\eea
To compare this with the discussion from Appendix \ref{SecAppSchw} without writing complicated formulas, we focus on the special case $\omega=m_i=0$, although similar relations hold in general. 

The first equation in (\ref{tempAA3}) leads to the expected result for the radial and temporal components of the gauge field,
\bea
A^{(mgn)}_r=A^{(mgn)}_{y_i}=0,
\eea
while the other projections require additional analysis presented in the Appendix \ref{AppStatic}. Here we just summarize the results. 

The non--rotating limit of the Myers--Perry solution is obtained by introducing the set of coordinates $\xi_i$ by 
\bea\label{DefYXimain}
y_k^2=b_k^2-(b_k^2-b_{k-1}^2)\xi_k^2,\quad b_0\equiv 0,
\eea
and taking a series of limits in the following order:
\bea\label{LimitElptcMain}
b_n\rightarrow b_{n-1},\quad b_{n-1}\rightarrow b_{n-2},\quad\dots\quad b_2\rightarrow b_1\equiv b\,.
\eea
In this limit, the relation (\ref{EplsdEven}) defining  ellipsoidal coordinates becomes
\bea
(\mu_j)^2=(1-\xi_{j+1}^2)\prod_{k=1}^j (\xi_j)^2\,.
\eea
In equation (\ref{tempAA3}), the limit (\ref{LimitElptcMain}) can be taken in several non--equivalent ways, and, as demonstrated in the Appendix \ref{AppStatic}, there are $2(n-1)$ discrete options,
\bea\label{STansPolrNuPmain}
\nu=\pm b_c:\ A^{(\pm),c}_{\xi_j}=\pm \frac{i}{b}N_{j,c}\,\d_{\xi_j}\Psi,\quad A^{\phi_p,c}
=-\sum_{j} \frac{1-\xi_j^2}{b r^2}
\left[\prod_{k<j} \frac{1}{\xi_k^2}\right] n_j N_{j,c}
L_{j,p}\,\xi_j\d_{\xi_j}\Psi,
\eea
and a family depending on a continuous parameter $\nu\ne \pm b$:
\bea\label{STansPolrNu0main}
A_{\xi_j}=\frac{i{\nu}}{b^2-\nu^2}\d_{\xi_j}\Psi,\quad
A^{\phi_p}=-\sum_{j}\frac{1-\xi_j^2}{b r^2}
\left[\prod_{k< j} \frac{1}{\xi_k^2}\right]\frac{b^2}{\nu^2-b^2}L_{j,p}\,
\xi_j\d_{\xi_j}\Psi\,.
\eea
Functions $(N_{j,c},L_{j,p},n_j)$ entering (\ref{STansPolrNuPmain})--(\ref{STansPolrNu0main}) are defined by (\ref{STdefNc}), (\ref{STdefLp}), (\ref{STdefNnJ}), and their explicit form will not play any role in our discussion. The label $c$ in (\ref{STansPolrNuPmain}) takes values $c=\{1,\dots,(n-1)\}$. We will now demonstrate that there are only $(2n-1)$ dynamical magnetic polarizations: they are given by (\ref{STansPolrNuPmain}) and by (\ref{STansPolrNu0main}) with $\nu=0$. Any magnetostatic configuration can be constructed by taking linear combinations of these independent separable solutions.

\bigskip

We begin with demonstrating that for $\nu\ne \{0,\pm b\}$, the polynomials containing separation constants  disappear from equations (\ref{EvenDimGenSpin}). The easiest way to see this is to observe that a gauge transformation with 
\bea
\Lambda=-\frac{i{\nu}}{b^2-\nu^2}\Psi\nonumber
\eea
leaves $A^{\phi_p}$ and $A_t$ unchanged, but leads to non--cyclic components
\bea 
A_{\xi_j}=0,\quad A_r=-\frac{i{\nu}}{b^2-\nu^2}\d_{r}\Psi,\quad \Psi=\Phi(r)\prod \Xi_j(\xi_j)
\eea
It is clear that Maxwell's equations completely determine all functions $\Xi_j(\xi_j)$, and there are no separation constants. The resulting solution is analogous to the configuration (\ref{aa11}) encountered in four dimensions. This argument breaks down for 
$\nu=\pm b_c$, i.e., for the $2(n-1)$ polarizations (\ref{STansPolrNuPmain}), and for $\nu=0$.  

\bigskip

Let us now discuss the configurations (\ref{STansPolrNuPmain}) with $\nu=\pm b_c$. To arrive at differential equations for various parts of $\Psi$ (specifically, for functions $\Phi(r)$ and $X_i(\xi_i)$), we begin with setting $\omega=m_i=0$ in (\ref{EvenDimGenSpin}) and sending $\la$ to zero:
\bea\label{STmain1}
&&r^2 \frac{d}{dr}\left[\frac{\Delta}{r^2} {\dot\Phi}\right]-
\frac{\beta}{\nu^2} r^{2n-2}\Phi=0,\\
&&(\nu^2-y_j^2) \frac{d}{dy_j}\left[\frac{\prod(b_k^2-y_j^2)}{\nu^2-y_j^2} \frac{dX_j}{dy_j}\right]+\frac{\nu^2-y_j^2}{\nu^2} Q_{n-2}[-y_j^2]X_j=0\nonumber
\eea
Here we defined
\bea
Q_{n-2}[-y_j^2]=\lim_{\la\rightarrow 0}\frac{1}{\la^{2n-2}}P_{n-2}[-\la^2 y_j^2],\quad
\beta=\frac{1}{r^{2(n-2)}}\lim_{\la\rightarrow 0} \frac{P_{n-2}[r^2]}{\la^2}\nonumber
\eea
The last expression agrees with the static limit (\ref{tempAA23}) of the polynomial $P_{n-2}[r^2]$ upon rescaling $\alpha$  as $\alpha=\la^2\beta$ to avoid singularities in the radial equation. Note that $Q_{n-2}$ is a homogeneous polynomial of degree $(n-2)$ in variables $(y_i^2,b_1^2,\dots,b_n^2)$, which also contains separation constants. As we have argued before, all such constants disappear when $y_j$ and $b_k$ go to the same value $b$ via (\ref{DefYXimain}), if $\nu\ne(0,\pm b)$. For $\nu=\pm b_c$, the limit (\ref{DefYXimain})--(\ref{LimitElptcMain}) in (\ref{STmain1}) leaves nontrivial separation constants, and the 
result should be compared to 
the general magnetic polarization in the Schwarzschild--Tangherlini geometry (\ref{TangMagn}) with 
$\omega=0$, 
\bea\label{tempAA26}
&&A^{(mgn)}_t=0,\quad A^{(mgn)}_r=0,\quad A^{(mgn)}_i=g(r)Y_i\nn
&&\frac{1}{\sqrt{h}}\d_i[\sqrt{h}h^{ij}Y_j]=0,\quad
\frac{1}{\sqrt{h}}\d_m[\sqrt{h}{\cal Y}^{mi}]=-\la_3h^{ij}Y_j,\\
&&\frac{1}{r^d}\d_r[r^{d-2}H\d_r g]-
\frac{\la_3g}{r^4}=0,\quad H=\frac{\Delta}{r^d}\,.\nonumber
\eea
Vector $Y_j$ for the most general configuration (\ref{tempAA26}) is presented in the Appendix \ref{SecAppSchw}. 
Identifying function $\Phi$ with $g$ and recalling that $d=2n$, we conclude that the radial equation in (\ref{STmain1}) is indeed reproduced. Then equations of motion guarantee that configuration (\ref{STansPolrNuPmain}) satisfies the remaining relations in (\ref{tempAA26}), and the corresponding vector $Y_j$ can be extracted from the static limit of (\ref{STansPolrNuPmain}). The resulting expressions are not very illuminating.

\bigskip

Finally, let us consider $\nu=0$. In this case the limit $m_i=\omega=0$ in equations (\ref{EvenDimGenSpin}) requires some care since one encounters $0/0$ ambiguity. Rather than analyzing such a limit, we just take the equations for 
$m_i=\omega=\mu=0$ directly from the Appendix \ref{SecAppMPeven}, where they were originally derived:
\bea\label{STmainNu0}
r^2 \frac{d}{dr}\left[\frac{\Delta}{r^2} \frac{d\Phi}{dr}\right]-
P_{n-1}[r^2]\Phi=0,\quad
x_j^2 \frac{d}{dx_j}\left[\frac{H_j}{x_j^2} \frac{dX_j}{dx_j}\right]+P_{n-1}[-x_j^2]X_j=0
\eea
Note that, in contrast to (\ref{EvenDimGenSpin}), these equations contain an arbitrary polynomial of degree $(n-1)$. Dimensional analysis ensures that, after the rescaling (\ref{abyLmbd}), this polynomial can be written as
\bea
P_{n-1}[z]=\alpha z^{n-1}+\sum_{k=1}^{n-1} c_k \la^{2k} z^{n-1-k}\, ,\nonumber
\eea
where coefficients $c_k$ depend on the values of $b_p$. In particular, the $\la=0$ limit in the radial equation in (\ref{STmainNu0}) gives
\bea
r^2 \frac{d}{dr}\left[\frac{\Delta}{r^2} \frac{d\Phi}{dr}\right]-
\alpha r^{2(n-1)}\Phi=0,
\eea
in the perfect agreement with the equation for $g$ from (\ref{tempAA26}).  Furthermore, it is clear that the configuration (\ref{STansPolrNu0main}) with $\nu=0$ satisfies the constraint $\nabla^j Y_j=0$ from (\ref{tempAA26}), then the equation for ${\cal Y}^{mi}$ follows from the ODEs on $X_j$. 

\bigskip

To summarize, in this subsection we have demonstrated that in the static limit, which involves taking $\la$ to zero followed by (\ref{DefYXimain})--(\ref{LimitElptcMain}), the separable solutions (\ref{AnsGenEvenElectr}), (\ref{AnsGenEvenMagn}) lead to $D-2$ non--equivalent branches. The limit (\ref{tempAA7}) reduces to the electric polarization (\ref{TangElectrTmp}), while the limits (\ref{STansPolrNuPmain}) and (\ref{STansPolrNu0main}) with $\nu=0$ reproduce all $2(n-1)+1=D-3$ magnetic polarizations (\ref{tempAA26}) constructed in the Appendix \ref{SecAppSchw}. Although we focused on even values of $D$, similar arguments are applicable to the odd--dimensional case as well, so solutions \{(\ref{AnsGenEvenElectr}), (\ref{AnsGenEvenMagn}), (\ref{EvenDimGenSpin})--(\ref{EvenDimGenSpinLmb})\} and (\ref{GenOddElectr})--(\ref{GenOddLmbda}) cover all 
$(D-2)$ polarizations of the electromagnetic field in an arbitrary number of dimensions.

\section{Discussion}

In this article we have demonstrated separability of the Maxwell's equations in the background of the Myers--Perry black hole and derived the systems of ODEs governing separable solutions. In four dimensions our ansatz differs from the classic solution by Teukolsky, and this modification allowed us to construct separable solutions for both polarizations of photons (\ref{4dHighSpinGuess})--(\ref{4dSmryAnstz}). In higher dimensions, we have constructed all independent polarizations of the electromagnetic waves, and our results are summarized in section \ref{SecKerrAdSMxwMP}. We have also clarified the relation between separation of variables in Maxwell's equations and symmetries encoded in the Killing(--Yano) tensors. 

This work has several implications. First and foremost, separation of Maxwell's equations should allow one to 
study electromagnetic excitations of higher dimensional black holes, both for understanding the scattering of waves from such objects and for getting new insights into Hawking radiation. By adding D--brane charges to the systems discussed in this article, one can also use the results derived here to get a better understanding of AdS/CFT correspondence for systems originating from rotating branes. It would also be very interesting to use the framework introduced this article for extending our results to particles with higher spin, in particular, to gravitational waves.

\section*{Acknowledgments}

This work is supported in part by the DOE grant DE\,-\,SC0017962.

\appendix
\section{Teukolsky's solution for the Kerr geometry}
\label{SecAppTeuk}
\label{SecAppChandr}
\renewcommand{\theequation}{A.\arabic{equation}}
\setcounter{equation}{0}

As discussed in section \ref{SecKerrReview}, equations for {\it some components} of the Maxwell field separate in the Kerr geometry, and the ansatz (\ref{TeukAnsatz}) results in equations (\ref{TeukEqn}) \cite{Teuk}. This appendix will present some details of the analysis leading to (\ref{TeukEqn}) and (\ref{KerrMaxwFrames1}), and we will mostly follow the nice pedagogical discussion of \cite{ChandraBook}.

To apply the Newman--Penrose formalism to Maxwell field in the Kerr geometry, it is convenient to 
introduce differential operators constructed from the frames (\ref{FramesKerr}):
\bea
&&\DD_n=\d_r+\frac{iK}{\Delta}+2n\frac{r-M}{\Delta},\quad
\DD^\dagger_n=\d_r-\frac{iK}{\Delta}+2n\frac{r-M}{\Delta},\nn
&&\LL_n=\d_\theta+Q+n\cot_\theta,\quad \LL^\dagger_n=\d_\theta-Q+n\cot_\theta,\\
&&K=-i(r^2+a^2)\d_t-ia\d_\phi\quad Q=-ia s_\theta\d_t-\frac{i}{s_\theta}\d_\phi\,.\nonumber
\eea
The relationship between operators with subscript zero and frames (\ref{FramesKerr}) is especially simple:
\bea\label{DoLoFrames}
\DD_0=l^\mu\d_\mu,\quad \DD_0^\dagger=-\frac{2\Sigma}{\Delta}n^\mu\d_\mu,\quad
\LL_0^\dagger=\sqrt{2}{\rho}m^\mu\d_\mu\,.
\eea
We will be interested in applying differential operators $(\DD_n,\LL_n)$ to functions with a specific dependence on the cyclic coordinates:
\bea
\Phi(r,\theta,t,\phi)=e^{i\omega t+im \phi}{\tilde\Phi}(r,\theta).
\eea
Then functions $K$ and $Q$ become
\bea
K=(r^2+a^2)\omega+am\quad Q=a\omega s_\theta+\frac{m}{s_\theta}.
\eea
Substituting the field strength (\ref{StrenFramesNP}) into Maxwell's equations ($dF=d\star F=0$) and contracting the results with frames, one finds \cite{ChandraBook}
\bea\label{ChandraMaxw1}
&&\left[\LL_1-\frac{ias_\theta}{\bar\rho}\right]\Phi_0=\left[\DD_0+\frac{1}{\bar\rho}\right]\Phi_1,\quad
\left[\LL_0^\dagger+\frac{ias_\theta}{\bar\rho}\right]\Phi_1=
-\Delta\left[\DD_1^\dagger-\frac{1}{\bar\rho}\right]\Phi_0,\\
\label{ChandraMaxw2}
&&\left[\LL_0+\frac{ias_\theta}{\bar\rho}\right]\Phi_1=\left[\DD_0-\frac{1}{\bar\rho}\right]\Phi_2,\quad
\left[\LL_1^\dagger-\frac{ias_\theta}{\bar\rho}\right]\Phi_2=
-\Delta\left[\DD_0^\dagger+\frac{1}{\bar\rho}\right]\Phi_1.
\eea
To make these equations more symmetric, some components of (\ref{StrenFramesNP}) were rescaled as
\bea
\Phi_0=\phi_0,\quad \Phi_1=\sqrt{2}\phi_1{\bar\rho},\quad \Phi_2=2\phi_2{\bar\rho}^2.
\eea
Commutativity of various operations appearing in (\ref{ChandraMaxw1}) allows one to eliminate $\Phi_1$ from these equations, then further simplifications lead to the final equation for $\Phi_0$ \cite{ChandraBook}:
\bea\label{ChandraTeuk0}
\left[\Delta\DD_1\DD_1^\dagger+\LL_0^\dagger\LL_1-2i\omega(r+iac_\theta)\right]\Phi_0=0.
\eea
Similar manipulations with equations (\ref{ChandraMaxw2}) give
\bea\label{ChandraTeuk2}
\left[\Delta\DD_0^\dagger\DD_0+\LL_0\LL_1^\dagger+2i\omega(r+iac_\theta)\right]\Phi_2=0.
\eea
Equations (\ref{ChandraTeuk0}), (\ref{ChandraTeuk2}) separate in the $(r,\theta)$ variables, and the ansatz 
(\ref{TeukAnsatz}) leads to the Teukolsky equations (\ref{TeukEqn}). However, equations  (\ref{ChandraMaxw1})--(\ref{ChandraMaxw2}) make it clear that modes $\Phi_0$ and $\Phi_2$ are not independent, so functions $(S_\pm,R_\pm)$ appearing in (\ref{TeukEqn}) are subject to various constraints. 
The relations between $(S_\pm,R_\pm)$ were worked out in a series of articles \cite{ChandPaper}, and the results read \cite{ChandraBook}
\bea\label{Starob}
\Delta\DD_0\DD_0 R_-=\mu\Delta R_+,\quad 
\Delta\DD^\dagger_0\DD^\dagger_0\Delta R_+=\mu R_-,\quad
\LL_0\LL_1 S_+=\mu S_-,\quad \LL^\dagger_0\LL^\dagger_1 S_-=\mu S_+\,.\nn
\eea
Here
\bea
\mu=\left[\la^2-4(a\omega)(a\omega+m)\right]^{1/2}\,.
\eea
Relations (\ref{Starob}) do not diminish the value of equations (\ref{TeukEqn}), they just mean that the modes with $s=\pm 1$ are not independent, and once a solution for $s=1$ is chosen, its counterpart for $s=-1$ is completely determined by (\ref{Starob}). In other words, relations (\ref{TeukEqn}), (\ref{Starob})  describe only one polarization of the electromagnetic wave, and to recover the second polarization, one must look at $\Phi_1$. Unfortunately, equation for this function does not separate. 

\bigskip

We conclude this appendix by quoting the expression for the gauge potential given by equations 
(8.90)--(8.93)
of \cite{ChandraBook}\footnote{We multiplied the entire gauge field by $\sqrt{2}$ to remove the unnecessary irrational factors.}:
\bea\label{ChandraPotential}
A_r&=&\frac{ia}{\Delta}\left[P_+ f_++P_-f_-\right]+\left[\DD_0H_++\DD^\dagger_0 H_-\right],\nn
A_\theta&=&-[g_+S_++g_-S_-]+[\LL_0^\dagger H_++\LL_0 H_-],\nn
A_t&=&\frac{ia}{|\rho|^2}[P_+f_+-P_-f_--s_\theta(g_+S_+-g_-S_-)]\nn
&&+\frac{1}{|\rho|^2}\left[\Delta(\DD_0H_+-\DD_0^\dagger H_-)+ia(\LL^\dagger_0 H_+-\LL_0 H_-)s_\theta\right]\,,\\
A_\phi&=&-\frac{i}{|\rho|^2}[a^2(P_+f_+-P_-f_-)s_\theta^2-(r^2+a^2)s_\theta(g_+S_+-g_-S_-)]\nn
&&-\frac{1}{|\rho|^2}\left[a s_\theta^2\Delta(\DD_0H_+-\DD_0^\dagger H_-)+
i(r^2+a^2)(\LL^\dagger_0 H_+-\LL_0 H_-)s_\theta\right]\,.\nonumber
\eea
Here 
\bea
P_-=R_-,\quad P_+=\Delta R_+,
\eea
and functions $(f_\pm,g_\pm,H_\pm)$ are determined by solving differential equations
\bea\label{ChndrEqnForDg}
\LL_0^\dagger f_+=c_\theta S_+,\quad
\LL_0 f_-=c_\theta S_-,\quad \Delta \DD_0 g_+=r P_+,\quad \Delta \DD^\dagger_0 g_-=r P_-\,,
\eea
and
\bea\label{ChndrEqnForH}
\DD^\dagger_0\frac{\Delta \DD_0 H_+}{{\bar\rho}^2}+\LL_1\frac{\LL^\dagger_0 H_+}{{\bar\rho}^2}
-\DD_0\frac{\Delta \DD^\dagger_0 H_-}{{\bar\rho}^2}-\LL_1^\dagger\frac{\LL_0 H_-}{{\bar\rho}^2}
=0.
\eea
The last equation does not appear to be separable. In section \ref{SecKerrReview} we rewrite the expressions (\ref{ChandraPotential}) in a more suggestive form (\ref{KerrMaxwFrames}), and in section \ref{SecKerrNewAns} we use this result as a motivation for a better ansatz that makes equations for all polarizations separable. 

\section{Derivation of the new equations for the Kerr geometry}
\renewcommand{\theequation}{B.\arabic{equation}}
\setcounter{equation}{0}
\label{SecAppKerrNew}

In section \ref{SecKerrNewAns} we introduced the new separable ansatz (\ref{KerrNewAnstz}) for the gauge field in four dimensions. While derivation of the resulting equations is rather straightforward in the non--cyclic case ($\omega=m=0$), extension to nontrivial time and angular dependence requires some work, and the details are presented in this appendix. As we saw in subsection \ref{SectSubKerrStat}, the separable solutions are divided into two branches already in the static case, so the same property must persist for nonzero $(\omega,m)$. The resulting ``electric'' and ``magnetic'' branches will be discussed in subsections \ref{SecAppKerrElectr} and \ref{SecAppKerrMagn}. Both analyses follow the logical steps outline on page \pageref{Steps4D}.

\subsection{Electric polarization}
\label{SecAppKerrElectr}

In this subsection we will derive the ```electric solution'' (\ref{KerrElectrFinal}) by starting with $\omega=m=0$ configuration (\ref{SpecialKerrElectr}) and adding the dependence on $(t,\phi)$ coordinates. The discussion will follow the steps outlined on page \pageref{Steps4D}. 

\bigskip

Although eventually we are interested in waves in the black hole geometry, it is instructive to begin with flat space in spheroidal coordinates:
\bea\label{KerrFlat}
ds^2=-dt^2+(r^2+a^2 c_\theta^2)\left[\frac{dr^2}{r^2+a^2}+d\theta^2\right]+
(r^2+a^2)s_\theta^2 d\phi^2.
\eea
This metric is obtained from (\ref{Kerr1}) by setting $M=0$. As we already observed in the static case (\ref{SpecialKerrElectr}), the mass appears only in the equation for the radial profile $R$, not in equation for $S$ or the prefactors 
$(F_\pm,G_\pm)$. We will see that this property persists for the general waves as well, so by solving Maxwell's equations in the metric (\ref{KerrFlat}) we will be able determine five out of six ingredients 
$(F_\pm,G_\pm, S,R)$ of the electromagnetic configuration. Then finding the last equation for $R$ would be rather straightforward. 

Although in principle one can repeat the analysis of section \ref{SectSubKerrStat} to argue for separation into two branches (the counterparts of (\ref{SpecialKerrElectr}) and  (\ref{SpecialKerrMagn})), such approach requires complicated algebraic manipulations. As an alternative, we observe that since the branches (\ref{SpecialKerrElectr}) and (\ref{SpecialKerrMagn}) are already distinct in the special case (\ref{OmMZero}), they must be disconnected for the generic values of $(m,\omega)$. The distinction between the two branches (\ref{SpecialKerrElectr}) and (\ref{SpecialKerrMagn}) becomes especially transparent at $a=0$, when the electric solution has $F_\pm=0$, and it is natural to insist on this property even for arbitrary $(\omega,m)$. In other words, we begin with imposing the ansatz
\bea\label{KerrElectrNonRot}
l^\mu A_\mu=G_+ {\hat l} \Psi,\quad
n^\mu A_\mu=G_-{\hat n}\Psi,\quad m^\mu A_\mu=
{\bar m}^\mu A_\mu=0,\quad \Psi=e^{i\omega t+im\phi}R(r)S(\theta)
\eea
in the flat space (\ref{KerrFlat}) with $a=0$. Defining the components of Maxwell's equations by (\ref{MaxwNotation}), we find
\bea\label{KerrA0SecOrder}
&&m_\mu \MM^\mu=-\frac{1}{2\sqrt{2}r}\left(S'-\frac{mS}{s_\theta}\right)\mathscr{N},
\quad
{\bar m}_\mu \MM^\mu=-\frac{1}{2\sqrt{2}r}\left(S'+\frac{mS}{s_\theta}\right)\mathscr{N},\nn
&&\mathscr{N}=
\frac{d}{dr}\left[(G_++G_-)\dot R\right]+\omega^2 (G_++G_-)R+i\omega({\dot G}_+-{\dot G}_-)R.
\eea
The remaining components are more complicated, so we are not writing them here. The Maxwell's equations require $\mathscr{N}$ to vanish\footnote{The alternative, $S=0$ leads to trivial gauge field.}, this allows one to express ${\ddot R}$ in terms of $(R,{\dot R},G_\pm,{\dot G}_\pm)$ and substitute the results into the remaining Maxwell's equations\footnote{The degenerate case, $G_-=-G_+$ requires a separate analysis, and the results are consistent with our final conclusion (\ref{KerrA0ElectSeqn}).}. Note that after such substitution $R$ and ${\dot R}$ must be treated as independent variables\footnote{Any relation between $R$ and ${\dot R}$ would lead to a first--order equation for the radial profile, which is more restrictive that the second order equation (\ref{SpecialKerrElectr})  which we have encountered in the special case.}, so we get two equations for every component of $\MM$. One combination looks especially simple:
\bea
(\Delta l_\mu-2\Sigma n_\mu)\MM^\mu\Big|_{R=0}&=&
-{\dot R}\, \frac{2i\omega r^2(G_+{\dot G}_--G_-{\dot G}_+)}{G_++G_-}S\nn
&&-{\dot R}(G_++G_-)
\left[\frac{1}{s_\theta}(s_\theta S')'-\frac{m^2 S}{s_\theta^2}\right].
\eea
Consistency of separation leads to a differential equation for $S$:
\bea\label{KerrA0ElectSeqn}
\frac{1}{s_\theta}(s_\theta S')'-\frac{m^2 S}{s_\theta^2}+\la_1 S=0,
\eea
which generalizes its counterpart from (\ref{SpecialKerrElectr}). Substitution of $S''$ into the remaining Maxwell's equations also eliminates $S'$, leading to a system of linear {\it algebraic} equations for 
$(R,{\dot R})$ with coefficients involving $(G_\pm,{\dot G}_\pm,{\ddot G}_\pm)$. Existence of solutions for $(R,{\dot R})$ implies that
\bea\label{KerrGpmIntCnstr}
G_-=-\frac{r}{1+C r},\quad G_+=\frac{\la_1 r}{\la_1+(2i\omega+C\la_1)r}
\eea
with an arbitrary integration constant $C$. We choose this constant to have some symmetry between $G_+$ and $G_-$:
\bea\label{KerrGpmFinal}
G_-=-\frac{\la_1 r}{\la_1-i\omega r},\quad G_+=\frac{\la_1 r}{\la_1+i\omega r}\,.
\eea 
Note that equation (\ref{KerrGpmIntCnstr}) with $\omega=0$ leads to relation $G_+=-G_-$ invalidating the steps following equation (\ref{KerrA0SecOrder})\footnote{This is especially clear from equation (\ref{KerrA0ElectSeqn}), where $(G_++G_-)$ appears in the denominator.}. A separate analysis of this degenerate case still leads to equation (\ref{KerrA0ElectSeqn}), although $G_+$ remains unconstrained. While it is possible to determine this function for arbitrary values of $m$ following the steps used in section \ref{SectSubKerrStat} for $m=\omega=0$, here we will focus on non--vanishing $\omega$ and recover the degenerate case by taking a limit. 

Substituting (\ref{KerrGpmFinal}) into (\ref{KerrA0SecOrder}), we find a differential equation for $R$:
\bea\label{KerrA0ElectReqn}
E_r\frac{d}{dr}\left[\frac{r^2}{E_r}{\dot R}\right]+\left[\la_1+(\omega r)^2-\frac{2\la_1^3}{E_r}\right]R=0.
\eea
Here we introduced a convenient notation
\bea\label{DefErSpecial}
E_r\equiv \la_1^2+(\omega r)^2,
\eea
which is used throughout this article. Before turning on the rotational parameter $a$, we observe that the solution (\ref{KerrElectrNonRot}), (\ref{KerrA0ElectSeqn}), (\ref{KerrGpmFinal}), (\ref{KerrA0ElectReqn}) works even for Schwarzschild geometry after a minor modification of the radial equation:
\bea\label{KerrElectrSchw}
&&l^\mu A^{(el)}_\mu=\frac{\la_1 r}{\la_1+i\omega r}{\hat l} \Psi,\quad
n^\mu A^{(el)}_\mu=-\frac{\la_1 r}{\la_1-i\omega r}{\hat n}\Psi,\quad m^\mu A^{(el)}_\mu=
{\bar m}^\mu A^{(el)}_\mu=0,\nn
&&\frac{1}{s_\theta}(s_\theta S')'-\frac{m^2 S}{s_\theta^2}+\la_1 S=0,\\
&&E_r\frac{d}{dr}\left[\frac{\Delta}{E_r}{\dot R}\right]+
\left[\la_1+(\omega r)^2-\frac{2\la_1^3}{E_r}+\frac{2M\omega^2 r^3}{\Delta}\right]R=0.\nonumber
\eea
The same pattern for introduction of mass will persist in the Kerr geometry and in its higher--dimensional counterparts: once the equations for massless case are found, $M$ can be added by a simple modification of the radial equation.

\bigskip

Let us now discuss the Maxwell's equations in the metric (\ref{KerrFlat}) containing the rotation 
parameter\footnote{We recall in spite of parameter $a$, the metric (\ref{KerrFlat}) describes flat space without rotation.} $a$. In section \ref{SectSubKerrStat} we have rigorously derived the solutions (\ref{SpecialKerrElectr}), (\ref{SpecialKerrMagn}), and in this appendix the derivation was extended to the full electric polarization (\ref{KerrElectrNonRot}), (\ref{KerrA0ElectSeqn}), (\ref{KerrGpmFinal}), (\ref{KerrA0ElectReqn}), even with no-zero $(\omega,m)$. This strongly suggests that even in the rotating case, there should be a unique ``electric'' configuration for every set  of $(\omega,m)$ and every allowed value of the separation constant $\la_1$. Assuming such uniqueness, we can use the result (\ref{KerrGpmFinal}) to guess the form of $(F_\pm,G_\pm)$ for the general case, and find the resulting equations for $R$ and $S$. A consistency of the final system will serve as a highly nontrivial confirmation of the guess.

Assuming that the structure (\ref{KerrGpmFinal}) is preserved even in the presence of rotation, we impose an ansatz
\bea\label{GpmElctrAnstz}
G_-=-\frac{r}{1-i\mu r},\quad G_+=\frac{r}{1+i\mu r}
\eea 
with constant $\mu$. Since we are no longer starting from equation (\ref{KerrA0ElectSeqn}), parameter $\la_1$ no longer plays a special role, so we replaced $\omega/\la_1$ by a new constant $\mu$ to simplify the relations 
(\ref{GpmElctrAnstz}). In the presence of the rotation parameter $a$, prefactors $(F_+,F_-)$ will be turned on as well (cf. the special solution (\ref{SpecialKerrElectr})), and to guess their form, we observe that the metric (\ref{KerrFlat}) is invariant under a $Z_2$ symmetry:
\bea\label{Z2Electr}
r\leftrightarrow  iac_\theta.
\eea
and its analog with a different sign. To make the ansatz (\ref{KerrNewAnstz}) covariant under this symmetry, the transformation (\ref{Z2Electr}) must interchange $(G_+,G_-)$ with $(F_+,F_-)$, so we find
\bea
F_+=-\frac{iac_\theta}{1+\nu ac_\theta},\quad F_-=\frac{iac_\theta}{1-\nu ac_\theta}\,.
\eea 
The symmetry predicts that $\nu=\pm\mu$, and the correct sign is determined from the consistency of Maxwell's equations:
\bea
\nu=\mu.
\eea
Next we look at a particular component of the Maxwell's equations $\MM^\mu=0$:
\bea\label{TempEqnJul17}
0&=&\sqrt{2}(\rho m_\mu+{\bar\rho}{\bar m}_\mu)\MM^\mu\nn
&=&
\frac{2i}{E_\theta}\left[\mu S'+\frac{a^2\Omega_\theta E_r s_{2\theta}}{2\Sigma}S\right]\frac{d}{dr}\left[\frac{r^2+a^2}{E_r}{\dot R}\right]+
R\FF (r,\theta,S,S',S'')\,,\\
\Omega_\theta&=&\omega+\frac{m}{as_\theta^2}\,.\nonumber
\eea
Here $E_r$ and $E_\theta$ are the analogs of the function (\ref{DefErSpecial}) encountered before:
\bea\label{KerrEEfuncEl}
E_r\equiv 1+(\mu r)^2,\quad E_\theta\equiv 1-(\mu a c_\theta)^2.
\eea
The second term ($R\FF$) in equation (\ref{TempEqnJul17}) is rather complicated, but it does not contain derivatives of $R$, then consistency of the relation (\ref{TempEqnJul17}) implies that
\bea\label{Jul17eqnR}
\frac{d}{dr}\left[\frac{r^2+a^2}{E_r}{\dot R}\right]+g(r)R=0
\eea
for some function $g(r)$. This is the main differential equation for $R$, which in special cases should reduce to (\ref{Jul17Reqn}) and (\ref{KerrA0ElectReqn}) discussed earlier.

Similar manipulations with Maxwell's equation
\bea
0&=&(\Delta l_\mu-2\Sigma n_\mu)\MM^\mu\nn
&=&-\frac{2i(r^2+a^2)}{s_\theta E_r}\left[\mu {\dot R}-\frac{\Omega_r E_\theta r}{\Sigma}R\right]\frac{d}{d\theta}\left[\frac{s_\theta}{E_\theta}{S'}\right]+
S{\tilde\FF} (r,\theta,R,{\dot R},{\ddot R})\,,\\
\Omega_r&=&\omega+\frac{am}{r^2+a^2}\,.\nonumber
\eea
lead to the main differential equation for $S$:
\bea\label{Jul17eqnS}
\frac{d}{d\theta}\left[\frac{s_\theta}{E_\theta}S'\right]+h(\theta)S=0.
\eea
Substitution of (\ref{Jul17eqnR}), (\ref{Jul17eqnS}) and their derivatives into Maxwell's equations leads to a system of {\it algebraic} equations for $(R,{\dot R},S,S')$, where all all four objects should be treated as independent variables. This gives many differential constraints on functions $(g,h)$ and their derivatives, and a priori the resulting over--defined system is not guaranteed to have nontrivial solutions. Remarkably, there is a unique solution, and this fact provides a highly nontrivial consistency check of our ansatz (\ref{KerrNewAnstz}).
Substituting the expressions for $g$ and $h$ into (\ref{Jul17eqnR}) and (\ref{Jul17eqnS}), we find the final expression for the photons with ``electric'' polarization in the flat geometry (\ref{KerrFlat}):
\bea\label{KerrElectrM0}
&&G_\pm=\pm\frac{r}{1\pm i\mu r},\quad
F_\pm=\mp\frac{i a c_\theta}{1\pm \mu ac_\theta},\quad
E_r=1+(\mu r)^2,\quad
E_\theta=1-(\mu a c_\theta)^2
\nn
&&\frac{E_\theta}{s_\theta}\frac{d}{d\theta}\left[\frac{s_\theta}{E_\theta} S'\right]
+\left\{-\frac{2\Lambda}{E_\theta}+
(a\omega c_\theta)^2-\frac{m^2}{s_\theta^2}-C\right\}S=0,
\\
&&
\Lambda=a\mu[m+a\omega-\frac{\omega}{a\mu^2}],\quad C=-{\Lambda}+2 am\omega+(a\omega)^2,\nonumber\\
&&{E_r}\frac{d}{dr}\left[\frac{r^2+a^2}{E_r} {\dot R}\right]
+\left\{\frac{2\Lambda}{E_r}+
(\omega r)^2+\frac{(am)^2}{r^2+a^2}+C\right\}R=0.\nonumber
\eea
To extend this result to the Kerr black hole, we observe that in the special cases (\ref{SpecialKerrElectr}) and (\ref{KerrElectrSchw}) the mass $M$ appears only in the differential equation for $R$, while all other relations remain the same as for $M=0$. Assuming that this property persists in the general case, we impose all relations in (\ref{KerrElectrM0}) with the exception of the last equation. The resulting system of Maxwell's equations turns out to be solvable for $R(r)$, and the unique result is given by (\ref{KerrElectrFinal}).

\subsection{Magnetic polarization}
\label{SecAppKerrMagn}

The analysis of the magnetic polarization follows the same steps as the electric case, so this subsection we will be very brief. The goal of this presentation is to stress the uniqueness of the magnetic polarization, the fact that will be very important in the discussion of black holes in higher dimensions. 

\bigskip

As in the electric case, we begin with analyzing the waves in the flat geometry (\ref{KerrFlat}), and our starting point is the application of the ansatz (\ref{KerrNewAnstz}) to the metric (\ref{KerrFlat}) with $a=0$. As in the electric case, expression (\ref{SpecialKerrMagn}) suggests that the non--rotating geometry would give 
$G_\pm=0$, thus in the present situation, the relation (\ref{KerrElectrNonRot}) is replaced by 
\bea\label{KerrMagnNonRot}
\hskip -0.5cm l^\mu A_\mu=
n^\mu A_\mu=0,\quad m^\mu A_\mu=F_+ {\hat m} \Psi,\quad 
{\bar m}^\mu A_\mu=F_-{\hat {\bar m}}\Psi,\quad \Psi=e^{i\omega t+im\phi}R(r)S(\theta).
\eea
Then it is natural to look at the components of the Maxwell's equations, which are complementary to (\ref{KerrA0SecOrder}):
\bea\label{KerrA0SecOrderMagn}
&&l_\mu \MM^\mu=-\frac{1}{2s_\theta^2}\left({\dot R}+i\omega R\right)\mathscr{N},
\quad
{\bar n}_\mu \MM^\mu=-\frac{1}{4s_\theta^2 \Sigma}\left({\dot R}-i\omega R\right)\mathscr{N},\nn
&&\mathscr{N}=
s_\theta\frac{d}{d\theta}\left[s_\theta(F_++F_-)S'\right]- m^2(F_++F_-)S-m({F}'_+-{F}'_-)s_\theta S.
\eea
Similar to the electric case, the projection
\bea
(\rho m_\mu+{\bar\rho}{\bar m}_\mu)\MM^\mu\Big|_{S=0}=0
\eea
leads to the equation for $R$, a counterpart of (\ref{KerrA0ElectSeqn}),
\bea\label{KerrA0MagnReqn}
{\ddot R}+\omega^2 R-\frac{\la_2}{r^2} R=0,
\eea
which generalizes the last equation in (\ref{SpecialKerrMagn}) (recall that we are working in the limit $a=M=0$). Repeating the steps which led to (\ref{KerrGpmIntCnstr}), we find
\bea
F_+=-\frac{\la_2}{\la_2 c_\theta+C},\quad F_-=\frac{\la_2}{\la_2 c_\theta-2m+C}\,.\nonumber
\eea
A symmetric choice of the integration constant $C$ gives the counterpart of (\ref{KerrGpmFinal}):
\bea\label{KerrGpmFinalMagn}
F_+=-\frac{\la_2}{\la_2 c_\theta+m},\quad F_-=\frac{\la_2}{\la_2 c_\theta-m}\,.
\eea 
As in the electric case, these functions lead to the unique differential equation for $S$, and we conclude the discussion of the $a=0$ limit by extending the result from flat space to the Schwarzschild geometry:
\bea\label{KerrMagnSchw}
&&l^\mu A_\mu=n^\mu A_\mu=0,\quad
m^\mu A_\mu=-\frac{\la_2}{\la_2 c_\theta+m}{\hat m} \Psi,\quad
{\bar m}^\mu A_\mu=\frac{\la_2}{\la_2 c_\theta-m}{\hat {\bar m}} \Psi,\nn
&&\frac{M_\theta}{s_\theta}\frac{d}{d\theta}\left[\frac{s_\theta}{M_\theta}S'\right]+
\left[\la_2-\frac{m^2}{s_\theta^2}+\frac{2\la_2 m^2}{M_\theta}\right]S=0,\quad
M_\theta=(\la_2 c_\theta)^2-m^2,\\
&&{\ddot R}+\omega^2 R-\frac{\la_2}{r^2} R+\frac{2M\omega^2 r}{\Delta}R=0.\nonumber
\eea
As before, we observe that $M$--dependence is introduced by a simple modification of the radial equation.

\bigskip

Let us now discuss the magnetic branch in the metric (\ref{KerrFlat}) containing a rotation 
parameter $a$. As in the electric case, we make a guess for $(F_\pm,G_\pm)$ and solve the resulting equations for $R$ and $S$. Specifically, we assume that function $F_+$ and $F_-$ maintain the structure (\ref{KerrGpmFinalMagn}) even in the rotating case:
\bea
F_+=-\frac{1}{c_\theta-\mu},\quad F_-=\frac{1}{c_\theta+\mu}\,.\nonumber
\eea
Then the symmetry (\ref{Z2Electr}) can be used to argue that functions $G_\pm$ have the form
\bea
G_+=\frac{ia}{r+ia\nu},\quad G_-=-\frac{ia}{r-ia\nu}\,.
\eea 
As in the electric case, the $Z_2$ symmetry symmetry (\ref{Z2Electr}) ensures that $\nu=\pm\mu$, and the correct sign is determined for the consistency of Maxwell's equations:
\bea
\nu=\mu.
\eea
Analyzing various components of Maxwell's equations, as in subsection \ref{SecAppKerrElectr}, we arrive at the counterparts of 
(\ref{Jul17eqnR}) and (\ref{Jul17eqnS}),
\bea\label{Jul17eqnRSmagn}
\frac{d}{dr}\left[\frac{r^2+a^2}{M_r}{\dot R}\right]+g(r)R=0,\quad
\frac{d}{d\theta}\left[\frac{s_\theta}{M_\theta}S'\right]+h(\theta)S=0\,,
\eea
with undetermined functions $g(r)$ and $h(\theta)$. Here $M_r$ and $M_\theta$ are the magnetic counterparts of the electric functions (\ref{KerrEEfuncEl}):
\bea\label{KerrMMfuncMagn}
M_r\equiv r^2+(\mu a)^2,\quad M_\theta\equiv c_\theta^2-\mu^2\,.
\eea
This is the main differential equation for $R$, which in special cases should reduce to (\ref{Jul17ReqnMgn}) and (\ref{KerrA0MagnReqn}) discussed earlier.

Substitution of relations (\ref{Jul17eqnRSmagn}) and their derivatives into Maxwell's equations leads to an over--constrained system of differential equations for $f$ and $g$, and as in the electric case, this system admits a unique solution, which provides a highly nontrivial consistency check of our ansatz (\ref{KerrNewAnstz}). 
The final expression for the photons with ``magnetic'' polarization in the flat geometry (\ref{KerrFlat}) reads
\bea\label{KerrMagnM0}
&&G_\pm=\pm\frac{ia}{r \pm i\mu a},\quad
F_\pm=\mp\frac{1}{c_\theta\mp \mu},\quad 
M_r= r^2+(\mu a)^2,\quad M_\theta= c_\theta^2-\mu^2,\nn
&&\frac{M_\theta}{s_\theta}\frac{d}{d\theta}\left[\frac{s_\theta}{M_\theta}\d_\theta S\right]
+\left\{-\frac{m^2}{s_\theta^2}-\frac{2\Lambda}{M_\theta}+(a\omega c_\theta)^2-C\right\}S=0,\nn
&&{M_r}\frac{d}{dr}\left[\frac{r^2+a^2}{M_r} R'\right]+\left\{
-\frac{2\Lambda a^2}{M_r}+\frac{(a m)^2}{r^2+a^2}+(r\omega)^2+C\right\}R=0,\\
&&
\Lambda=\mu\left[a\omega+m-{a\omega\mu^2}\right],\quad
C=\frac{\Lambda}{\mu}+a\omega\left[a\omega+2m\right].\nonumber
\eea
The extension to the Kerr black hole is accomplished by modifying the radial equation, and the final result is given by (\ref{KerrMagnFinal}). 

\section{Maxwell's equations in five dimensions}
\renewcommand{\theequation}{C.\arabic{equation}}
\setcounter{equation}{0}
\label{SecApp5D}

In this appendix we will derive the systems of ordinary differential equations associated with separable solutions of Maxwell's equations in the background of a five--dimensional rotating black hole. Although the logical steps will be very similar to the one encountered in the Appendix \ref{SecAppKerrNew}, the resulting four-- and five--dimensional solution will have very different structures. The main goal of this appendix is to demonstrate that the five--dimensional solutions (\ref{FullElectr5d}), (\ref{FullMagn5d}) are unique, and in section \ref{SecWaveMP} the results are extended to black holes in all odd dimensions. 

\subsection{Electric polarization}
\label{SecApp5DElectr}

In this subsection we will derive the ```electric solution'' (\ref{FullElectr5d}) by starting with $\omega=m=n=0$ configuration (\ref{Special5DElectr}) and adding dependence on $(t,\phi,\psi)$ coordinates. The discussion will follow the steps outlined on page \pageref{Steps4D}. 

\bigskip

Although eventually we are interested in waves in black hole geometry, as in the four--dimensional case, we begin with flat space in spheroidal coordinates\footnote{This metric is obtained from (\ref{Metr5D}) by setting $M=0$.}:
\bea\label{KerrFlat5D}
ds^2&=&-dt^2+\Sigma\left[\frac{r^2dr^2}{\Delta_0}+d\theta^2\right]+
(r^2+a^2)s_\theta^2 d\phi^2+(r^2+b^2)c_\theta^2 d\psi^2,\\
\Sigma&=&r^2+a^2 c_\theta^2+b^2 s_\theta^2,\quad \Delta_0=(r^2+a^2)(r^2+b^2).\nonumber
\eea
As demonstrated in section \ref{SectSub5dStat}, in the special case (\ref{Special5D}) the electromagnetic field splits into two distinct branches, (\ref{Special5DElectr}) and (\ref{Special5dMagn}), then continuity implies that these polarizations remain separate for generic values of $(m,n,\omega)$. In this subsection we focus on generalizing the electric solution (\ref{Special5DElectr}).

\bigskip

Comparing the special solution (\ref{Special5DElectr}) with the general ansatz (\ref{5dNewAnstz}), 
we observe that in the 
non--rotating case ($a=b=0$) functions $F_\pm(\theta)$ and constant $\la$ vanish. By relaxing the assumption (\ref{Special5D}), we add dimensionless parameters $(m,n)$ and frequency $\omega$ that has a dimension of inverse length. Since it is impossible to build $F_\pm(\theta)$, which has dimension of length, from these objects, we conclude that $F_\pm(\theta)=0$ if $a=b=0$, even when the assumption (\ref{Special5D}) is relaxed. The same dimensional analysis implies that $\la=0$, so the electric solution for $a=b=0$ must have the form
\bea\label{Kerr5DElectrNonRot}
l_\pm^\mu A_\mu=G_\pm(r) {\hat l}_\pm \Psi,\quad
m_\pm^\mu A_\mu=0,\quad n^\mu A_\mu=0,
\quad \Psi=e^{i\omega t+im\phi+i n\psi}R(r)S(\theta).
\eea
More explicitly, we find
\bea
A=\frac{e^{i\omega t+im\phi+i n\psi}}{2}\left[(G_++G_-)({\dot R}dr+i\omega R dt)+
(G_+-G_-)({\dot R}dt+i\omega R dr)\right]S.
\eea
Defining the components of Maxwell's equations by (\ref{MaxwNotation5d}), we find
\bea\label{5dA0SecOrder}
&&\hskip -1cm (m_+^\mu+m_-^\mu) \MM_\mu=-\frac{s_{2\theta}}{2}S'\mathscr{N},\quad
(m_+^\mu-m_-^\mu) \MM_\mu=\frac{amc^2_\theta-bns^2_\theta}{\Theta}S\mathscr{N},\nn
&&\hskip -1cm \mathscr{N}=
r^2\frac{d}{dr}\left[r(G_++G_-)\dot R\right]+\omega^2r^3 (G_++G_-)R+i\omega r^2\frac{d}{dr}[r({\dot G}_+-{\dot G}_-)]R.
\eea
Note that even though $a$ and $b$ appear in these expressions, the equations work only in the limit where $a$ goes to zero while $a/b$ is kept fixed. We also recall that 
\bea
\Theta=\sqrt{(a c_\theta)^2+(b s_\theta)^2}\,.
\eea
Solving equation $\mathscr{N}=0$ and substituting the result into the remaining Maxwell's equations, we can eliminate second and third derivatives of $R$. Then $R$ and ${\dot R}$ should be treated as independent variables, and we look at a particular combination of Maxwell's equations:
\bea
l_+^\mu\MM_\mu\Big|_{R=0}=r^4 G_+{\dot R}\left[\frac{1}{2}\frac{d}{d\theta}[s_{2\theta}S']-m^2 \cot_\theta^2 S-n^2 \tan_\theta^2 S\right]-f(r) s_{2\theta}S{\dot R}=0,
\eea
where $f(r)$ is some complicated combination of functions $G_\pm(r)$ and their derivatives. 
Consistency of the last relation leads to an ODE for function $S$:
\bea\label{5dA0ElectSeqn}
\frac{1}{s_{2\theta}}\frac{d}{d\theta}[s_{2\theta} S']+\left[\la_1-\frac{m^2}{s_\theta^2}-
\frac{n^2}{c_\theta^2}\right] S=0\,.
\eea
Substitution of $S''$ into the remaining Maxwell's equations also eliminates $S'$, leading to a system of linear {\it algebraic} equations for $(R,{\dot R})$ with coefficients involving $(G_\pm,{\dot G}_\pm,{\ddot G}_\pm)$. One projection looks especially simple\footnote{The case $G_+=-G_-$ should be considered separately, and it does not lead to nontrivial solutions unless $\omega=0$.}:
\bea
(l_+^\mu+l_-^\mu)\MM_\mu&=&\frac{i({\dot R}+i\omega R)r^4}{2(G_++G_-)}s_{2\theta}S\left[
2r^2\omega {\dot G}_+G_-+(r\omega+i\la_1)G_+G_--(r\omega-i\la_1)G_+^2
\right]\nn
&-&\frac{i({\dot R}-i\omega R)r^4}{2(G_++G_-)}s_{2\theta}S\left[
2r^2\omega {\dot G}_-G_++(r\omega-i\la_1)G_+G_--(r\omega+i\la_1)G_-^2
\right].\nonumber
\eea
The square brackets in both lines of this expression must vanish separately leading to two differential equations for functions $(G_+,G_-)$. The first line gives an algebraic equation for $G_-$ in terms of 
$(G_+,{\dot G}_+)$, and substitution of the result in the second line leads to a linear ODE for $g=G_+^{-1}$:
\bea
r^2(\la_1+i\omega r)\ddot g+\la_1 r {\dot g}-\la_1 g=0.
\eea
The most general solution of this equation is
\bea
G_\pm=\pm \frac{C_1 r}{\la_1\pm 2i\omega r+C_2 r^2}\,.\nonumber
\eea
Rescaling of the gauge field leads to simpler expressions:
\bea\label{5DKerrGpm}
G_\pm=\pm \frac{r}{1\pm i\mu r+\nu r^2}\,.
\eea
Although we focused on $M=0$ to simplify the intermediate expressions, the same derivation applies to the the five--dimensional Schwarzschild black hole, where the ansatz (\ref{Kerr5DElectrNonRot}) leads to (\ref{5DKerrGpm}) and to differential equations
\bea\label{Kerr5DeqnSchw}
&&\frac{1}{s_{2\theta}}\frac{d}{d\theta}[s_{2\theta} S']+\left[\frac{2\omega}{\mu}-\frac{m^2}{s_\theta^2}-
\frac{n^2}{c_\theta^2}\right] S=0,\nn
&&\frac{E_r}{r}\frac{d}{dr}\left[\frac{\Delta}{rE_r}{\dot R}\right]+\left[-\frac{2\omega}{\mu}+(\omega r)^2+\frac{2\mu\omega r^2(1-\nu r^2)}{E_r}+\frac{M\omega^2 r^4}{\Delta}\right]R=0.
\eea
Here we defined
\bea\label{DefErSpecial5d}
E_r\equiv (1+\nu r^2)^2+(\mu r)^2,\quad \Delta=r^4-M r^2.
\eea
Note that equations (\ref{Kerr5DeqnSchw}) contain one separation constant, $\omega/\mu$, while $\nu$ is a free parameter that does not affect the spectrum. 
\bigskip

Let us now discuss Maxwell's equations in the metric (\ref{KerrFlat5D}) containing rotation 
parameters $a$ and $b$. We begin with rewriting the metric in terms of a new coordinate 
$\Theta=\sqrt{(ac_\theta)^2+(bs_\theta)^2}$:
\bea
ds^2&=&-dt^2+\Sigma\left[\frac{r^2dr^2}{\Delta_0}-\frac{\Theta^2 d\Theta^2}{\Xi}\right]+
\frac{(r^2+a^2)(\Theta^2-a^2)}{b^2-a^2} d\phi^2+\frac{(r^2+b^2)(\Theta^2-b^2)}{a^2-b^2} d\psi^2\nn
\Sigma&=&r^2+\Theta^2,\quad \Delta_0=(r^2+a^2)(r^2+b^2),\quad
\Xi=(\Theta^2-a^2)(\Theta^2-b^2)\,.\nonumber
\eea
It is clear that this geometry has a symmetry
\bea\label{Z2Kerr5}
r\rightarrow -i\Theta,\quad \Theta\rightarrow i r,
\eea
so in the presence of rotation parameters the factors (\ref{5DKerrGpm}) and their counterparts $F_\pm$ are\footnote{The symmetry determines $F_\pm$ up to a sign, which is fixed by the electrostatic solution (\ref{Special5DElectr}).}
\bea
G_\pm=\pm \frac{r}{1\pm i\mu r+\nu r^2}\,,\qquad
F_\pm=\mp \frac{i\Theta}{1\pm \mu \Theta-\nu \Theta^2}\,.\nonumber
\eea
To determine the value of the parameter $\nu$, we look at equation $n_\mu\MM^\mu=0$. It contains only $(S,S'S'',\Phi,{\dot\Phi},{\ddot\Phi})$, and separation of variables leads to second order ODEs for 
$S$ and $\Phi$. Substitution of the results into the remaining Maxwell's equations leads to the relation 
$\nu=0$. This argument breaks down for $a=b=0$, when some of the equations disappear, then the parameter $\nu$ remains undetermined, as in the non--rotating case. 

To summarize, in the presence of rotation, the factors $(G_\pm,F_\pm)$ must be given by
\bea
G_\pm=\pm \frac{r}{1\pm i\mu r}\,,\qquad
F_\pm=\mp \frac{i\Theta}{1\pm \mu \Theta}\,,
\eea
and our next task is to find the differential equations for $(S,\Phi)$. We begin with looking at equation $n_\mu\MM^\mu=0$. Explicit calculations give
\bea\label{tempAA85}
\hskip -0.5cm
n_\mu\MM^\mu=-\frac{\mu r^2(ab\omega+an+bm)-2 ab}{2}\frac{d}{dr}\left[\frac{\Delta_0 {\dot\Phi}}{r E_r}\right]s_{2\theta}S+\Phi\FF (r,\theta,S,S',S'')=0.
\eea
To simplify this and subsequent expressions, we define ``electric factors''
\bea
E_r\equiv 1+(\mu r)^2,\quad E_\theta\equiv 1-(\mu\Theta)^2\,.
\eea
Function $\FF$ entering (\ref{tempAA85}) is rather complicated, but even without seeing its explicit form, we conclude the consistency of this equation leads to an ODE for function $\Phi$:
\bea
\frac{d}{dr}\left[\frac{\Delta_0 {\dot\Phi}}{r E_r}\right]+g(r)\Phi=0
\eea
with some function $g(r)$. A different arrangement of terms in $n_\mu\MM^\mu=0$ leads to an alternative form of this equation:
\bea\label{tempAA85a}
n_\mu\MM^\mu=(ab+\mu \alpha s_\theta^2+\mu \beta c_\theta^2)r\Phi\frac{d}{d\theta}\left[\frac{s_{2\theta}}{E_\theta}{S'}\right]+
S{\tilde\FF} (r,\theta,R,{\dot R},{\ddot R})=0\,,\nonumber
\eea
where $\alpha$ and $\beta$ are complicated combinations of $(a,b,\omega,m,n)$. Consistency of the last equation leads to an ODE for $S(\theta)$:
\bea
\frac{d}{d\theta}\left[\frac{s_{2\theta}}{E_\theta}{S'}\right]+h(\theta)S=0.
\eea
Maxwell's equations give an over--constrained system for two unknown functions, $(g(r),h(\theta))$, and existence of solution is a highly nontrivial confirmation of our ansatz. Straightforward but tedious manipulations lead to the {\it unique} final answer for the ``electric'' polarization in the flat geometry (\ref{KerrFlat5D}):
\bea\label{FullElectr5dM0}
&&l_\pm^\mu A_\mu=\pm\frac{r}{1\pm i\mu r}{\hat l}_\pm\Psi ,\quad
m_\pm^\mu A_\mu=\mp\frac{i\Theta}{1\pm \mu\Theta}{\hat m}_\pm\Psi,\quad 
n^\mu A_\mu=0,\nn
&&\frac{E_\theta}{s_{2\theta}}\frac{d}{d\theta}\Big[\frac{s_{2\theta}}{E_\theta}S'\Big]+
\Big[\frac{2\Lambda}{E_\theta}+\omega^2\Theta^2-\frac{n^2}{c^2_\theta}-
\frac{m^2}{s_\theta^2}+C\Big]S=0,\\
&&\frac{E_r}{r}\frac{d}{dr}\Big[\frac{\Delta}{r E_r}{\dot\Phi}\Big]+
\Big[-\frac{2\Lambda}{E_r}+(\omega r)^2+\frac{m^2(a^2-b^2)}{r^2+a^2}+\frac{n^2(b^2-a^2)}{r^2+b^2}
-C\Big]\Phi=0.\nonumber
\eea
The expressions for $\Lambda$ and $C$ are given by (\ref{Electr5dLmbd}).

To extend this result to the black hole geometry, we observe that in the special cases (\ref{Special5DElectr}) and (\ref{Kerr5DeqnSchw}) the mass $M$ appears only in the differential equation for $R$, while all other relations remain the same as for $M=0$. Direct calculation shows that this feature persists in the general case, and the final answer for the electric polarization of the electromagnetic field is given by (\ref{FullElectr5d}).

\subsection{Magnetic polarization}
\label{SecApp5DMagn}

Let us now discuss the magnetic polarization. 
As in the electric case, we begin with analyzing the waves in the flat geometry (\ref{KerrFlat5D}), and our starting point is the application of the ansatz (\ref{KerrNewAnstz}) to the metric (\ref{KerrFlat5D}) with $a=b=0$. As in the electric case, expression (\ref{SpecialKerrMagn}) and dimensional analysis imply that the non--rotating geometry would give 
$G_\pm=0$ even for arbitrary $(\omega,m,n)$, thus in the present situation, the relation (\ref{Kerr5DElectrNonRot}) is replaced by 
\bea\label{Kerr5DMagnNonRot}
l_\pm^\mu A_\mu=0,\quad
m_\pm^\mu A_\mu=F_\pm(\theta) {\hat m}_\pm \Psi,\quad n^\mu A_\mu=\la \Psi,
\quad \Psi=e^{i\omega t+im\phi+i n\psi}\Phi(r)S(\theta).
\eea
Note that as $a$ and $b$ go to zero, parameter $\la$ should scale like $a$. The gauge potential becomes
\bea
A&=&\frac{e^{i\omega t+im\phi+i n\psi}}{\Theta^2}\left[\la(bs_\theta^2 d\phi+a c_\theta^2 d\psi)+
{\tilde F}_+\Big\{\Theta^2 S' d\theta+iS(amc_\theta^2-bn s_\theta^2)(a d\phi-b d\psi)\Big\}\right.\nn
&&\left.-
\Theta{\tilde F}_-\Big\{ (amc_\theta^2-bn s_\theta^2)\frac{S}{s_\theta c_\theta} d\theta+is_\theta c_\theta S'(a d\phi-b d\psi)\Big\}\right]R,
\eea
where
\bea
{\tilde F}_\pm=\frac{F_+\pm F_-}{2}\,.\nonumber
\eea
One component of Maxwell's equations looks especially simple:
\bea
n^\mu\MM_\mu=\la r^2 s_\theta c_\theta S[\frac{d}{dr}(r{\dot \Phi})+r\omega^2\Phi]+r\Phi\FF(S,S',S'',F_+,F_-,\theta)=0.
\eea
As discussed in section \ref{SectSub5dStat}, parameter $\la$ is not equal to zero even for vanishing 
$(\omega,m,n)$, this leads to an ordinary differential equation for $\Phi(r)$:
\bea
r\frac{d}{dr}(r{\dot \Phi})+\left[(r\omega)^2+\la_1\right]\Phi=0.
\eea
Here $\la_1$ is a separation constant. Solving this equation for $\ddot\Phi$ and substituting the result on Maxwell's equations, we find an over--constrained system of {\it algebraic} relations between $\Phi$ and $\dot\Phi$. Requiring the coefficients to vanish, we arrive at a system of ODEs for $S$ and $F_\pm$. Manipulations with this system lead to a counterpart of (\ref{5DKerrGpm}):
\bea\label{5DKerrGpmMagn}
F_\pm=\pm \frac{i}{\Theta\mp\mu}\,.
\eea
However, $\mu$ is no longer a free parameter, but rather it is determined in terms of other ingredients:
\bea
\mu=\frac{an+bm}{\la}\,.
\eea
Maxwell's equations also lead to an ODE for function $S$, and we conclude the discussion of $a=b=0$ case by quoting the full solution for the five--dimensional Schwarzschild geometry: 
\bea\label{FullMagn5dSchw}
&&l^\mu_\pm A_\mu=0,\quad
m^\mu_\pm A_\mu=\pm\frac{i}{\Theta\mp \mu}{\hat m}_\pm \Psi,\quad 
n^\mu A_\mu=\frac{an+bm}{\mu} \Psi,\nn
&&\frac{M_\theta}{s_{2\theta}}\frac{d}{d\theta}\left[\frac{s_{2\theta}}{M_\theta}S'\right]+
\left[\left(\frac{an+bm}{\mu}\right)^2+\frac{\alpha}{M_\theta}-\frac{m^2}{s_\theta^2}-\frac{n^2}{c_\theta^2}\right]S=0,\\
&&r\frac{d}{dr}\Big[\frac{\Delta}{r^3}\dot\Phi\Big]+
\Big[-\left(\frac{an+bm}{\mu}\right)^2+
(\omega r)^2+\frac{Mr^4 \omega^2}{\Delta}\Big]\Phi=0\nonumber
\eea
Here we defined
\bea\label{Magn5dMfuncApp}
M_\theta=\Theta^2-\mu^2,\quad \alpha=ab\frac{an+bm}{\mu}-\mu[am+bn].
\eea

Addition of the rotation parameters follows the pattern familiar from sections \ref{SecAppKerrElectr}, \ref{SecAppKerrMagn}, \ref{SecApp5DElectr}. First we use the symmetry (\ref{Z2Kerr5}) to determine functions $(F_\pm,G_\pm)$:
\bea
F_\pm=\pm \frac{i}{\Theta\mp\mu}\,,\quad 
G_\pm=\pm \frac{1}{r\pm i\mu}\,.\nonumber
\eea
Then rearranging terms in the Maxwell's equation $n_\mu\MM^\mu=0$ as in (\ref{tempAA85}) and (\ref{tempAA85a}), we arrive at ODEs for $\Phi$ and $S$:
\bea
\frac{M_r}{r}\frac{d}{d\theta}\left[\frac{\Delta_0\dot\Phi}{rM_r}S'\right]+g(r)\Phi=0,\qquad 
\frac{M_\theta}{s_{2\theta}}\frac{d}{d\theta}\left[\frac{s_{2\theta}}{M_\theta}S'\right]+h(\theta)S=0.
\eea
Here
\bea
M_\theta=\Theta^2-\mu^2,\quad M_r=-(r^2+\mu^2),
\eea
and $(g,h)$ are undetermined functions. Substitution into the remaining Maxwell's equations produces  an over--constrained system of differential equations for these functions, and eventually leads to the final answer for system describing the magnetic polarization in the flat geometry (\ref{KerrFlat5D}):
\bea\label{FullMagn5dApp}
&&l^\mu_\pm A_\mu=\pm \frac{1}{r\pm i \mu}{\hat l}_\pm \Psi,\quad
m^\mu_\pm A_\mu=\pm\frac{i}{\Theta\mp \mu}{\hat m}_\pm \Psi,\quad 
n^\mu A_\mu=\la \Psi\nn
&&\frac{M_\theta}{s_{2\theta}}\frac{d}{d\theta}\left[\frac{s_{2\theta}}{M_\theta}S'\right]+
\left[\frac{2\Lambda}{M_\theta}+\omega^2\Theta^2-\frac{m^2}{s_\theta^2}-\frac{n^2}{c_\theta^2}+
C\right]S=0,\\
&&\frac{M_r}{r}\frac{d}{dr}\Big[\frac{\Delta}{r M_r}\dot\Phi\Big]+
\Big[-\frac{2\Lambda}{M_r}-C+\frac{m^2(a^2-b^2)}{r^2+a^2}+\frac{n^2(b^2-a^2)}{r^2+b^2}+
(\omega r)^2\Big]\Phi=0\nonumber
\eea
Here constants $(\Lambda,C,\mu)$ are given by (\ref{Magn5dLmbd}) and (\ref{Magn5dLmbd1}). As we have seen before, extension from the flat space (\ref{KerrFlat5D}) to the black hole geometry is straightforward: it is accomplished by adding an extra term in the radial equation, and the final answer is given by (\ref{FullMagn5d}).

\section[Electromagnetic waves in the Schwarzschild--Tangherlini geometry]{
Electromagnetic waves in the\\ Schwarzschild--Tangherlini geometry}
\renewcommand{\theequation}{D.\arabic{equation}}
\setcounter{equation}{0}
\label{SecAppSchw}
Although the main goal of this article is the study of equations describing electromagnetic waves in the background of rotating black holes, to get some intuition, it is useful to review the separation of variables in the geometries produced by static black holes in arbitrary dimensions. The details of such construction are discussed in this appendix, and the main outcome is the explicit form of $D-2$ polarizations in $D$ dimensions. 

\bigskip

In this appendix we consider a static Schwarzschild--Tangherlini black hole in $D=d+2$ dimensions \cite{Tang}:
\bea
ds^2=-Hdt^2+\frac{dr^2}{H}+r^2 d\Omega_d^2,\qquad H=1-\frac{M}{r^{d-1}}.
\eea
Since the radial coordinate plays a very special role, it is convenient to impose the gauge $A_r=0$.
We are looking for separable solutions forming a complete basis, and as a first step we separate $r$ and $t$ from the coordinates on the sphere $S^d$:  
\bea\label{SchwTngAnstz}
A_t=e^{i\omega t}f(r)Y,\quad A_r=0,\quad A_i=e^{i\omega t}g(r)Y_i\,.
\eea
Here $Y$ and $Y_i$ are functions of the coordinates $x_k$ on the sphere, and latin indices are going over $d$ values. Without committing to specific coordinates we write the metric on the sphere as
\bea
d\Omega_d^2=h_{ij}dx^idx^j.
\eea
Defining various components of Maxwell's equations by
\bea
\MM^\mu=\frac{e^{-i\omega t}}{\sqrt{-G}}\d_\nu[\sqrt{-G}F^{\mu \nu}],
\eea
we find\footnote{Recall that the square root of the determinant of the $D$--dimensional metric is 
$\sqrt{-G}=r^d \sqrt{h}$. We used $G$ to avoid confusion with the profile function $g$ in (\ref{SchwTngAnstz}).}
\bea\label{TangMaxw}
\MM^t&=&\frac{1}{r^d}\d_r[r^d Y\d_r f]+
\frac{f}{r^2 H\sqrt{h}}\d_i[\sqrt{h}h^{ij}\d_j Y]-\frac{i\omega g}{r^2 H\sqrt{h}}\d_i[\sqrt{h}h^{ij}Y_j],
\nn
\MM^r&=&
\frac{g'H}{r^2\sqrt{h}}\d_m[\sqrt{h}h^{mj}Y_j]-i\omega f' Y,\\
-\MM^i&=&\frac{1}{r^d}\d_r[r^{d-2}Hh^{ij} Y_j\d_r g]+
\frac{g}{r^4\sqrt{h}}\d_m[\sqrt{h}h^{mj}h^{ik}{\cal Y}_{jk}]+\frac{i\omega}{Hr^2}h^{ij}[f\d_j Y-i\omega g Y_j].\nonumber
\eea
Here
\bea
{\cal Y}_{jk}=\d_j Y_k-\d_k Y_j
\eea
is the field strength associated with potential $Y_i$.

Separation of variables in equations $\MM^\mu=0$ implies several relations between functions on the sphere:
\bea\label{la2eqn}
&&\frac{1}{\sqrt{h}}\d_i[\sqrt{h}h^{ij}\d_j Y]=-\la_1 Y,\quad
\frac{1}{\sqrt{h}}\d_i[\sqrt{h}h^{ij}Y_j]=-\la_2 Y,
\\
&&\frac{1}{\sqrt{h}}\d_m[\sqrt{h}h^{mj}h^{ik}{\cal Y}_{jk}]=-\la_3h^{ij}Y_j
-\la_4h^{ij}\d_jY.\nonumber
\eea
Here $(\la_1,\la_2,\la_3,\la_4)$ are separation constants, which also enter radial equations. Such equations will be discussed below. It is convenient to decompose the vector field $Y_i$ into a scalar $Z$ and a ``transverse mode'' ${\tilde Y}_i$ as
\bea\label{la2eqn1}
Y_j=\d_j Z+{\tilde Y}_j,\quad \frac{1}{\sqrt{h}}\d_i[\sqrt{h}h^{ij}{\tilde Y}_j]=0.
\eea
Such decomposition always exists, but it is not unique: $Z$ can be shifted by a harmonic function on the sphere. Substituting the expression for $Y_i$ into the second equation in (\ref{la2eqn}) and comparing the result with the first equation, we conclude that function $\la_1 Z-\la_2 Y$ must be harmonic, so without loss of generality we can set\footnote{This argument breaks down only for $\la_1=0$, which does not lead to interesting solutions. Indeed, if $\la_1=0$, then function $g$ can be eliminated from equations $\MM^t=0$, $\MM^r=0$ to yield a second order ODE for $f'$ which does not contain separation parameters. Then function $g$ is determined in terms of $f$.} $\la_1 Z-\la_2 Y=0$. Then equations (\ref{la2eqn}) and (\ref{la2eqn1}) become
\bea\label{la2eqn2}
&&\frac{1}{\sqrt{h}}\d_i[\sqrt{h}h^{ij}\d_j Y]=-\la_1 Y,\quad 
Y_j=\frac{\la_2}{\la_1}\d_j Y+{\tilde Y}_j,\quad 
\frac{1}{\sqrt{h}}\d_i[\sqrt{h}h^{ij}{\tilde Y}_j]=0,
\nn
&&\frac{1}{\sqrt{h}}\d_m[\sqrt{h}{\tilde{\cal Y}}^{mi}]=-\la_3h^{ij}{\tilde Y}_j
-(\la_4+\frac{\la_2\la_3}{\la_1})h^{ij}\d_jY.
\eea
The indices are raised and lowered using the metric $h_{ij}$. Definition (\ref{la2eqn1}) implies that 
${\tilde{\cal Y}}_{ij}={\cal Y}_{ij}$. Substituting (\ref{la2eqn2}) into Maxwell's equations 
$\MM^\mu=0$ with $\MM^\mu$ given by (\ref{TangMaxw}), we find several relations between functions of $r$:
\bea\label{TangMaxwRad}
&&\left[\frac{1}{r^d}\d_r[r^d\d_r f]-
\frac{\la_1 f}{r^2 H}+\frac{i\omega\la_2 g}{r^2 H}\right]Y=0,\qquad
\left[\frac{\la_2g'H}{r^2}+i\omega f'\right]Y =0,\\
&&({\tilde Y}_j+\frac{\la_2}{\la_1}\d_j Y)\left[\frac{1}{r^d}\d_r[r^{d-2}H\d_r g]-
\frac{\la_3 g}{r^4}+\frac{\omega^2 g}{Hr^2}\right]
+\d_j Y\left[-
\frac{\la_4g}{r^4}+\frac{i\omega f^2}{Hr^2}\right]=0.\nonumber
\eea
We kept functions $(Y,{\tilde Y}_i)$ to stress that some radial equations disappear in special cases, for example,
the first two equations disappear if $Y= 0$. Let us consider several options depending on the values of $Y$ and 
${\tilde{\cal Y}}_{ij}$.

\bigskip

If we assume that both ${\tilde{\cal Y}}_{ij}$ and $Y$ are nontrivial, then coefficients in front of ${\tilde Y}_j$ and $\d_j Y$ must vanish independently, in particular, this leads to two relations
\bea
\frac{\la_2g'H}{r^2}+i\omega f'=0,\qquad -\frac{\la_4 g H}{r^2}+{i\omega f}=0.
\eea
Integrability condition\footnote{We are interested only in solutions where the profiles of $f$ and/or $g$ depend on separation parameters. We also assume that $\omega\ne 0$ during the derivation, but the resulting system (\ref{TangMagn}) works for $\omega=0$ as well.} implies that $\la_4 g=0$, then $f=0$.
To have a nontrivial solution, we must require $\la_2=\la_4=0$, arriving at the system describing the ``magnetic'' polarizations:
\bea\label{TangMagn}
&&A^{(mgn)}_t=0,\quad A^{(mgn)}_r=0,\quad A^{(mgn)}_i=e^{i\omega t}g(r)Y_i\nn
&&\frac{1}{\sqrt{h}}\d_i[\sqrt{h}h^{ij}Y_j]=0,\quad
\frac{1}{\sqrt{h}}\d_m[\sqrt{h}{\cal Y}^{mi}]=-\la_3h^{ij}Y_j,\\
&&\frac{1}{r^d}\d_r[r^{d-2}H\d_r g]-
\frac{\la_3g}{r^4}
+\frac{\omega^2 g}{Hr^2}=0.\nonumber
\eea
Note that the scalar function $Y$ does not appear in this system, so one arrives at the same answer by setting $Y=0$ in the beginning. 
If both $Y$ and ${\tilde{\cal Y}}_{ij}$ are trivial, then either $Y_j=0$ or $\la_3=0$ in the last system, and the resulting configurations don't contain separation constants.

The only remaining option is ${\tilde{\cal Y}}_{ij}=0$, then equations (\ref{la2eqn}) imply that
\bea
Y_j=\mu \d_j Y 
\eea
for some constant $\mu$. The angular equations (\ref{la2eqn}) become
\bea
\la_3\mu+\la_4=0,\quad \la_1\mu-\la_2=0,\quad \frac{1}{\sqrt{h}}\d_i[\sqrt{h}h^{ij}\d_j Y]=-\la_1 Y.
\eea
The Maxwell's equations $\MM^\mu=0$ with $\MM$ given by (\ref{TangMaxw}) reduce to a system of ODEs
\bea
&&\frac{1}{r^d}\d_r[r^d \d_r f]-
\frac{\la_1 f}{r^2 H}+\frac{i\la_1\mu\omega g}{r^2 H}=0,
\nn
&&\frac{\la_1\mu g'H}{r^2}+i\omega f'=0,\nn
&&\frac{\mu}{r^d}\d_r[r^{d-2}H\d_r g]
+\frac{\mu\omega^2 g}{Hr^2}
+\frac{i\omega f}{Hr^2}=0,\nonumber
\eea
and the last equation follows from the first two. Furthermore, a rescaling $g\rightarrow \mu^{-1}g$ removes parameter 
$\mu$ from the differential equations and from the expression (\ref{SchwTngAnstz}) for the gauge field. Since neither 
$\mu=0$ nor $\mu=\infty$ lead to nontrivial solutions, we can set $\mu=1$ without loss of generality. We also observe that for $\omega=0$, function $g$ describes a pure gauge, so it is convenient to introduce another rescaling $g\rightarrow i\omega g$. This leads to the final form of the system describing the ``electric'' polarization:
\bea\label{TangElectr}
&&A^{(el)}_t=e^{i\omega t}f(r)Y,\quad A^{(el)}_r=0,\quad A^{(el)}_i=i\omega e^{i\omega t} g(r)\d_i Y,\nn
&&\frac{1}{\sqrt{h}}\d_i[\sqrt{h}h^{ij}\d_j Y]=-\la_1 Y,\\
&& g'=-\frac{r^2}{\la_1 H}f',\quad
\frac{1}{r^d}\d_r[r^d \d_r f]-
\frac{\la_1 f}{r^2 H}-\frac{\la_1\omega^2 g}{r^2 H}=0.\nonumber
\eea
The system of coupled ODEs for functions $(f,g)$ is governed by a single second order differential equation for $F\equiv f'$:
\bea
\d_r\left[\frac{H}{r^{d-2}}\d_r[r^d F]\right]-\la_1 F+
\frac{\omega^2 r^2}{H}F=0,\quad f'=F,\quad g'=-\frac{r^2}{\la_1 H}F.
\eea
We used the labels ``magnetic'' and ``electric'' to distinguish between polarizations (\ref{TangMagn}) and (\ref{TangElectr}), as well as between their counterparts discussed in the rest of the paper. The names originate from the fact that in the static ($\omega=0$) limit, the systems  (\ref{TangMagn}) and (\ref{TangElectr}) describe pure magnetic and pure electric fields. Of course, the wave with $\omega\ne 0$ has both.

\bigskip

Let us briefly discuss the structure of functions $(Y,Y_i)$. Although scalar and vector spherical harmonics are well--known, to count polarizations and to connect with discussion of the rotating black holes, it is convenient to use an explicit representation of $(Y,Y_i)$ in terms of symmetric polynomials. We begin with observing that the spherical harmonics appearing in the systems 
(\ref{TangMagn}) and (\ref{TangElectr}), as well as the separation constants $(\la_1,\la_3)$, cannot depend on the values of $\omega$ and $M$, so to construct these functions one can look at static problems in flat space (i.e., set $\omega=M=0$). As usual, the spherical harmonics can be constructed using either the fields decaying at infinity or configurations regular at the origin, and we choose the second option. Introducing Cartesian coordinates $y_a=r f_a(x)$ in flat space, we can expand  the gauge fields (\ref{TangMagn}) and (\ref{TangElectr}) in the Taylor series near the origin:
\bea\label{PolynSchwTemp1}
A^{(el)}=dt\sum_{p=0}^\infty A^{(p)}_{a_1\dots a_p}y_{a_1}\dots y_{a_p},\quad
A^{(mgn)}=\sum_{p=0}^\infty  dy^a B^{(p)}_{a;a_1\dots a_p}y_{a_1}\dots y_{a_p}\,.
\eea
To avoid the $A^{(mgn)}_r$ component, coefficients $B^{(p)}_{aa_1\dots a_p}$ must satisfy a constraint
\bea
B^{(p)}_{a;a_1\dots a_p}y_ay_{a_1}\dots y_{a_p}=0.
\eea
Each $A^{(p)}$ and $B^{(p)}$ leads to separation between the sphere and $r$ coordinate, and parameters 
$(\la_1,\la_3)$ can be found by substituting the appropriate $f$ and $g$ in the radial equation:
\bea
&&Y^{(p)}=\frac{1}{r^p} A^{(p)}_{a_1\dots a_p}y_{a_1}\dots y_{a_p},\quad
f=r^p,\quad \la_1=p(p+d-1),
\nn
&&Y^{(p)}_i= \frac{1}{r^{p}}  B^{(p)}_{i;a_1\dots a_p}y_{a_1}\dots y_{a_p},\quad g=r^{p+1},\quad
\la_3=(p+1)(p+d-2).
\eea

Maxwell's equations for potentials (\ref{PolynSchwTemp1}) in {\it flat} space impose certain restrictions on coefficients $A^{(p)}$ and $B^{(p)}$, for example, the full solution in the electric case is
\bea\label{PolynSchwTempElec}
\la_1=p(p+d-1),\ Y^{(p)}=\frac{1}{r^p} A^{(p)}_{a_1\dots a_p}y_{a_1}\dots y_{a_p}, \
\delta^{ab}A^{(p)}_{ab a_3\dots a_p}=0
\eea
This system describes a single electric polarization. The magnetic case is governed by the system
\bea\label{PolynSchwTempMag}
\hskip -0.7cm
B^{(p)}_b=B^{(p)}_{b;a_1\dots a_p}y_{a_1}\dots y_{a_p},\quad Y^{(p)}_i= \frac{B^{(p)}_i}{r^{p}},
\quad  
B^{(p)}_a y_a=0,\ \d_a B^{(p)}_a=0,\ \d_a\d_a B^{(p)}_b=0.
\eea
The corresponding eigenvalue is $\la_3=(p+1)(p+d-2)$. Although index $b$ in $B^{(p)}_b$ takes $D-1=d+1$ values, the constraints $B^{(p)}_a y_a=0$ and $\d_a B^{(p)}_a=0$ ensure that there are only $D-3$ independent magnetic polarizations, as expected for the electromagnetic field. 

\bigskip

To summarize, in this appendix we have demonstrated that separable solutions of Maxwell's equations in the Schwarzschild--Tangherlini geometry must reduce to one of the two systems, (\ref{TangMagn}) or (\ref{TangElectr}). The second option describes one electric polarization, and the corresponding spherical harmonics can be constructed in terms of symmetric polynomials as in (\ref{PolynSchwTempElec}). The ``magnetic'' option  (\ref{TangMagn}) describes $d-1$ polarizations with spherical harmonics given by (\ref{PolynSchwTempMag}). In section \ref{SecSubCmprSchw} these results are compared with separation of variables in Maxwell's equations in the background of rotating black holes.

\section{Separation of variables in the Myers--Perry geometry}
\label{SecAppMP}
\renewcommand{\theequation}{E.\arabic{equation}}
\setcounter{equation}{0}

The main result of this article is separation of variables in the Maxwell's equations in the background of the Myers--Perry black hole, and in this appendix we will present some details of the calculations leading to the final answers (\ref{EvenDimGenSpin})--(\ref{EvenDimGenSpinLmb}) and (\ref{GenOddElectr})--(\ref{OddDimGenSpinLmb}). Since the ``master equations'' (\ref{EvenDimGenSpin}) 
and (\ref{OddMPmaster}) cover massless scalars along with electromagnetism, it is instructive to look at the wave equation first. We will do this in section \ref{SecAppMPWave} before focusing on electromagnetic field in sections \ref{SecAppMPeven} and \ref{SecAppMPodd}.

\subsection{Wave equation}
\label{SecAppMPWave}

As discussed in section \ref{SecMyersPerry}, the frames associated with the Myers--Perry black holes have very different structures in even and odd dimensions. Thus it is convenient to discuss these two cases separately. 

\bigskip

In even dimensions the Myers--Perry black hole is described by the metric (\ref{MPeven}), and the corresponding frames are given by (\ref{AllFramesMP}). Then, as discussed in section \ref{SecMPwave}, the wave equation can be written as (\ref{Jul19temp}):
\bea\label{Jul19tempApp}
\sqrt{\prod {d_i}}\,\d_\mu[{\tilde g}_r^{\mu\nu}\d_\nu\Psi]+
\sum_i \frac{FR \sqrt{\prod {d_k}}}{d_i(r^2+x^2_i)}\d_\mu[{\tilde g}_i^{\mu\nu}\d_\nu\Psi]=0,
\eea
where the individual pieces of the inverse metric are defined by (\ref{gUpEven}):
\bea
{\tilde g}_r^{\mu\nu}\d_\mu\d_\nu&=&
[R-Mr]\d_r^2-\frac{R^2}{R-Mr}\left[\d_t-\sum_k\frac{a_k}{a_k^2+r^2}\d_{\phi_k}\right]^2\,,\nn
{\tilde g}_i^{\mu\nu}\d_\mu\d_\nu&=&H_i\d_{x_i}\d_{x_i}+H_i\left[\d_t-
\sum_k\frac{a_k}{a_k^2-x_i^2}\d_{\phi_k}
\right]^2\,.
\eea
A separable ansatz for function $\Psi$,
\bea\label{SeparFuncAppWave}
\Psi=E\Phi(r)\left[\prod X_i(x_i)\right],\qquad E=e^{i\omega t+i\sum m_i\phi_i}\,,
\eea
leads to ordinary differential equations for functions $(X_i,\Phi)$, and the goal of this appendix is to find such ODEs and to identify the separation constants. 

We begin with observing that $\sqrt{\prod {d_i}}$ is a polynomial of degree $n-1$ in all $(x_k)^2$, and 
\bea
\frac{FR \sqrt{\prod {d_k}}}{d_i(r^2+x^2_i)}\nonumber
\eea
is a polynomial of degree $n-1$ in $r^2$ and in all $(x_k)^2$ with the exception of $k=i$. Then consistency of equation (\ref{Jul19tempApp}) implies a system of ODEs
\bea\label{SeparWaveRadODE}
\d_\mu[{\tilde g}_r^{\mu\nu}\d_\nu (E\Phi)]=P^{(0)}_{n-1}[r^2]E\Phi,\qquad
\d_\mu[{\tilde g}_j^{\mu\nu}\d_\nu (E X_j)]=P^{(j)}_{n-1}[-x_j^2]E\Phi,
\eea
where $P^{(k)}_{n-1}$ are arbitrary polynomials of degree $n-1$. We will now demonstrate that all $n$ functions $P^{(k)}_{n-1}[y]$ must be the same. 

Using relations (\ref{SeparWaveRadODE}) to remove derivatives from the wave equation (\ref{Jul19tempApp}), we find a very stringent constraint on the polynomials:
\bea\label{tempE5}
\frac{P^{(0)}_{n-1}[r^2]}{\prod (r^2+x_k^2)}-\sum \frac{P_{n-1}^{(j)}[-x_j^2]}{(r^2+x_j^2)\prod' (x_k^2-x_j^2)}=0.
\eea
To make this expression more symmetric, we introduce $x_0=ir$. Then the wave equation becomes
\bea\label{PolynPone}
S=0,\qquad S\equiv \sum_{j=0}^n  \frac{P_{n-1}^{(j)}[-x_j^2]}{\prod_{k\ne j} (x_k^2-x_j^2)}\,.
\eea
The restrictions on $P_{n-1}^{(j)}$ come from studying the analytical structure of the meromorphic function $S$.  
As $x_i$ approaches $x_{i+1}$, only two terms in the sum become singular ($j=i,i+1$), and the singularities must cancel. Computing the residue at the pole $\frac{1}{x_i-x_{i+1}}$, we conclude that the two adjacent polynomials must be identical. 
\bea
P_{n-1}^{(i+1)}[-x_i^2]=P_{n-1}^{(i)}[-x_i^2]\,.\nonumber
\eea
Since this relation must hold for all values of $i$, all polynomials $P_{n-1}^{(i+1)}$ are completely determined by $P_{n-1}\equiv P_{n-1}^{(0)}$, i.e.,
\bea\label{PolynPtwo}
P^{(j)}_{n-1}[-x_j^2]\equiv P_{n-1}[-x_j^2].
\eea
To summarize, we have demonstrated that equation (\ref{PolynPone}) implies the relation (\ref{PolynPtwo}), and now we will show that relation (\ref{PolynPtwo}) is also a sufficient condition guaranteeing (\ref{PolynPone}). In other words, there are no restrictions on coefficients of $P_{n-1}$.

Using relation (\ref{PolynPtwo}), function $S$ defined in equation (\ref{PolynPone}) can be rewritten as
\bea\label{PolynPthree}
S=\sum_{j=0}^n  \frac{P_{n-1}[-x_j^2]}{\prod' (x_k^2-x_j^2)}\,.
\eea
Clearly $S[x_0]$ is a meromorphic function  with potential poles at $x_0=\pm x_j$. Evaluating the residues, we conclude that $S[x_0]$ is regular everywhere in the complex plane and it approaches zero as $x_0$ goes to infinity. The only analytic function with such properties is $S=0$, so the wave equation is trivially satisfied  for any polynomial $P_{n-1}$. A more explicit form of equations (\ref{SeparWaveRadODE}) is given by (\ref{SeparWaveEvenFull}). 

\bigskip

In the odd--dimensional case the logic is very similar, although the result is somewhat different. Starting from the metric (\ref{MPodd}) and frames (\ref{AllFramesMPOdd}), we arrive at the counterpart of equation (\ref{Jul19tempApp}):
\bea
\sqrt{\prod {d_i}}\,r\d_\mu[\frac{1}{r}{\tilde g}_r^{\mu\nu}\d_\nu\Psi]+
\sum_i \frac{FR \sqrt{\prod {d_k}}}{x_i d_i(r^2+x^2_i)}\d_\mu[x_i{\tilde g}_i^{\mu\nu}\d_\nu\Psi]+
\frac{\sqrt{\prod {d_i}}}{r^2[\prod x_i^2]}
{\tilde g}_\psi^{\mu\nu}\d_\mu\d_\nu=0.
\eea
We used the expansion (\ref{gUpOdd}) for inverse metric as well as the expression (\ref{OddDetMet}) for the determinant. We will also need more explicit expressions for the components of the reduced metric:
\bea
g^{\mu\nu}\d_\mu\d_\nu&=&\frac{1}{FR}{\tilde g}_r^{\mu\nu}\d_\mu\d_\nu+
\sum \frac{1}{{d_i(r^2+x^2_i)}}{\tilde g}_i^{\mu\nu}\d_\mu\d_\nu+\frac{1}{r^2[\prod x_i^2]}
{\tilde g}_\psi^{\mu\nu}\d_\mu\d_\nu\,,\nonumber\\
{\tilde g}_r^{\mu\nu}\d_\mu\d_\nu&=&[R-Mr^2]\d_r^2+\frac{R^2}{R-Mr^2}\left[ \omega
-\sum_k\frac{m_ka_k}{r^2+a_k^2}\right]^2,\\
{\tilde g}_i^{\mu\nu}\d_\mu\d_\nu&=&
{\frac{H_i}{x_i^2}}[\d_{x_i}]^2
-{\frac{H_i}{x^2_i}}\left[\omega-\sum_k\frac{a_k m_k}{a_k^2-x_i^2}
\right]^2,\quad
{\tilde g}_\psi^{\mu\nu}\d_\mu\d_\nu=-\left[{\prod a_i}\right]^2\left[\omega-\sum_k\frac{m_k}{a_k}
\right]^2.\nonumber
\eea
Here we assumed that all differential operators act on a function that has the form (\ref{SeparFuncAppWave}). The counterparts of equations (\ref{SeparWaveRadODE}) are
\bea\label{SeparWaveOddFull}
&&r\frac{d}{dr}\left[\frac{R-Mr^2}{r}\frac{d\Phi}{dr}\right]+\frac{R^2}{R-Mr^2}
\left[\omega-
\sum_k\frac{a_k n_k}{r^2+a_k^2}\right]^2\Phi=P^{(0)}_{n}[r^2]\Phi,\nn
&&{x_i}\frac{d}{dx_i}\left[\frac{H_i}{x_i}\frac{dX_i}{dx_i}\right]-
H_i\left[\omega-
\sum_k\frac{a_k n_k}{a_k^2-x_i^2}\right]^2X_i=-P^{(i)}_{n}[-x_i^2]X_i.
\eea
Substitution into the wave equation leads to an algebraic relation
\bea
&&\frac{P^{(0)}_{n-1}[r^2]}{FR}+\frac{P^{(i)}_{n-1}[-x_i^2]}{x_i^2 d_i(r^2+x_i^2)}-\frac{B}{r^2\prod x_k^2}
=0,\quad B\equiv \left[{\prod a_i}\right]^2\left[\omega-\sum_k\frac{n_k}{a_k}
\right]^2.\nonumber
\eea
As before, introducing $x_0=ir$, we find
\bea
S=0,\qquad S\equiv\sum_{j=0}^n  \frac{P^{(j)}_{n}[-x_j^2]}{x_j^2\prod_{k\ne j} (x_k^2-x_j^2)}-\frac{B^2}{\prod x_k^2}.
\eea
The pole structure of $S$ ensures that all polynomials $P^{(j)}_{n-1}$ are the same:
\bea\label{tmpPPone}
P^{(j)}_{n}[-x_j^2]= P_{n}[-x_j^2].
\eea
The last remaining limit, $x_j\rightarrow 0$ implies that 
\bea\label{tmpPPtwo}
P_n[0]=B^2=\left[{\prod a_i}\right]^2\left[\omega-\sum_k\frac{n_k}{a_k}
\right]^2.
\eea
Once the relations (\ref{tmpPPone}) and (\ref{tmpPPtwo}) are imposed, equation $S=0$ becomes an identity. 

\subsection{Maxwell's equations in even dimensions}
\label{SecAppMPeven}

In this and next subsections we will discuss separation of variables in Maxwell's equations. To avoid cumbersome formulas, we will focus on configurations which don't depend on the cyclic coordinates 
$(t,\phi_i)$, although the final results  (\ref{AnsGenEvenElectr})-- (\ref{AnsGenEvenMagn}) and (\ref{GenOddElectr})--(\ref{OddDimGenSpinLmb}) work in the general case. Due to significant differences between the even-- and odd--dimensional cases, these situations will be analyzed separately, and in this subsection we will focus on even dimensions.

\bigskip

In even dimensions we impose the ansatze (\ref{AnsGenEvenElectr}), (\ref{AnsGenEvenMagn}) inspired by the gauge potentials (\ref{4dSmryAnstz}) in the Kerr geometry, and in the absence of $(t,\phi_i)$--dependence, the electric polarization of 
$A_\mu$ is determined by the following projections:
\bea\label{MaxEvenOne}
&&[m^{(j)}_\pm]^\mu A^{(el)}_\mu=\pm i{x_j}{\hat m}^{(j)}_\pm \Psi,\quad 
l^\mu_\pm A^{(el)}_\mu=\pm{r}{\hat l}_\pm \Psi,
\eea
where
\bea\label{SeparPsiAAA}
\Psi=\Phi(r)\prod X_i(x_i).\nonumber
\eea
The expressions for $(m^{(j)}_\pm,l_\pm)$ are given by (\ref{GenFramesEvenD}). Note that since differential operators ${\hat m}^{(j)}_\pm$ and ${\hat l}_\pm$ act on functions of $(r,x_i)$, the ansatz (\ref{MaxEvenOne}) guarantees that the nontrivial components of the gauge can point only along cyclic directions:
\bea
A=A_t dt+\sum A_{\phi_i}d\phi_i.\nonumber
\eea
In the absence of $(t,\phi_i)$--dependence it is convenient to introduce real frames 
$({\hat e}^{(j)},{\hat f}^{(j)})$ instead of the complex objects $(m^{(j)}_\pm,l_\pm)$:
\bea\label{fgFrames}
{\hat m}^{(j)}_\pm={\hat e}^{(j)}\pm i {\hat f}^{(j)},\quad 
{\hat l}_\pm={\hat e}^{(0)}\pm {\hat f}^{(0)}.
\eea
In particular, index $\mu$ in the frames ${\hat f}^{(j)}_\mu$ can point only along cyclic directions. 
In terms of the real frames, the ansatz (\ref{MaxEvenOne}) becomes
\bea\label{MaxEvenTwo}
{f}^{(j)}_\mu A_{(el)}^\mu=x_j {\hat e}^{(j)}\Psi,\quad 
{f}^{(0)}_\mu A_{(el)}^\mu=r {\hat e}^{(0)}\Psi.
\eea
The arguments presented below will be applicable to the magnetic case as well, and to put the two polarizations on the same footing, we write the last set of relations as
\bea\label{tempElctrEven}
{f}^{(j)}_\mu A_{(el)}^\mu=S_j {\hat e}^{(j)}\Psi,\quad 
{f}^{(0)}_\mu A_{(el)}^\mu=S_r {\hat e}^{(0)}\Psi,
\quad S_j=x_j,\quad S_r=r.
\eea

Next we define the nontrivial components of Maxwell's equations\footnote{We recall that 
$q^{AC}= f^A_\mu {f}^{C}_\la g^{\mu\la}$ is a nontrivial coordinate--dependent matrix since ${\hat f}^{(j)}_\mu$ are constructed from {\it rescaled} frames $({\hat l}_\pm,{\hat m}^{(j)}_\pm)$ defined by (\ref{GenFramesEvenD}).}. 
\bea\label{aaMaxwComp}
\MM_C=\frac{1}{\sqrt{-g}}{f}_{C}^\la g_{\mu\la}\d_i[\sqrt{-g}q^{AB}f_A^\mu f_B^\nu g^{ij}\d_j A_\nu]={f}_{C}^\la g_{\mu\la} \MM^\mu
\eea
Direct evaluation for the configuration (\ref{tempElctrEven}) gives
\bea\label{EvenMPeqnNC}
{\MM_C}{}&=&\frac{\sqrt{H_C}}{x_C}\left[\sum_{i'} \frac{x^2_C-x^2_i}{q_i}
\frac{x_i}{S_i}\left\{\d_C\frac{S_C x_C}{x_C^2-x_i^2}\right\}
\d_i\left\{\frac{H_i S_i}{x_i}\d_i\right\}\right.\\
&&\left.+
\frac{r^2+x^2_C}{FR}\frac{r}{S_r}
\left\{\d_C\frac{S_C x_C}{r^2+x_C^2}\right\}
\d_r\left\{\frac{R_M S_r}{r}\d_r\right\}
+\d_C\left\{\frac{x^2_C}{q_C}\d_C [\frac{H_C S_C}{x_C}\d_C]\right\}
\right]\Psi,\nn
{\MM_0}{}&=&\frac{\sqrt{R}}{r}\left[\sum_{i} \frac{r^2+x^2_i}{q_i}\frac{x_i}{S_i}
\left\{\d_r\frac{S_r r}{r^2+x_i^2}\right\}
\d_i\left\{\frac{H_i S_i}{x_i}\d_i\right\}
+\d_r\left\{\frac{r^2}{FR}\d_r [\frac{\Delta S_r}{r}\d_r]\right\}
\right]\Psi.\nonumber
\eea
Here 
\bea\label{temp21}
q_i=(r^2+x_i^2)d_i=(r^2+x_i^2)\prod(x_k^2-x_i^2),\quad \Delta\equiv R-Mr.
\eea
Before analyzing the system (\ref{EvenMPeqnNC}), we observe that a similar set of equations emerges from the magnetic ansatz 
\bea
&&m^\mu_\pm A_\mu=\mp \frac{i}{x_j}{\hat m}_\pm \Psi,\quad 
l^\mu_\pm A_\mu=\pm\frac{1}{r}{\hat l}_\pm \Psi.
\eea
A counterpart of (\ref{tempElctrEven}) for this case is
\bea\label{tempMagnEven}
{\hat f}^{(j)}_\mu A^\mu=-S_j {\hat e}^{(j)}\Psi,\quad 
{\hat f}^{(0)}_\mu A^\mu=-S_r {\hat e}^{(0)}\Psi,
\quad S_j=\frac{1}{x_j},\quad S_r=-\frac{1}{r},
\eea
and Maxwell's equations still reduce to (\ref{EvenMPeqnNC}), although with different functions $(S_j,S_r)$.
Note for genetic factors $(S_j,S_r)$ the Maxwell's equations are much more complicated than (\ref{EvenMPeqnNC}) and they don't admit separation of variables.

\bigskip

Let us now discuss the system (\ref{EvenMPeqnNC})  and demonstrate that it is satisfied for the ansatz (\ref{SeparPsiAAA}), as long as $\Phi$ and $X_j$ obey certain ordinary differential equations. Note that equations (\ref{EvenMPeqnNC}) contain third derivatives of some functions, while we expect to have second order ODEs. Motivated by separation of the wave equation discussed in section \ref{SecMPwave}, we impose equations
\bea\label{SeparEqn6D}
\hskip -0.5cm \d_i\left\{\frac{H_i S_i}{x_i}\d_i X_i\right\}
=\frac{S_i}{x_i^3}\left[\sum_k\la_k x_i^{2k}\right]X_i,\quad
\frac{r}{S_r}\d_r\left\{\frac{\Delta S_r}{r}\d_r\right\}\Phi=\frac{1}{r^2}\left[\sum_k\la_k (ir)^{2k}\right]\Phi,
\eea
where $\la_k$ are arbitrary constants. Substituting relations (\ref{SeparEqn6D}) into equations (\ref{EvenMPeqnNC}) and focusing on the coefficient in front of $\la_k$, we find
\bea\label{TempMCeven}
{\MM_C}{}\Big|_{\la_k}&=&\frac{\sqrt{H_C}}{x_C}\left[\sum_{i'} \frac{x^2_C-x^2_i}{q_i}
\left\{\d_C\frac{S_C x_C}{x_C^2-x_i^2}\right\}
\frac{x_i^{2k}}{x_i^2}\right.\nn
&&\left.+
\frac{r^2+x^2_C}{FR}\frac{(ir)^{2k}}{r^2}
\left\{\d_C\frac{S_C x_C}{r^2+x_C^2}\right\}
+\d_C\left\{\frac{S_C}{x_C q_C}x_C^{2k}\right\}
\right]\Psi\\
&=&\frac{\sqrt{H_C}}{x_C}\d_C\left\{S_Cx_C\left[\sum_{i\ne C} \frac{x_i^{2k-2}}{q_i}
-
\frac{(ir)^{2k-2}}{FR}
+\frac{x_C^{2k-2}}{q_C}
\right]\Psi\right\}.\nonumber
\eea
We used the $x_C$--independence of the ratios (see definitions (\ref{MisclEllipticEvev}) and (\ref{temp21}))
\bea
\frac{x^2_C-x^2_i}{q_i}\quad\mbox{and}\quad \frac{r^2+x^2_C}{FR}\,.
\eea
To demonstrate that the right--hand side of (\ref{TempMCeven}) vanishes, it is sufficient to show the square bracket appearing in the last line is equal to zero. Using the definition of $q_i$, we find
\bea
\left[\sum_{i} \frac{x_i^{2k-2}}{q_i}
-
\frac{(ir)^{2k-2}}{FR}
\right]=\sum \frac{x_j^{2k-2}}{(r^2+x_j^2)\prod' (x_p^2-x_j^2)}-
\frac{(ir)^{2k-2}}{\prod (r^2+x_k^2)}.
\eea
The right--hand side of the last expression vanishes, as a special case of equation (\ref{tempE5}), then we conclude that 
\bea
{\MM_C}=\sum_k{\MM_C}{}\Big|_{\la_k}\la_k=0.
\eea
The relation ${\MM_0}=0$ for the remaining component of  (\ref{EvenMPeqnNC}) can be verified in the same way. 

\subsection{Maxwell's equations in odd dimensions}
\label{SecAppMPodd}

In odd dimensions, equations for the electric and magnetic polarizations are very different, so they have to be discussed separately. In the absence of $(t,\phi_i)$--dependence, the ansatz for the electric polarization of the gauge field is 
\bea
m^\mu_\pm A_\mu=\pm{ix_j}{\hat m}_\pm \Psi,\quad 
l^\mu_\pm A_\mu=\pm{r}{\hat l}_\pm \Psi,\quad n^\mu A_\mu=0.
\eea
Introducing the frames $(e^{(j)}, f^{(j)})$ defined in (\ref{fgFrames}), we find
\bea
{f}^{(j)}_\mu A^\mu=x_j {\hat e}^{(j)}\Psi,\quad {f}^{(0)}_\mu A^\mu=r {\hat e}^{(0)}\Psi,\quad
n^\mu A_\mu=0.
\eea
As before, the gauge field has only cyclic components, so ${e}^{(j)}_\mu A^\mu=0$.
Substitution of the ansatz into Maxwell's equations (\ref{aaMaxwComp}) leads to a counterpart of (\ref{EvenMPeqnNC})\footnote{Expression for $q_i$ is still given by (\ref{temp21}), but now $\Delta=R-M r^2$. Functions $(H_i,F,R)$ are defined by (\ref{MiscElliptic}), (\ref{EplsdOdd}).}
\bea\label{OddMPeqnNCelc}
{\MM_C}{}&=&-\frac{\sqrt{H_C}}{x_C}\left[\sum_{i\ne C} \frac{x^2_C-x^2_i}{x_iq_i}
\left\{\d_C\frac{x^2_C}{x_C^2-x_i^2}\right\}
\d_i\left\{\frac{H_i}{x_i}\d_i\right\}\right.\\
&&\left.-
\frac{r^2+x^2_C}{rFR}
\left\{\d_C\frac{x^2_C}{r^2+x_C^2}\right\}
\d_r\left\{\frac{\Delta}{r}\d_r\right\}
+\d_C\left\{\frac{x_C}{q_C}\d_C [\frac{H_C}{x_C}\d_C]\right\}
\right]\Psi,\nn
{\MM_0}{}&=&\frac{\sqrt{R}}{r}\left[-\sum_{i} \frac{r^2+x^2_i}{x_iq_i}
\left\{\d_r\frac{r^2}{r^2+x_i^2}\right\}
\d_i\left\{\frac{H_i}{x_i}\d_i\right\}
+\d_r\left\{\frac{r}{FR}\d_r [\frac{\Delta}{r}\d_r]\right\}
\right]\Psi.\nonumber
\eea
We also find a new component:
\bea\label{OddEltrcMPspec}
n_\mu \MM^\mu=2\left[-\sum_j \frac{1}{x_j q_j}\d_j\left\{\frac{H_j}{x_j}\d_j\right\}+
\frac{1}{rFR}\d_r\left\{\frac{\Delta}{r}\d_r\right\}\right]\Psi.
\eea
Introducing a separable ansatz,
\bea\label{MPoddElctrSepAns}
\Psi=\Phi(r)\prod X_j(x_j),
\eea
we find that equation $n_\mu \MM^\mu=0$ is inconsistent unless functions $(\Phi,X_j)$ satisfy a system of ODEs
\bea\label{SeparEqnMPtemt}
\frac{1}{x_j}\d_j\left\{\frac{H_j}{x_j}\d_j X_j\right\}
=P^{(j)}_{n-1}[-x_j^2]X_j,\quad
\frac{1}{r}\d_r\left\{\frac{R_M}{r}\d_r\right\}\Phi=P^{(0)}_{n-1}[r^2]\Phi,
\eea
where $P^{(k)}_{n-1}$ are some polynomial of degree $n-1$. Substitution into (\ref{OddEltrcMPspec}) gives
\bea\label{AAeqnBB}
n_\mu \MM^\mu=-2\left[\sum_{j=1}^{n-1} \frac{P^{(j)}_{n-1}[-x_j^2]}{q_j}-
\frac{P^{(0)}_{n-1}[r^2]}{FR}\right]=0.
\eea
To make the last expression more symmetric, we introduce a new coordinate $x_0=ir$ and recall that
\bea\label{TempQandFR}
&&q_i=(r^2+x_i^2)\prod_{k\ne 0,i}(x_k^2-x_i^2)=-\prod_{k\ne i}(x_k^2-x_i^2),\nn
&&FR=\prod_{k\ne 0}(r^2+x_k^2)=\prod_{k\ne 0}(x_k^2-x_0^2).
\eea
Then equation (\ref{AAeqnBB}) becomes
\bea\label{AAeqnBB1}
-2\sum_{j=0}^{n-1} \frac{P^{(j)}_{n-1}[-x_j^2]}{\prod_{k\ne i}(x_k^2-x_i^2)}=0.
\eea
Analyzing the poles and residues of the last expression as in section \ref{SecAppMPeven} (i.e., taking limits $x_j\rightarrow x_{j+1}$), we conclude that all polynomials $P^{(j)}_{n-1}$ must be the same. After this condition is imposed, the left hand side of (\ref{AAeqnBB1}) becomes a regular function in the complex $x_0$--plane that vanishes at infinity, so equation (\ref{AAeqnBB1}) is trivially satisfied.

Thus we have shown that application of the ansatz (\ref{MPoddElctrSepAns}) to equation (\ref{OddEltrcMPspec}) leads to ODEs (\ref{SeparEqnMPtemt}) with {\it one} independent polynomial $P_{n-1}$:
\bea\label{SeparEqnMPtemt1}
\frac{1}{x_j}\d_j\left\{\frac{H_j}{x_j}\d_j X_j\right\}
=P_{n-1}[-x_j^2]X_j,\quad
\frac{1}{r}\d_r\left\{\frac{R_M}{r}\d_r\right\}\Phi=P_{n-1}[r^2]\Phi.
\eea
Substitution of these equations into (\ref{OddMPeqnNCelc}) gives
\bea\label{OddMPeqnNCelc1}
{\MM_C}{}&=&-\frac{\sqrt{H_C}}{x_C}\left[\sum_{i\ne C} \frac{x^2_C-x^2_i}{q_i}P_{n-1}[-x_i]
\left\{\d_C\frac{x^2_C}{x_C^2-x_i^2}\right\}\right.\nn
&&\left.-
\frac{r^2+x^2_C}{FR}P_{n-1}[r^2]
\left\{\d_C\frac{x^2_C}{r^2+x_C^2}\right\}
+\d_C\left\{\frac{x_C^2}{q_C}P_{n-1}[-x_C^2]\right\}
\right]\Psi\nn
&=&-\frac{\sqrt{H_C}}{x_C}\d_C\left\{x_C^2\left[\sum_{i\ne C} \frac{P_{n-1}[-x_i]}{q_i}
-
\frac{P_{n-1}[r^2]}{FR}
+\frac{P_{n-1}[-x_C^2]}{q_C}
\right]\Psi\right\}\,.
\eea
To go to the last line we used the relations
\bea
\d_C \frac{x^2_C-x^2_i}{q_i}=0,\quad \d_C \frac{r^2+x^2_C}{FR}=0.
\eea
The expression in the square brackets of (\ref{OddMPeqnNCelc1}) is proportional to the right hand side of (\ref{AAeqnBB}) (recall than $P^{(j)}_{n-1}[y]=P_{n-1}[y]$), so ${\MM_C}=0$. The remaining equation 
in (\ref{OddMPeqnNCelc}), $\MM_0=0$, is verified in a similar way.

To summarize, Maxwell's equations for the electric polarization in odd dimensions work in the same way as the even--dimensional relations discussed in section \ref{SecAppMPeven}, and the new equation 
$n_\mu \MM^\mu=0$ makes the derivation even more straightforward since one no longer has to {\it assume} an existence of second order ODEs, such equations (\ref{SeparEqnMPtemt1}) are {\it derived}.

\bigskip

Let us now discuss the magnetic polarization. The ansatz for the gauge potential is
\bea
m^\mu_\pm A_\mu=\mp\frac{i}{x_j}{\hat m}_\pm \Psi,\quad 
l^\mu_\pm A_\mu=\pm\frac{1}{r}{\hat l}_\pm \Psi,\quad n^\mu A_\mu=\la\Psi,
\eea
and in terms of the frames $(e^{(j)}, f^{(j)})$ defined in (\ref{fgFrames}) it becomes
\bea
{\hat f}^{(j)}_\mu A^\mu=-\frac{1}{x_j} {\hat e}^{(j)}\Psi,\quad 
{\hat f}^{(0)}_\mu A^\mu=\frac{1}{r} {\hat e}^{(0)}\Psi,\quad n^\mu A_\mu=\la\Psi.
\eea
Substitution into the components of the Maxwell's equations (\ref{aaMaxwComp}) leads to the magnetic counterparts of (\ref{OddMPeqnNCelc}) and (\ref{OddEltrcMPspec})
\bea\label{OddMPeqnNCmgn}
{\MM_C}{}&=&\frac{\sqrt{H_C}}{x_C}\left[\sum_{i\ne C} \frac{x^2_C-x^2_i}{x_iq_i}x_i^2
\left\{\d_C\frac{1}{x_C^2-x_i^2}\right\}
\d_i\left\{\frac{H_i}{x^3_i}\d_i\right\}-
\frac{2\la x_C^2\prod a_i^2}{r^2\prod x_k^2}\d_C\frac{1}{x^2_C}\right.\nn
&&\left.-
\frac{r^2+x^2_C}{rFR}r^2
\left\{\d_C\frac{1}{r^2+x_C^2}\right\}
\d_r\left\{\frac{R_M}{r^3}\d_r\right\}
+\d_C\left\{\frac{x_C}{q_C}\d_C [\frac{H_C}{x^3_C}\d_C]\right\}
\right]\Psi,\\
{\MM_0}{}&=&\frac{\sqrt{R}}{r}\left[-\sum_{i} \frac{r^2+x^2_i}{x_iq_i}x_i^2
\left\{\d_r\frac{1}{r^2+x_i^2}\right\}
\d_i\left\{\frac{H_i}{x^3_i}\d_i\right\}
+\d_r\left\{\frac{r}{FR}\d_r [\frac{R_M}{r^3}\d_r]\right\}\right.\nn
&&\left.+\frac{2\la \prod a_i^2}{\prod x_k^2}\d_r\frac{1}{r^2}
\right]\Psi,\nonumber\\
n_\mu \MM^\mu&=&\left[\sum_j \frac{2-\la x_j^2}{x_j q_j}\d_j\left\{\frac{H_j}{x^3_j}\d_j\right\}+
\frac{2+\la r^2}{rFR}\d_r\left\{\frac{R_M}{r^3}\d_r\right\}
-\frac{4\la \prod a_i^2}{r^2\prod x_k^2}\left[\sum\frac{1}{x_j^2}-\frac{1}{r^2}\right]\right]\Psi.
\nonumber
\eea 
As in the electric case, we begin solving this system by looking at equation $n_\mu \MM^\mu=0$. Consistency of this relation for the ansatz (\ref{MPoddElctrSepAns}) leads to a system of ODEs:
\bea\label{SeparEqnMPtemt2}
\frac{1}{x_j}\d_j\left\{\frac{H_j}{x^3_j}\d_j X_j\right\}
=\frac{Q^{(j)}[-x_j^2]}{x_j^4}X_j,\quad
\frac{1}{r}\d_r\left\{\frac{R_M}{r^3}\d_r\right\}\Phi=-\frac{Q^{(0)}[r^2]}{r^4}\Phi
\eea
with some undetermined polynomials $Q^{(k)}$. Introducing $x_0\equiv r$ and using the identities 
(\ref{TempQandFR}), we find
\bea\label{AA729}
n_\mu \MM^\mu=\left[-\sum_{j=0}^{n-1} \frac{(2-\la x_j^2)Q^{(j)}[-x_j^2]}{
x_j^4\prod_{k\ne j}(x_k^2-x_j^2)}
+\frac{4\la \prod a_i^2}{\prod x_k^2}\sum_{j=0}^{n-1}\frac{1}{x_j^2}\right]\Psi.
\eea
As before, the analysis of poles and residues at $x_j=x_{j+1}$ leads to the conclusion that equation 
$n_\mu \MM^\mu=0$ can be satisfied only if all functions $Q^{(j)}$ are the same:
\bea
Q^{(j)}[y]=Q[y].\nonumber
\eea
Focusing on such configurations and looking at the right--hand side of (\ref{AA729}) in the vicinity of $x_0=0$, we find 
\bea
n_\mu \MM^\mu\sim \left[
\frac{-(2-\la x_0^2)Q[-x_0^2]+4\la \prod a_i^2}{\prod_{k\ne 0} x_k^2}\left[\frac{1}{x_0^4}+
\sum_{j=1}^{n-1}\frac{1}{x_0^2 x_j^2}\right]+regular\right]\Psi.\nonumber
\eea
To avoid singularities in the right--hand side of the last expression, we must requite
\bea
Q[y]=\la (2-\la y)\prod a_i^2+y^2 P_{n-2}[y],
\eea
where $P_{n-2}$ is an arbitrary polynomial of degree $n-2$. With such function $Q$, the right--hand side of (\ref{AA729}) is a regular function in the complex $x_0$--plane, and it vanishes at infinity, then  
$n_\mu \MM^\mu=0$. 

Substituting relations (\ref{SeparEqnMPtemt2}) into the first expression in (\ref{OddMPeqnNCmgn}), we find
\bea
{\MM_C}{}&=&\frac{\sqrt{H_C}}{x_C}\d_C\left\{\left[\sum_{i\ne C} 
\frac{Q[-x_i^2]}{q_i x_i^2}-
\frac{2\la \prod a_i^2}{r^2\prod x_k^2}+\frac{Q[r^2]}{FR r^2}
+\frac{Q[-x_C^2]}{x_C^2q_C}
\right]\Psi\right\}\,.
\eea 
Introducing $x_0=ir$, we can rewrite the last expression in a more symmetric form,
\bea
{\MM_C}{}&=&\frac{\sqrt{H_C}}{x_C}\d_C\left\{\left[-\sum_{i=0}^{n-1} 
\frac{Q[-x_i^2]}{x_i^2\prod_{k\ne i}(x_k^2-x_i^2)}+
\frac{2\la \prod a_i^2}{r^2\prod x_k^2}
\right]\Psi\right\}\,.
\eea
The expression in the square brackets is a meromorphic function of $x_0$, which has no poles, and which vanishes at infinity. Then the square bracket must vanish, and $\MM^C=0$. Equation $\MM_0=0$ is verified in the same way.

\bigskip

\bigskip

To summarize, in this appendix we have demonstrated that the ansatze (\ref{AnsGenEvenElectr}), (\ref{AnsGenEvenMagn}), and (\ref{GenOddElectr}) lead to separable solutions for function $\Psi$, and the resulting ODEs are given by (\ref{EvenDimGenSpin})--(\ref{EvenDimGenSpinLmb}) in even, and by (\ref{OddMPmaster})--(\ref{OddDimGenSpinLmb}) in odd dimensions. Although we focused on configurations without dependence on cyclic coordinates, the general case can be verified in the same way. The relevant calculations were performed using Mathematica, but we did not present the intermediate expressions due to their complexity. The final results are given by equations  (\ref{EvenDimGenSpin})--(\ref{EvenDimGenSpinLmb}) and (\ref{OddMPmaster})--(\ref{OddDimGenSpinLmb}).

\section{Reduction to static configurations}
\label{AppStatic}
\renewcommand{\theequation}{F.\arabic{equation}}
\setcounter{equation}{0}

In this appendix we will take the static limit of the solutions discussed in section \ref{SecWaveMP} and demonstrate that the resulting configurations reproduce $(D-3)$ magnetic polarizations in the Schwarzschild--Tangherlini geometry, which were reviewed in the Appendix \ref{SecAppSchw}. The results derived here are summarized in section \ref{SecSubCmprSchw}, which also discusses the electric polarization. To avoid unnecessary complications, we focus on solutions with $\omega=m_i=0$, although similar arguments are applicable in the general case.

\bigskip

The static limit of the Myers--Perry geometry (\ref{MPeven}), (\ref{AllFramesMP})\footnote{In this appendix we are focusing on 
$D=2n+2$ dimensions, and the odd--dimensional case can be treated in a similar way.} is obtained by performing a rescaling
\bea\label{ReclLmbApp}
a_i=\la b_i,\quad x_i=\la y_i
\eea
and sending $\la$ to zero, while keeping $b_i$ and $y_i$ fixed. The resulting metric describes the Schwarzschild--Tangherlini geometry, but the sphere $S^{D-2}$ is written in unusual variables, which are inherited from ellipsoidal coordinates. As discussed in section \ref{SecSubCmprSchw} this unusual parameterization does not affect the analysis of the electric polarization, and the static limit $\la\rightarrow 0$ of the system (\ref{AnsGenEvenElectr}) reproduces the relevant equations for the Schwarzschild--Tangherlini geometry. The situation with magnetic polarization is more interesting: an ambiguity in taking the static limit allows one to recover all $(D-3)$ polarizations from a single system (\ref{AnsGenEvenMagn}), and this appendix is dedicated to the detailed analysis of this phenomenon.   

The magnetic polarizations for the static geometry are obtained by applying the rescaling (\ref{ReclLmbApp}) to the ansatz (\ref{AnsGenEvenMagn}) and sending $\la$ to zero. To arrive at a well--defined limit, we also have to multiply all components of the gauge field by $\la$ and rescale the parameter $\mu$ as
\bea
\mu=\la\nu. 
\eea
This leads to the final result
\bea\label{MagnSTlimit}
l^\mu_\pm A^{(mgn)}_\mu=0,\quad
[{\tilde m}_\pm^{(j)}]^\mu A^{(mgn)}_\mu=\mp\frac{i}{y_j\pm \nu}{\hat {\tilde m}}^{(j)}_\pm \Psi\,,
\eea
where frames $[{\tilde m}_\pm^{(j)}]^\mu$ are defined by (\ref{STframesLimit}). As expected, the gauge field (\ref{MagnSTlimit}) has vanishing radial and temporal components:
\bea
A^{(mgn)}_t=A^{(mgn)}_r=0.
\eea
The remaining components of the gauge field can be naturally separated into the cyclic and non--cyclic projections, and the relevant parts of the equation (\ref{MagnSTlimit}) give
\bea\label{MagnSTlimit2}
A_{y_j}^{(mgn)}=\frac{i{\nu}}{y_j^2-{\nu}^2}\d_{y_j}\Psi,\quad
\sum_k\frac{b_k }{b_k^2-y_j^2}A^{(mgn)}_{\phi_k}=\frac{y_j}{y_j^2-{\nu}^2}\d_{y_j} \Psi.
\eea
The rest of this appendix is dedicated to the analysis of the field (\ref{MagnSTlimit2}), so to avoid unnecessary clutter, we will suppress the label $(mgn)$. 

To proceed, we need to establish the relation between variables $y_j$ and the standard spherical coordinates. The ellipsoidal coordinates $x_j$ are defined by equation (\ref{EplsdEven}),
\bea\label{TempST1}
(b_i\mu_i)^2=\frac{1}{\prod (b_i^2-b_k^2)}\prod (b_i^2-y_k^2),
\eea
while the spherical coordinates $\xi_k$ are given by
\bea\label{TempST2sph}
\mu_1=\xi_1\sqrt{1-\xi_2^2},\quad \mu_2=\xi_1\xi_2\sqrt{1-\xi_3^2},\quad \dots
\eea
To obtain the last relation from equation (\ref{TempST1}), we write 
\bea\label{TempST2e}
y_k^2=b_k^2-(b_k^2-b_{k-1}^2)\xi_k^2,\quad b_0\equiv 0,
\eea
and take limits in the following order:
\bea\label{LimitElptc}
b_n\rightarrow b_{n-1},\quad b_{n-1}\rightarrow b_{n-2},\quad\dots\quad b_2\rightarrow b_1\equiv b\,.
\eea
To see that this limit gives the desired result, we first rewrite (\ref{TempST1}) as
\bea\label{TempST2}
(b_i\mu_i)^2=\frac{1}{\prod (b_i^2-b_k^2)}\prod \Big[b_i^2-b_k^2+(b_k^2-b_{k-1}^2)\xi_k^2\Big]\,,
\eea
and then take the limit (\ref{LimitElptc}) in the following ratios:
\bea\label{TempST2a}
k>i+1:&&\frac{b_i^2-b_k^2+(b_k^2-b_{k-1}^2)\xi_k^2}{b_i^2-b_k^2}\rightarrow 
\frac{b_i^2-b_k^2}{b_i^2-b_k^2}=1,\nn
k=i+1:&&\frac{b_i^2-b_k^2+(b_k^2-b_{i}^2)\xi_k^2}{b_i^2-b_k^2}\rightarrow 
(1-\xi_{i+1}^2),\nn
k=i:&&\frac{(b_k^2-b_{k-1}^2)\xi_k^2}{b_{i}^2-b_{k-1}^2}\rightarrow \xi_k^2,\\
1<k<i:&&\frac{b_i^2-b_k^2+(b_k^2-b_{k-1}^2)\xi_k^2}{b_i^2-b_{k-1}^2}\rightarrow \xi_k^2,\nn
k=1:&&b_i^2-b_1^2(1-\xi_1)^2\rightarrow b^2\xi_1^2.\nonumber
\eea
The right--hand side of equation (\ref{TempST2}) is the product of the expressions (\ref{TempST2a}), so the limit of (\ref{TempST2}) reproduces relations (\ref{TempST2sph}) for the spherical coordinates:
\bea\label{STmuJ}
(\mu_j)^2=(1-\xi_{j+1}^2)\prod_{k=1}^j (\xi_j)^2\,.
\eea
We can now take the limit (\ref{LimitElptc}) in the expressions (\ref{MagnSTlimit2}) for the gauge field. It is convenient to analyze $A_y$ and $A_\phi$ separately.

\bigskip

\noindent
{\bf $\xi$--components of the gauge field}

We begin with discussing the $y$--components of (\ref{MagnSTlimit2}):
\bea
A_{y_j}=\frac{i{\nu}}{y_j^2-{\nu}^2}\d_{y_j}\Psi\,.
\eea
Performing the change of variables (\ref{TempST2e}), we find
\bea
A_{\xi_j}=\frac{i{\nu}}{b_j^2-(b_j^2-b_{j-1}^2)\xi_j^2-{\nu}^2}\d_{\xi_j}\Psi\,.
\eea
For $\nu\ne b$, the limit (\ref{LimitElptc}) of this expression gives
\bea\label{TempSTgauge1}
A_{\xi_j}=\frac{i{\nu}}{b^2-\nu^2}\d_{\xi_j}\Psi\,.
\eea
To obtain the remaining $\xi$--polarizations, we have to send $\nu$ to $b$ at various rates consistent with the hierarchy (\ref{LimitElptc}). For example, setting $\nu=\pm b_n$, we find that 
\bea
A_{\xi_j}\propto \frac{1}{b_j^2-b_{j-1}^2},
\eea
so to have a well--defined limit, the gauge field must be rescaled: 
\bea\label{tempST10}
A_{\xi_j}=\frac{i{\nu}}{b^2}\left[\frac{b_{n-1}^2-b_{n}^2}{y_j^2-b_n^2}\right]\d_{\xi_j}\Psi\rightarrow
\frac{i{\nu}}{b^2}N_{j,n}\d_{\xi_j}\Psi,\qquad
N_{j,n}=\left\{
\begin{array}{ll}
0,&j\le n-1\\
\frac{1}{\xi_{n}^2},&j=n
\end{array}
\right.
\eea
Such constant rescaling does not affect Maxwell's equation due to their linearity. 

To construct the polarizations with $\nu=\pm b_c$ for $c<n$, we again consider a rescaled gauge field
\bea\label{STresldAxi}
A_{\xi_j}=\frac{i{\nu}}{b^2}\left[\frac{b_c^2-b_{c+1}^2}{y_j^2-b_c^2}\right]\d_{\xi_j}\Psi
\eea
The expression in the square brackets will be encountered for the $A_\phi$ polarizations as well, so it is convenient to introduce a special notation $N_{j,c}$:
\bea\label{TempSTgauge11}
N_{j,c}\equiv\lim \frac{b_{c+1}^2-b_{c}^2}{y_j^2-b_c^2}=
\lim \frac{b_{c+1}^2-b_{c}^2}{b_j^2-(b_j^2-b_{j-1}^2)\xi_j^2-b_c^2}
\eea
Here and below the symbol ``$\lim$'' refers to the limit (\ref{LimitElptc}). Direct evaluation gives
\bea\label{STdefNc}
N_{j,c}=\left\{
\begin{array}{ll}
0,&j\le c\\
\frac{1}{1-\xi_{c+1}^2},&j=c+1\\
1,&j>c+1
\end{array}
\right.\,.
\eea
Although we focused on $\nu=\pm b_c$, it is clear that all values of 
$|\nu|$ that lie between $b_c$ and $b_{c+1}$ lead to solutions described by $N_{j,c}$, with a possible exception of $N_{c+1,c}$\footnote{Specifically, 
${\cal N}_{j}\equiv\lim \frac{b_{c+1}^2-\nu^2}{y_j^2-\nu^2}$ for $\nu^2=b^2_c+\sigma (b^2_{c+1}-b_c^2)$ reproduces $N_{j,c}=0,1$ from (\ref{STdefNc}), but gives 
${\cal N}_{c+1}=\frac{1-\sigma}{1-\sigma-\xi_{c+1}^2}$.}. Furthermore, $N_{j,n-1}$ is proportional to $N_{j,n}$ from (\ref{tempST10}), so we can keep only (\ref{TempSTgauge1}) and (\ref{TempSTgauge11})--(\ref{STdefNc}) with $c=\{1,\dots,(n-1)\}$.

To summarize, in the static limit (\ref{LimitElptc}), the $\xi$--components of the gauge field are described by (\ref{TempSTgauge1}) and additional $2(n-1)$ polarizations:
\bea
\nu=\pm b_c:\qquad A^{(\pm),c}_{\xi_j}=\pm \frac{i}{b}N_{j,c}\,\d_{\xi_j}\Psi.
\eea
Although $A^{(+),c}_{\xi_j}=-A^{(-),c}_{\xi_j}$, such simple relations do not persist for the other components of the gauge field, so polarizations $A^{(+),c}$ and $A^{(-),c}$ are linearly independent. 
We now turn to the discussion of the $\phi$--components of the gauge field. 

\bigskip

\noindent
{\bf $\phi$--components of the gauge field}

Although the expressions for $A_{\phi_k}$ can be obtained by inverting the second set of equations in (\ref{MagnSTlimit2}), the results are rather complicated. This problem can be traced to a more involved form of the frame components $[{\tilde m}_\pm^{(j)}]_\mu$ in comparison with $[{\tilde m}_\pm^{(j)}]^\mu$. This suggests that the expressions for $A^{\phi_k}$ could be more transparent, and the indices can be lowered after taking the limit (\ref{LimitElptc}). Recall that the relevant part of the metric is
\bea
ds_{\phi}^2=r^2\sum \mu_j^2 d\phi_j^2\nonumber
\eea
with $\mu_j$ given by (\ref{STmuJ}). 

To proceed, it is convenient to decompose $l_\pm$ and 
${\tilde m}_\pm^{(j)}$ as in (\ref{fgFrames}):
\bea
{\hat{\tilde m}}^{(j)}_\pm={\hat e}^{(j)}\pm i {\hat f}^{(j)}\,,\quad 
{\hat l}_\pm={\hat e}^{(0)}\pm {\hat f}^{(0)}\,,
\eea
and to write the $\la=0$ limit of the metric (\ref{ST533}) as
\bea
g^{\mu\nu}=\frac{1}{r^{2n}}\left[e^{(0)\mu}e^{(0)\nu}-f^{(0)\mu}f^{(0)\nu}\right]+
\sum_{j=1}^n \left[e^{(j)\mu}e^{(j)\nu}+f^{(j)\mu}f^{(j)\nu}\right]\,.
\eea
Recall that, according to (\ref{STframesLimit}),
\bea
e^{(j)\mu}\d_\mu&=&\frac{1}{r}\left[\sqrt{\frac{H_j}{\la^2 d_j}}\right]_{\la=0}{\hskip -0.5cm}\d_{y_j}=
\frac{1}{r}\left[(b_j^2-y_j^2)\prod_{k\ne j}\frac{a_k^2-y_j^2}{y_k^2-y_j^2}\right]^{1/2}\d_{y_j}\,,
\nn
f^{(j)\mu}\d_\mu&=&-\frac{1}{r}\left[(b_j^2-y_j^2)\prod_{k\ne j}\frac{a_k^2-y_j^2}{y_k^2-y_j^2}\right]^{1/2}\sum_p\frac{b_p}{b_p^2-y_j^2}\d_{\phi_p}\,.
\eea
Vectors $f^{(j)\mu}$ form a basis for the $\phi$--components of the gauge potential, so ignoring  
$A_{y_j}$  components for a moment, we can write\footnote{We use notation $A_{(\phi)}^\mu$ in (\ref{TempST7}) to stress that this relation applies 
only to $\phi$--components of the gauge field.}:
\bea\label{TempST7}
A_{(\phi)}^\mu=\sum {\cal A}_a f^{(a)\mu}\,.
\eea
Coefficients ${\cal A}_a$ are determined by multiplying the last relation by $f_{(b)\mu}$ and using (\ref{MagnSTlimit}):
\bea
{\cal A}_b=f_{(b)}^\mu A_\mu=\sum_j \frac{y_j}{y_j^2-\nu^2}e_{(b)}^i \d_j\Psi\,.
\eea
Combining the last two equations, we find
\bea
A_{(\phi)}^\mu\d_\mu=-\sum_{j} \frac{1}{r^2}\left[(b_j^2-y_j^2)\prod_{k\ne j}\frac{b_k^2-y_j^2}{y_k^2-y_j^2}\right] \frac{y_j}{y_j^2-\nu^2}\sum_p\frac{b_p}{b_p^2-y_j^2}[\d_{y_j}\Psi]\d_{\phi_p}\,.\nonumber
\eea
Index $\mu$ in the left hand side goes only over $\phi_k$ coordinates.  The last relation can be rewritten as a set of expressions for $A^{\phi_p}$:
\bea\label{TempSTAphi}
A^{\phi_p}=-\sum_{j} \frac{y_j^2-b_j^2}{r^2}\left[\prod_{k\ne j}\frac{b_k^2-y_j^2}{y_k^2-y_j^2}\right] \frac{y_j}{\nu^2-y_j^2}\frac{b_p}{b_p^2-y_j^2}\d_{y_j}\Psi\,.
\eea
The gauge potentials for various polarizations are obtained by taking the limit (\ref{LimitElptc}) in (\ref{TempSTAphi}). As in the case of the $A_{y_k}$ polarizations discussed earlier, the limits depend on the scaling of $\nu$ relative to the sequence  (\ref{LimitElptc}). Before discussing individual cases, we make some general simplifications in (\ref{TempSTAphi}). 

The product appearing in the square bracket in (\ref{TempSTAphi}) does not depend on $\nu$, so we begin with evaluating this bracket. Using relations (\ref{TempST2e}), we find
\bea\label{TempST8a}
\prod_{k\ne a} \frac{y_a^2-b_k^2}{y_a^2-y_k^2}&=&\prod_{k\ne a} 
\frac{b_a^2-b_k^2-(b_a^2-b_{a-1}^2)\xi_a^2}{b_a^2-b_k^2-(b_a^2-b_{a-1}^2)\xi_a^2+(b_k^2-b_{k-1}^2)\xi_k^2}
\nn
&\sim&
\frac{(b_a^2-b_{a-1}^2)(1-\xi_a^2)}{(b_{a-1}^2-b_{a-2}^2)\xi_{a-1}^2}
\prod_{k< a-1} 
\frac{b_a^2-b_k^2}{b_a^2-b_k^2+(b_k^2-b_{k-1}^2)\xi_k^2}\\
&\sim&
\frac{(b_a^2-b_{a-1}^2)(1-\xi_a^2)}{(b_{a-1}^2-b_{a-2}^2)\xi_{a-1}^2}
\prod_{k< a-1} 
\frac{b_{k+1}^2-b_k^2}{(b_k^2-b_{k-1}^2)\xi_k^2}
=(1-\xi_a^2)\prod_{k< a} 
\frac{b_{k+1}^2-b_k^2}{(b_k^2-b_{k-1}^2)\xi_k^2}
\nn
&\sim&\frac{(1-\xi_a^2)(b_a^2-b_{a-1}^2)}{b_1^2-b_0^2}\prod_{k< a} 
\frac{1}{\xi_k^2}\,.\nonumber
\eea
Symbol $\sim$ here refers to the leading contribution in the limit (\ref{LimitElptc}). Furthermore, conversion from $\d_{y_j}$ to $\d_{\xi_j}$ also does not depend on $\nu$:
\bea\label{TempST8b}
\frac{y_j^2-b_j^2}{y_j}\d_{y_j}=\xi_j\d_{\xi_j},\quad j>1;\qquad
\frac{y_1^2-b_1^2}{y_1}\d_{y_1}=-b^2\xi_1^2\frac{1}{y_1}\d_{y_1}=\xi_1\d_{\xi_1}\,.
\eea
Substitution of (\ref{TempST8a}) and (\ref{TempST8b}) into (\ref{TempSTAphi}) leads to the final expression for $A^{\phi_p}$, which is applicable to all values of $\nu$:
\bea\label{TempSTAphi2}
A^{\phi_p}=-\sum_{j} \frac{(1-\xi_j^2)(b_j^2-b_{j-1}^2)}{b^2 r^2}
\left[\prod_{k<j} \frac{1}{\xi_k^2}\right] \frac{y_j^2}{\nu^2-y_j^2}
\frac{b_p}{b_p^2-y_j^2}\xi_j\d_{\xi_j}\Psi\,.
\eea
Let us now construct various polarizations by taking the limit (\ref{LimitElptc}) in (\ref{TempSTAphi2}), while scaling $\nu$ in an appropriate fashion.

As in the case of $A_\xi$ components, we begin with polarizations $A^{\phi_p}$ for $\nu\ne b$, the cyclic counterpart of (\ref{TempSTgauge1}),
\bea\label{STpolrzPhiNu0}
A^{\phi_p}=-\sum_{j}\frac{1-\xi_j^2}{b r^2}
\left[\prod_{k< j} \frac{1}{\xi_k^2}\right]\frac{b^2}{\nu^2-b^2}L_{j,p}\,
\xi_j\d_{\xi_j}\Psi\,.
\eea
Here coefficients $L_{j,p}$ are defined by
\bea
L_{j,p}\equiv\lim \frac{b_j^2-b_{j-1}^2}{b_p^2 -y_j^2}\,,
\eea
and a direct evaluation gives
\bea\label{STdefLp}
L_{j,p}=\left\{
\begin{array}{ll}
\frac{1}{\xi_j^2},&j\le p;\\
\frac{1}{\xi_{j}^2-1},&j=p+1;\\
0,&j>p+1.
\end{array}
\right.
\eea

Additional polarizations are obtained by sending $\nu$ to $b$ at various rates. In particular, for 
$\nu=\pm b_c$, the gauge field has to be rescaled by $\frac{1}{b^2}(b_c^2-b_{c+1}^2)$ as in (\ref{STresldAxi}):
\bea\label{STpolrzPhiNuP}
A^{\phi_p}&=&-\lim \sum_{j} \frac{1-\xi_j^2}{b r^2}
\left[\prod_{k<j} \frac{1}{\xi_k^2}\right] \frac{b_c^2-b_{c+1}^2}{b_c^2-y_j^2}\frac{y_j^2}{b^2}
L_{j,p}\,\xi_j\d_{\xi_j}\Psi\nn
&=&-\sum_{j} \frac{1-\xi_j^2}{b r^2}
\left[\prod_{k<j} \frac{1}{\xi_k^2}\right] n_j N_{j,c}
L_{j,p}\,\xi_j\d_{\xi_j}\Psi\,.
\eea
Here $N_{j,c}$ is given by (\ref{STdefNc}), and we also defined
\bea\label{STdefNnJ}
n_{j}\equiv\lim \frac{y_j^2}{b^2}=\left\{
\begin{array}{ll}
1,&j>1;\\
1-\xi_1^2,&j=1.
\end{array}
\right.
\eea
This concludes evaluation of the gauge field in the static limit, let us now summarize the results.

\bigskip

\noindent
{\bf Summary}

In this appendix we have demonstrated that the static limit (\ref{LimitElptc}) of the separable solution (\ref{AnsGenEvenMagn}) leads to several magnetic polarizations in the Schwarzschild-Tangherlini geometry. The physical consequences of this fact are discussed in section \ref{SecSubCmprSchw}, here we just summarize the technical results. Keeping $\nu\ne b$ in the limit (\ref{LimitElptc}), one arrives at the polarization (\ref{TempSTgauge1}), (\ref{STpolrzPhiNu0}):
\bea\label{STansPolrNu0}
A_{\xi_j}=\frac{i{\nu}}{b^2-\nu^2}\d_{\xi_j}\Psi,\quad
A^{\phi_p}=-\sum_{j}\frac{1-\xi_j^2}{b r^2}
\left[\prod_{k< j} \frac{1}{\xi_k^2}\right]\frac{b^2}{\nu^2-b^2}L_{j,p}\,
\xi_j\d_{\xi_j}\Psi\,.
\eea
Alternatively, setting $\nu=\pm b_c$ and performing an appropriate rescaling of the gauge potential, one arrives at polarizations (\ref{STpolrzPhiNuP})
\bea\label{STansPolrNuP}
\hskip -0.5cm
\nu=\pm b_c:\ A^{(\pm),c}_{\xi_j}=\pm \frac{i}{b}N_{j,c}\,\d_{\xi_j}\Psi,\ A^{\phi_p,c}
=-\sum_{j} \frac{1-\xi_j^2}{b r^2}
\left[\prod_{k<j} \frac{1}{\xi_k^2}\right] n_j N_{j,c}
L_{j,p}\,\xi_j\d_{\xi_j}\Psi\,.
\eea
Functions $(N_{j,c},L_{j,p},n_j)$ entering (\ref{STansPolrNu0})--(\ref{STansPolrNuP}) are defined by (\ref{STdefNc}), (\ref{STdefLp}), (\ref{STdefNnJ}), and label $c$ takes values $c=\{1,\dots,(n-1)\}$.

\end{document}